\documentclass[twocolumn,showpacs,nofootinbib,superscriptaddress]{revtex4-1}

\usepackage[nointlimits]{amsmath}
\usepackage{hyperref}
\usepackage{amsfonts}
\usepackage{amssymb}
\usepackage{amsmath}
\usepackage{amsthm}
\usepackage{latexsym}
\usepackage{graphicx}
\usepackage{float}
\usepackage{color}
\usepackage{layout}
\usepackage[english]{babel}

\usepackage{wrapfig}

\def\LL{\mathcal{L}}
\def\x{\mathbf{x}}
\def\y{\mathbf{y}}
\def\n{\mathbf{n}}
\def\vv{\mathbf{v}}
\def\a{\mathbf{a}}
\def\I{\mathcal{I}}
\def\J{\mathcal{J}}
\def\F{\mathcal{F}}

\def\eps{\varepsilon}
\def\OO{{\cal O}}
\def\VV{{\cal V}}

\begin{document}

\title{Equivalence principle violation in nonminimally coupled gravity and constraints from Lunar Laser Ranging}

\author{Riccardo March}
\email{r.march@iac.cnr.it}
\affiliation{Istituto per le Applicazioni del Calcolo, CNR, Via dei Taurini 19, 00185 Roma, Italy}
\affiliation{INFN - Laboratori Nazionali di Frascati (LNF), Via E. Fermi 40, Frascati 00044 Roma, Italy}

\author{Orfeu Bertolami}
\email{orfeu.bertolami@fc.up.pt}
\affiliation{Departamento de F\'isica e Astronomia,\\Faculdade de Ci\^encias da Universidade do Porto, Rua do Campo Alegre 687, 4169-007 , Porto, Portugal}
\affiliation{Centro de F\'isica das Universidades do Minho e do Porto, Rua do Campo Alegre 687, 4169-007, Porto, Portugal}

\author{Marco Muccino}
\email{Marco.Muccino@lnf.infn.it}
\affiliation{INFN - Laboratori Nazionali di Frascati (LNF), Via E. Fermi 40, Frascati 00044 Roma, Italy}

\author{Simone Dell'Agnello}
\email{simone.dellagnello@lnf.infn.it}
\affiliation{INFN - Laboratori Nazionali di Frascati (LNF), Via E. Fermi 40, Frascati 00044 Roma, Italy}

\date{today}

\begin{abstract}
We analyze the dynamics of the Sun-Earth-Moon system in the context of a particular class of theories of gravity where curvature and matter are nonminimally coupled (NMC). These theories can potentially violate the Equivalence Principle as they give origin to a fifth force and a extra non-Newtonian force that may imply that Earth and Moon fall differently towards the Sun. We show, through a detailed analysis, that consistency with the bound on Weak Equivalence Principle arising from 48 years of Lunar Laser Ranging data, for a range of parameters of the NMC gravity theory, can be achieved via the implementation of a suitable screening mechanism. 
\end{abstract}

\maketitle

\section{Introduction}

General Relativity (GR)  can account for astrophysical and cosmological phenomena such as the flattening of the rotation curves of galaxies and the accelerated expansion of the Universe provided about $95\%$ of the content of the Universe is composed of dark energy and dark matter. In principle, this somewhat puzzling situation can be circumvented without these dark components in the context of some alternative theories of gravity.  Some alternatives include, $f(R)$ gravity \cite{Capoz-1,Carroll,Capoz-2,DeFTs}, in which  the linear Ricci curvature scalar, $R$, in the Einstein-Hilbert action is replaced by a more general function $f (R)$ and theories where a nonminimally coupling (NMC) between curvature and matter is introduced \cite{BBHL}. In the latter, the Einstein-Hilbert action is replaced by two functions of curvature, $f^1(R)$ and $f^2(R)$. The function $f^1(R)$ is analogous to $f(R)$ gravity theory, while the function
$f^2(R)$ multiplies the matter Lagrangian density,  which couples nontrivially geometry and matter \cite{BBHL}. 
This theoretical route has been extensively examined in what concerns dark matter \cite{drkmattgal}, dark energy \cite{curraccel}, inflation \cite{c}, energy density fluctuations \cite{d}, gravitational waves \cite{e}, cosmic virial theorem \cite{f} and black holes \cite{g}. This modified theory has also been examined through the Newton-Schrodinger approach \cite{h,i}.

In a previous work \cite{MPBD} the case of functions $f^1(R),f^2(R)$ analytic at $R=0$ was considered, and
implications of the NMC model were examined via the perturbations to the perihelion precession
by using data from observations of Mercury's orbit. 

It turns out that NMC gravity modifies gravity as it introduces both a Yukawa type fifth force and an extra force which depends on the spatial gradient of the Ricci scalar.
While the Yukawa force is typical also of $f(R)$ gravity, the existence of the extra force is specific of NMC gravity \cite{BBHL,BLP},
as the nonminimal coupling induces a non-vanishing covariant derivative of the energy-momentum tensor. On its hand, the fifth force can give rise to static solutions even though in the absence of pressure \cite{i}.

Constraints to the NMC gravity model with analytic $f^1,f^2$ functions have been computed using the results
of a geophysical experiment in Ref. \cite{MBMBD}. The idea was to consider deviations from Newton's inverse square law in the ocean \cite{Zum}. It was found that the presence of the extra force in a fluid such as seawater imposes more stringent constraints on the NMC gravity model
than the observation of both Mercury's perihelion precession and lunar geodetic precession. 
Hence for the NMC gravity model, 
Solar System constraints are weaker than geophysical constraints.

In  the present paper we look for meaningful Solar System constraints to NMC gravity, for a function $f^2(R)$
which contains a term proportional to $R^\alpha$, with $\alpha<0$, so that $f^2(R)$ is not analytic at $R=0$.
The resulting model has been used in Ref. \cite{drkmattgal} to predict the flattening of the galaxy rotation curves,
and to predict the current accelerated expansion of the Universe \cite{curraccel}.

In Ref. \cite{MPBD}  the method based on the $1/c$ expansion was used to study the NMC gravity model. However, since the $f^2(R)$ function is not analytic, a different nonlinear approach has to be employed. It turns out that in Solar System the above NMC model exhibits a screening mechanism, which is a version of the so called {\it chameleon mechanism} \cite{KW} adapted to the NMC gravity and used to obtain Solar System constraints \cite{MBMGDeA}.

In Ref. \cite{MBMGDeA} the complications of the NMC where considered. These generalise the nonlinear computations of Ref. \cite{HS} for the chameleon mechanism for the gravitational field of the Sun and the corresponding calculation for the the case of $f(R)$ gravity \cite{HS}. Spherical symmetry was considered and the constraints arising from the Cassini measurement of PPN parameter $\gamma$ \cite{Cassini} were used to constrain the parameters of the NMC gravity model. It was shown that the chameleon solution in NMC gravity turns out to be close to GR inside a \emph{screening radius} $r_s$ that has to be large enough,
so that $r_s$ either lies inside the solar convection zone, close to the top of the zone, or it is larger. Deviations from GR are sourced by the fraction of solar
mass, including solar atmosphere, contained in the region with radii $r>r_s$, so that if $r_s$ lies in the top of the convection zone then such deviations are essentially sourced
by a \emph{thin shell} of mass in the convection zone. This is a typical feature of the chameleon mechanism \cite{KW}.

In the present work we extend the nonlinear computations made in Ref. \cite{MBMGDeA} and evaluate the contribution of the interactions in the NMC gravity in the interior of the Sun, Earth and Moon. It is shown that in order to satisfy Solar System constraints for the NMC gravity, the solution of the chameleon mechanism is a suitable scalar function that
must remain close to the minimizer of an effective potential $V_{\rm eff}$ in most of the interior of massive astronomical bodies (Sun, Earth and Moon), so that GR is approximately satisfied \cite{KW,HS}. More specifically, for each astronomical body, the solution has to be close to the minimizer of $V_{\rm eff}$ inside a critical radius, 
the screening radius, which must be determined. If the screening radius is close to the radius of the astronomical body, then the thin shell condition is satisfied and deviations from GR are screened. The potentials of the metric tensor are then expressed in terms of the scalar function.

The chameleon solution for the three-body system is computed by taking into account the appropriate boundary conditions at the boundaries of the
screened zones, and we develop a method of solution based on different linear approximations of the field equations in different zones. The Earth and Moon are modeled by means of
layers of constant density, we solve a Yukawa equation in each layer, a Poisson equation in interplanetary space outside of the screened zones of the three astronomical bodies,
and Laplace equation in the solar neighborhood of the Galaxy.
An analytic solution of the Poisson equation with Dirichlet conditions at the boundaries of the screened zones is computed by means of Green's function which is in turn approximated
by using an extension of the method of images to a system of three spheres. The screening radii of the bodies are computed by solving a system of integral equations which
result from Neumann boundary conditions. 

We compute the equations of motion for the centers of mass of Earth and Moon in the gravitational field of the Sun from first principles, by taking the covariant derivative of the field equations,
then solving the resulting stressed-matter equations of motion.
The Earth and Moon are treated as layered spheres of matter characterized by the energy-momentum tensor of
continuous bodies in a hydrostatic state of stress. The equations of motion exhibit the presence of both the fifth force and the extra force which give rise to deviations from GR.
Such deviations are sourced by the masses contained in the thin shells of the bodies which in turn depend on the density profiles of the bodies themselves, 
so that the Earth and Moon fall towards the Sun with different accelerations giving rise to a violation of the Weak Equivalence Principle (WEP). 
Such a violation takes place in modified gravity theories which exhibit the chameleon mechanism \cite{Hui-NS,Kraiselburd}. 
The WEP violation in the Sun-Earth-Moon system makes it possible to constrain the parameters of the NMC gravity model by means of Lunar Laser Ranging (LLR) measurements,
which is the result achieved in the present paper by resorting to most recent LLR data \cite{Visw-Fienga}.

The paper is organized as follows. In Section II the NMC gravity model is presented. In Section III we consider the field equations inside and around the astronomical bodies (Sun, Earth, Moon) for the chameleon mechanism. In Section IV we compute the solutions for the interior of the astronomical bodies and in the outskirts of the Solar System. 
In Section V we determine the boundary conditions for the chameleon solution at the boundaries of the screened zones, then the gravitational field of the astronomical bodies
is evaluated using Green's function and the method of images for a system of spheres. The integral equations which determine the screening radii are also found.
In Section VI the dynamics of the continuous bodies is considered in order to compute the fifth force and the extra force inside the bodies and to evaluate a yielding jump in the pressure. Both the fifth force and the extra force are shown to be negligible inside the screening radii. In Section VII the acceleration of Earth and Moon due to the fifth force is computed. In Section VIII the acceleration of Earth and Moon due to the extra NMC force is computed. 
In Section IX the potential violation of the Equivalence Principle is quantified and in section X the bound on the WEP arising from the LLR data is used to constrain the parameters of the NMC gravity model. It is shown that the screening mechanism is successful in ensuring that the bound on the WEP can be respected for a suitable range of model parameters. 
Finally, our conclusions are presented in Section XI. Appendices A to E contain the technical details of the calculations needed to obtain the various results of the paper.

\section{Nonminimally coupled gravity}

The action functional of NMC gravity theory here considered is of the form \cite{BBHL},
\begin{equation}\label{action-funct}
S = \int \left[\frac{1}{2}f^1(R) + [1 + f^2(R)] \LL_m \right]\sqrt{-g} \, d^4x,
\end{equation}
where $f^i(R)$ (with $i=1,2$) are functions of the Ricci curvature scalar $R$, $\LL_m$ is the Lagrangian
density of matter, and $g$ is the metric determinant.
The Einstein-Hilbert action of GR is recovered by taking
\begin{equation}
f^1(R) = \frac{c^4}{8\pi G}R, \qquad f^2(R) = 0,
\end{equation}
where $G$ is Newton's gravitational constant. We work in the Jordan frame throughout this paper.

The first variation of the action functional with respect to the metric $g_{\mu\nu}$ yields the field equations:
\begin{eqnarray}\label{field-eqs}
& &\left(f^1_R + 2f^2_R \LL_m \right) R_{\mu\nu} - \frac{1}{2} f^1 g_{\mu\nu}  \\
& &= \left(\nabla_\mu \nabla_\nu -g_{\mu\nu} \square \right) \left(f^1_R + 2f^2_R \LL_m \right)
+ \left(1 + f^2 \right) T_{\mu\nu}, \nonumber
\end{eqnarray}
where $f^i_R \equiv df^i\slash dR$. The trace of the field equations is given by
\begin{eqnarray}\label{trace}
& &\left( f^1_R + 2f^2_R \LL_m \right) R - 2f^1 + 3\square f^1_R +
6\square\left( f^2_R \LL_m \right)  \nonumber\\
& &= \left( 1 + f^2 \right) T,
\end{eqnarray}
where $T$ is the trace of the energy-momentum tensor $T_{\mu\nu}$.

In NMC gravity the energy-momentum tensor of matter is not covariantly conserved \cite{multiscalar,Sotiriou1}: applying the Bianchi identities to Eq. (\ref{field-eqs}), it follows
\begin{equation}\label{covar-div-1}
\nabla_\mu T^{\mu\nu} = \frac{f^2_R }{ 1 + f^2} ( g^{\mu\nu} \LL_m - T^{\mu\nu} ) \nabla_\mu R,
\end{equation}
a property which gives rise to an extra force which is added to the fifth force which is typical of $f(R)$ gravity theory. We will find that this extra force
has a negligible effect on the motion of Earth and Moon for values of NMC gravity parameters of astrophysical and cosmological interest, 
while such a force is expected to have important effects at the galactic scale.

\subsection{Metric and energy-momentum tensors}\label{sec:metric-stress-tensor}

We use the following notation for indices of tensors:
Greek letters denote space-time indices ranging from 0 to 3,
whereas Latin letters denote spatial indices ranging from 1 to 3.
Cartesian three-vectors are indicated by boldface type and scalar product is indicated by a dot.
The signature of the metric tensor is $(-,+,+,+)$.

The Sun is modeled as a static spherically symmetric distribution of matter, while the Earth and Moon are modeled as orbiting spherically symmetric bodies.
The metric tensor which describes the spacetime in the Sun-Earth-Moon system is given, in the Newtonian gauge, by
\begin{eqnarray} \label{metric}
ds^2= &-& \left[1-2\Phi(\x,t)+2\Psi(\x,t)\right]c^2dt^2 \nonumber\\
&+& \left[1+2\Phi(\x,t)\right] \delta_{ij}dx^idx^j,
\end{eqnarray}
where the potentials $\Phi$ and $\Psi$ are perturbations of the Minkowski metric of order $\OO(1/c^2)$.

The Sun is considered as a perfect fluid in hydrostatic equilibrium, while the Earth and Moon are approximately described as continuous bodies
in a hydrostatic state of stress, {\em i.e.}, the normal stresses are equal to the pressure and shear stresses are neglected \cite{Turcotte}.
In this approximation the components of the energy-momentum tensor, to the relevant order for our computations, for all the astronomical bodies
are given by (Ref. \cite{Wi}, Chapter 4.1):
\begin{eqnarray}
T^{00} &=& \rho c^2 + \OO\left(1\right), \label{T-00}\\
T^{0i} &=& \rho c v^i + \OO\left(\frac{1}{c}\right), \label{T-0i}\\
T^{ij} &=& \rho v^i v^j + p\delta_{ij} + \OO\left(\frac{1}{c^2}\right), \label{T-ij}
\end{eqnarray}
where matter is characterized by density $\rho$, velocity field $v^i$, and pressure $p$.
The trace of the energy-momentum tensor is
\begin{equation}
T = -\rho c^2 + \OO\left(1\right).
\end{equation}
In the present paper we use $\LL_m = -\rho c^2+\OO(1)$ for the Lagrangian density of matter \cite{BLP}.

\subsection{Choice of functions $f^1(R)$ and $f^2(R)$}

We choose the following functions:
\begin{equation}\label{f1-f2-specific}
f^1(R) = \frac{c^4}{8\pi G} R, \qquad f^2(R) = q R^\alpha,\quad \alpha<0,
\end{equation}
where the function $f^1(R)$ corresponds to GR and $q,\alpha$ are real numbers that have to be considered as parameters of the NMC model of gravity.

The functions (\ref{f1-f2-specific}) have been used in Ref. \cite{drkmattgal} to model the rotation curves of galaxies,
and in Ref. \cite{curraccel} to model the current accelerated expansion of the Universe. 

\section{Approximation of the field equations}

We approximate the field equations (\ref{field-eqs}) and (\ref{trace}) taking into account that
the metric potentials $\Phi$ and $\Psi$ are small perturbations of the Minkowski metric, so that we neglect 
the higher order terms that include products of potentials or their derivatives, and cross-products between their derivatives and the potentials.
Moreover, velocities of bodies are negligible with respect to $c$.
By computing the Ricci tensor and the Ricci curvature scalar $R$ from the metric (\ref{metric}), it then follows that the functions $\Phi$ and $\Psi$ satisfy the following equations:
\begin{eqnarray}
& &\nabla^2(\Phi+\Psi) = -\frac{R}{2}, \label{Phi-Psi-equation}\\
& &\nabla^2\Psi = \frac{1}{2}\left( R_{00} - \frac{R}{2} \right), \label{Psi-equation}
\end{eqnarray}
where $\nabla^2$ denotes Laplace operator in flat three-dimensional space.
We introduce the scalar field $\eta$ which is a function of curvature $R$ also explicitly depending on spacetime coordinates $(\x,t)$ through mass density:
\begin{equation}\label{eta-definition}
\eta = \eta(\x,t,R) = f^1_R -2f^2_R \, \rho(\x,t)c^2.
\end{equation}
Using the metric (\ref{metric}) and Eqs. (\ref{Phi-Psi-equation}-\ref{Psi-equation}), the time-time component of the field equations (\ref{field-eqs}) at leading order is given by
\begin{eqnarray}\label{time-time-eq}
& &\eta(-\nabla^2\Phi+\nabla^2\Psi)+\frac{f^1}{2}(1-2\Phi+2\Psi)\nonumber\\
&=&\nabla_0\nabla_0\eta+(1-2\Phi+2\Psi)\square\eta+(1+f^2)\rho c^2,
\end{eqnarray}
and the trace (\ref{trace}) of the field equations becomes
\begin{equation}\label{trace-rewrit}
\eta R -2f^1 + 3\square\eta = (1+f^2)T.
\end{equation}
Eliminating $\square\eta$ in Eq. (\ref{time-time-eq}) by means of the trace equation, and $\nabla^2\Phi$ by means of Eq. (\ref{Phi-Psi-equation}), we obtain
\begin{eqnarray}\label{Psi-equation-eta}
& &\eta\left(2\nabla^2\Psi+\frac{5}{6}R\right) - \frac{f^1}{6}(1-2\Phi+2\Psi)\\
&=&\nabla_0\nabla_0\eta+\frac{2}{3}\eta R(\Phi-\Psi) + \frac{2}{3}(1+f^2)\rho c^2(1+\Phi-\Psi).\nonumber
\end{eqnarray}
We require the functions $f^1(R)$ and $f^2(R)$ to satisfy the following conditions:
\begin{equation}\label{cond-f1-f2}
\left\vert \frac{8\pi G}{c^4} \, \frac{f^1}{R} - 1 \right\vert \ll 1, \qquad \left\vert f^2 \right\vert \ll 1,
\end{equation}
and the following condition on the derivatives of $f^1$ and $f^2$ with respect to $R$,
\begin{equation}\label{cond-f1R-f2R}
\left\vert \frac{8\pi G}{c^4}\eta -1 \right\vert \ll 1.
\end{equation}
The conditions (\ref{cond-f1-f2}) mean that the Lagrangian density in Eq. (\ref{action-funct}) is
a small perturbation of the Lagrangian of GR. While the first of conditions (\ref{cond-f1-f2}) is trivial for the choice (\ref{f1-f2-specific}) of function $f^1$,
the other two conditions will be verified a posteriori.
Using such conditions Eq. (\ref{Psi-equation-eta}) is approximately given by:
\begin{eqnarray}
\nabla^2\Psi+\frac{R}{3} &=& \frac{8\pi G}{3c^2}\rho + \frac{1}{3}\left(\frac{R}{2} + \frac{8\pi G}{c^2}\rho\right)(\Phi-\Psi) \nonumber\\
&+&\frac{4\pi G}{c^4}\nabla_0\nabla_0\eta,
\end{eqnarray}
from which, keeping terms of order $\OO(1/c^2)$, we find
\begin{equation}\label{Psi-equation-approx}
\nabla^2\Psi = \frac{1}{3}\left(\frac{8\pi G}{c^2}\rho - R\right),
\end{equation}
and, using Eq. (\ref{Phi-Psi-equation}),
\begin{equation}\label{Phi-equation-approx}
\nabla^2\Phi = -\frac{4\pi G}{c^2}\rho + \frac{1}{6}\left( \frac{8\pi G}{c^2}\rho - R \right).
\end{equation}

\subsection{Equation for the scalar field $\eta$}\label{sec:scalar-eta-equation}

Equations (\ref{Psi-equation-approx}) and (\ref{Phi-equation-approx}) have to be completed with an equation for the scalar field $\eta$.
Neglecting cross-products between the potentials $\Phi,\Psi$ and their derivatives we have at leading order
\begin{eqnarray}
\square\eta &=& (1-2\Phi)\nabla^2\eta - \frac{1}{c^2}(1+2\Phi-2\Psi)\frac{\partial^2\eta}{\partial t^2} + \frac{\partial\Psi}{\partial x_i}\frac{\partial\eta}{\partial x^i} \nonumber\\
&\approx& \nabla^2\eta.
\end{eqnarray}
Using conditions (\ref{cond-f1-f2}-\ref{cond-f1R-f2R}) the trace equation (\ref{trace-rewrit}) is approximately given by
\begin{equation}\label{trace-approx}
\nabla^2\eta = \frac{c^4}{24\pi G}\, R - \frac{1}{3}\rho c^2.
\end{equation}
Note that the equations (\ref{Psi-equation-approx}), (\ref{Phi-equation-approx}) and (\ref{trace-approx}) are formally the same as the ones found in Ref. \cite{HS} for $f(R)$ gravity
in the special case of spherical symmetry, with the difference that the scalar field $\eta$, defined in (\ref{eta-definition}), depends explicitly on $(\x,t)$ through the multiplication
by $\rho(\x,t)$ due to the nonminimal coupling. Such a dependence on $\rho(\x,t)$ will be exploited in the sequel.

By introducing a potential function $V=V(\x,t,\eta)$ and an effective potential $V_{\rm eff}$ as in Refs. \cite{KW,HS},
\begin{equation}\label{potential-V-def}
\frac{\partial V}{\partial\eta}= \frac{c^4}{24\pi G}\,\omega(\eta,\rho), \qquad V_{\rm eff}=V-\frac{1}{3}\rho c^2\eta,
\end{equation}
where the function $\omega(\eta,\rho)$ is obtained by solving the equation (\ref{eta-definition}) with respect to $R$, the equation for the scalar field $\eta$ becomes
\begin{equation}\label{eta-equation-potential}
\nabla^2 \eta=\frac{\partial V_{\rm eff}}{\partial\eta}.
\end{equation}
Note that for the choice (\ref{f1-f2-specific}) of functions $f^1(R),f^2(R)$ the function $\omega(\eta,\rho)$ exists and it is unique.
The effective potential has an extremum which corresponds to the GR solution
\begin{equation}\label{high-curv-sol}
R=\omega(\eta,\rho) = \frac{8\pi G}{c^2}\rho,
\end{equation}
and we require that such an extremum is a minimum \cite{KW,HS}, which yields the condition
\begin{equation}
\frac{\partial^2 V_{\rm eff}}{\partial\eta^2} = \frac{c^4}{24\pi G}\,\frac{1}{\eta_R} \geq 0,
\end{equation}
with $\eta_R = f^1_{RR}-2f^2_{RR}\rho c^2$ and $R=\omega(\eta,\rho)$, 
the double subscript in $f^i_{RR}$ denoting second derivative with respect to $R$.
For the choice (\ref{f1-f2-specific}) of $f^1,f^2$ such a minimum condition requires
\begin{equation}\label{NMC-stability-condition}
\alpha(\alpha-1)q[\omega(\eta,\rho)]^{\alpha-2} \leq 0,
\end{equation}
from which, for $q\neq 0$ and $R=\omega(\eta,\rho)>0$, it follows
\begin{equation}\label{q-sign}
\alpha <0 \implies q<0,
\end{equation}
which is an application of a general stability condition against Dolgov-Kawasaki instability in NMC gravity found in Refs.
\cite{Faraoni,BeSeq}.

At the minimum of the effective potential $V_{\rm eff}$ we set
\begin{equation}\label{lambda-definition}
\frac{\partial^2 V_{\rm eff}}{\partial\eta^2} = \frac{1}{\lambda^2} >0, \quad \mbox{with}
\quad \omega(\eta,\rho) = \frac{8\pi G}{c^2}\rho,
\end{equation}
where $\lambda=\lambda(\rho)>0$ has dimension of length and depends on mass density.
For the choice (\ref{f1-f2-specific}) of functions of curvature we have
\begin{equation}\label{lambda-expression}
\lambda^2 = 6 q \alpha(1-\alpha) \left(\frac{8\pi G}{c^2}\rho\right)^{\alpha-1},
\end{equation}
and the function $\lambda(\rho)$ decreases as density $\rho$ increases. In the next sections we consider only the choice
(\ref{f1-f2-specific}) of functions $f^1(R),f^2(R)$, and we compute an analytic approximate solution of the equation
for the scalar field $\eta$.

\section{Solution in the interior of bodies and in interplanetary space}\label{sec:interior-bodies-solut}

Equation (\ref{eta-equation-potential}) for the scalar field $\eta$ is typical of chameleon theories of gravity \cite{KW,HS}. The difference with respect to
other chameleon theories such as $f(R)$ gravity consists in the explicit dependence of $\partial V/\partial\eta$ on $\rho(\x,t)$ due to the nonminimal coupling
(see the discussion in Ref. \cite{MBMGDeA}).

In order to satisfy the stringent bounds from Solar System experiments on modified gravity, a chameleon theory requires the solution $\eta$ to remain close
to the minimizer of the effective potential $V_{\rm eff}$ in most of the interior of massive astronomical bodies such as the Sun, Earth and Moon, so that GR
is approximately satisfied \cite{KW,HS}. More precisely, in each body $\eta$ has to be close to the minimizer of $V_{\rm eff}$ inside a critical radius, 
called the screening radius, that has to be determined. If the screening radius is close to the radius of the astronomical body for each body, then
the thin shell condition is satisfied and deviations from GR are screened.
In the following we denote by $r_S,r_E$ and $r_M$ the screening radii of the Sun, Earth and Moon, respectively.

For each body we compute a solution inside the screening radius by using the approximation of spherical symmetry around the center.
In Sec. \ref{sec:screen-radii} we will match the interior solution with the solution outside of the screening radii and we will show that the approximation of interior
spherical symmetry can be used, provided that the distances between the astronomical bodies are much greater
than the radii, a condition which is satisfied for the Sun-Earth-Moon system.

For a spherically symmetric solution the finiteness of $\nabla^2\eta$ imposes the boundary condition
\begin{equation}\label{bound-cond-0}
\frac{d\eta}{dr} = 0, \quad\mbox{at } r=0,
\end{equation}
at the center of each astronomical body, where the variable $r$ is distance from the center.

\subsection{Solution in the Sun's interior}

The Sun is modeled as a static spherically symmetric distribution of matter with density $\rho_S=\rho_S(r)$ where $r$ is distance from the center.
A model of mass density profile for the Sun has been used in Ref. \cite{MBMGDeA}  and the parts of the model that will be used in the sequel
are reported in Section \ref{sec:Sun-density-model} of Appendix A.

Since the effective potential has an extremum which corresponds to the GR solution, $\partial V_{\rm eff}/\partial\eta=0$, then expression
(\ref{high-curv-sol}) of curvature yields an exact solution of the equation (\ref{trace-approx}) only
if $\nabla^2\eta=0$. Under spherical symmetry, the only harmonic function which satisfies
the boundary condition (\ref{bound-cond-0}) is a constant, however, one can check that the solution $\eta=constant$ implies
that density $\rho_S(r)$ must also be constant, which is not the case for the Sun's interior \cite{MBMGDeA}. 

Though the GR solution is not an exact solution of Eq. (\ref{eta-equation-potential}), it is an
an approximate solution if the following consistency condition is satisfied for $r<r_S$ \cite{HS,MBMGDeA}:
\begin{equation}\label{necess-cond}
\left\vert \nabla^2\left(\eta(r,R=8\pi G\rho_S(r)\slash c^2)\right) \right\vert \ll \frac{1}{3}\rho_S c^2,
\end{equation}
with $r_S<R_\odot$, where $R_\odot$ is the Sun's radius.
Computing the Laplacian of $\eta$ according to (\ref{eta-definition}), the consistency condition has been evaluated in
Ref. \cite{MBMGDeA} and reads
\begin{equation}\label{consist-condition}
\lambda^2 \left\vert \frac{\alpha}{1-\alpha}\nabla^2\rho_S - \frac{\alpha+1}{\rho_S}\left(\frac{d\rho_S}{dr}\right)^2 \right\vert \ll \rho_S.
\end{equation}
Then, using definition (\ref{eta-definition}) of $\eta$, the GR expression of curvature (\ref{high-curv-sol}), 
and formula (\ref{lambda-expression}) of $\lambda^2$, we have the following approximate solution:
\begin{eqnarray}\label{eta-inner-interior-complete}
\eta &\approx& \frac{c^4}{8\pi G} -2 \alpha q\left(\frac{8\pi G}{c^2}\rho_S\right)^{\alpha-1} \rho_S c^2 \nonumber\\
&=& \frac{c^4}{8\pi G} -\frac{\lambda^2(\rho_S)}{3(1-\alpha)} \rho_S c^2,
\end{eqnarray}
which holds for $r<r_S$. The boundary condition (\ref{bound-cond-0}) is satisfied provided that the Sun's density model has the property
$d\rho_S/dr=0$ at the center.

Eventually, by using the Sun density profile in Section \ref{sec:Sun-density-model} of Appendix A, we find $\lambda(\rho_S) \ll R_\odot$,
particularly $\lambda(\rho_S)$ turns out to be a negligible quantity for $|\alpha|$ not too close to zero.

\subsection{Solution in the Earth's interior}\label{sec:sol-inside-Earth}

Differently from the Sun, the density profile of the Earth's interior is conveniently modeled by resorting to density discontinuities, detected by seismology,
such as the Mohorovi\v{c}i\'c discontinuity, or Moho, at the boundary between the crust and the mantle. On the two sides of a density discontinuity
the values of $\eta$ that minimize $V_{\rm eff}$ are different, however, since the function $\eta$ and its gradient have to be continuous in order to
guarantee the existence of $\nabla^2\eta$, then at the discontinuity the solution $\eta$ has to interpolate between the two minimizers so that it cannot
be close to both minimizers. Hence, the solution locally deviates from GR and one has to check if such a deviation is small enough in such a way that
screening takes place anyway.

The Earth is modeled as a spherically symmetric distribution of matter with density $\rho_E=\rho_E(r)$, where $r$ is distance from the center,
and axial rotation of Earth is neglected. An average Earth model \cite{PREM} is considered, and the planet is divided into four homogeneous regions
separated by spherical surfaces of density discontinuities: ocean layer, crust, mantle and core. Numerical values of density and radii of 
discontinuity surfaces are reported in Section \ref{sec:Earth-density-model} of Appendix A.

Since, for $r<r_E$, the solution $\eta$ has to remain close to the minimizer of $V_{\rm eff}$, then
inside each region of constant density the derivative of the potential $V$ is approximated by
\begin{equation}
\frac{\partial V}{\partial\eta}(\eta,\rho_E) \approx \frac{\partial V}{\partial\eta}(\eta_{\rm min},\rho_E) + \frac{\partial^2 V}{\partial\eta^2}(\eta_{\rm min},\rho_E)(\eta-\eta_{\rm min}),
\end{equation}
where $\eta_{\rm min}$ is the minimizer of $V_{\rm eff}$ in the considered region. Since
\begin{equation}
\frac{\partial V}{\partial\eta}(\eta_{\rm min},\rho_E) = \frac{1}{3}\rho_E c^2,
\end{equation}
Eq. (\ref{eta-equation-potential}) for $\eta$ becomes
\begin{equation}\label{Yukawa-equation}
\nabla^2\eta \approx \frac{1}{\lambda^2(\rho_E)}(\eta-\eta_{\rm min}).
\end{equation}
Since in each region the value of density $\rho_E$ is constant, using (\ref{lambda-expression}) $\lambda^2(\rho_E)$ is also constant,
then Eq. (\ref{Yukawa-equation}) admits a closed-form solution that, for a given screening radius $r_E$ of Earth,
is completely determined by conditions of continuity of $\eta$ and its radial derivative at the discontinuity surfaces,
and by the following condition at $r_E$ given in Ref. \cite{TaTsu}:
\begin{equation}
\frac{1}{\lambda^2(\rho_E)}\left[\eta(r_E)-\eta_{\rm min}\right] = -\frac{1}{3}\rho_E c^2,
\end{equation}
where $r_E<R_\oplus$, with $R_\oplus$ the Earth's radius, and $\rho_E$ is evaluated at the uppermost layer of the Earth's interior model, {\it i.e.}, the ocean layer.

Note that the Sun's interior is modeled by means of a continuously varying density profile (see Section \ref{sec:Sun-density-model} of Appendix A),
hence Eq. (\ref{Yukawa-equation}) in the case of Sun does not admit a closed-form solution, so that we have used a different approximation of the solution $\eta$
given by Eq. (\ref{eta-inner-interior-complete}). Conversely, the presence of discontinuity surfaces in the Earth's interior
prevents us from using a consistency condition of the type (\ref{consist-condition}) where derivatives of density are involved, so that we resorted to
a piecewise constant density profile which permits us to compute an analytic solution that can be proved to be close to the minimizer of the effective potential,
except at the density discontinuities.

In the sequel $R_{E,c}$ is the outer radius of the crust, $R_{E,m}$ is the outher radius of the mantle, and $R_{E,n}$
is the radius of the core (nucleus). The mass densities of ocean, crust, mantle and core are denoted by $\rho_{E,w},\rho_{E,c},\rho_{E,m}$ and $\rho_{E,n}$, respectively.
The corresponding values of $\lambda(\rho_E)$ are denoted by $\lambda_{E,w},\lambda_{E,c},\lambda_{E,m}$ and $\lambda_{E,n}$. Analogously, the values of
$\eta$ minimizing the effective potential in the various layers are denoted by $\eta_{E,w},\eta_{E,c},\eta_{E,m}$ and $\eta_{E,n}$, and they are given by
\begin{equation}\label{eta-minimizer-Earth}
\eta_{\rm min} =  \frac{c^4}{8\pi G} -\frac{\lambda^2(\rho_E)}{3(1-\alpha)} \rho_E c^2.
\end{equation}
By using the numerical values of density of the various Earth's layers, we find in Section \ref{sec:Earth-density-model} of Appendix A $\lambda(\rho_E) \ll R_\oplus$,
particularly $\lambda(\rho_E)$ turns out to be a completely negligible quantity for $|\alpha|$ not too close to zero.

The expression of the analytic solution of Eq. (\ref{Yukawa-equation}) in the various Earth's layers is cumbersome, nevertheless,
it admits a manageable approximation, that guarantees continuity of $\eta$ and approximate continuity of its derivative, and it is given in the following.

{\bf Ocean layer}. The ocean and seas cover $70.8\%$ of the surface of the Earth, so that we approximate the uppermost layer with seawater.
The approximate solution for $R_{E,c}\leq r \leq r_E$ is given by
\begin{eqnarray}\label{eta-sol-Earth-ocean}
& &\eta(r) \approx \eta_{E,w} - \frac{1}{r}\left[ \frac{1}{3}r_E \lambda^2_{E,w}\rho_{E,w}c^2\exp\left(\frac{r-r_E}{\lambda_{E,w}}\right) \right. \nonumber\\
&-& \left. \frac{\lambda_{E,w}\slash\lambda_{E,c}}{1+\lambda_{E,w}\slash\lambda_{E,c}} R_{E,c}(\eta_{E,c}-\eta_{E,w})\exp\left(\frac{R_{E,c}-r}{\lambda_{E,w}}\right)\right]. \nonumber\\
\end{eqnarray}
If $r_E$ is close to $R_\oplus$ then the thin shell condition for Earth is satisfied, and $r_E-R_{E,c}\gg\lambda_{E,w}$.
It turns out that the difference $|\eta(r)-\eta_{E,w}|$ between $\eta$ and the minimizer is exponentially suppressed for 
$r_E-r \gg \lambda_{E,w}$ and $r-R_{E,c}\gg \lambda_{E,w}$, hence in most of the ocean layer, due to the smallness of $\lambda_{E,w}$.
The value of $\eta$ at the boundary between seawater and the oceanic crust is
\begin{equation}\label{eta-seaw-crust}
\eta(R_{E,c}) \approx \eta_{E,w} + \frac{\lambda_{E,w}\slash\lambda_{E,c}}{1+\lambda_{E,w}\slash\lambda_{E,c}}(\eta_{E,c}-\eta_{E,w}),
\end{equation}
and we see that the solution $\eta$ interpolates between the minimizers $\eta_{E,w}$ and $\eta_{E,c}$, hence deviating from GR in a thin shell of thickness
of order $\lambda_{E,w}$. In Secs. \ref{sec:fifth-force-inside-sr} and \ref{sec:extra-force-inside-sr} we will prove that the resulting perturbation of the Newtonian gravitational force,
hence of hydrostatic equilibrium, is negligible.

{\bf Crust}. The approximate solution for $R_{E,m}\leq r \leq R_{E,c}$ is given by
\begin{eqnarray}
& &\eta(r) \approx \eta_{E,c} - \frac{1}{r}\left[ \frac{R_{E,c}(\eta_{E,c}-\eta_{E,w})}{1+\lambda_{E,w}\slash\lambda_{E,c}}\exp\left(\frac{r-R_{E,c}}{\lambda_{E,c}}\right) \right. \nonumber\\
&-& \left. \frac{\lambda_{E,c}\slash\lambda_{E,m}}{1+\lambda_{E,c}\slash\lambda_{E,m}} R_{E,m}(\eta_{E,m}-\eta_{E,c})\exp\left(\frac{R_{E,m}-r}{\lambda_{E,c}}\right)\right]. \nonumber\\
\end{eqnarray}
Again, it turns out that the difference between $\eta$ and the minimizer $\eta_{E,c}$ is exponentially suppressed in most of the crust.
The value of $\eta$ at Moho, the discontinuity between the crust and the mantle, is
\begin{equation}
\eta(R_{E,m}) \approx \eta_{E,c} + \frac{\lambda_{E,c}\slash\lambda_{E,m}}{1+\lambda_{E,c}\slash\lambda_{E,m}}(\eta_{E,m}-\eta_{E,c}),
\end{equation}
a formula analogous to Eq. (\ref{eta-seaw-crust}). In the crust the solution $\eta$ deviates from GR in thin shells of thickness of order $\lambda_{E,c}$
adjacent to the upper boundary at $r=R_{E,c}$ and to the lower boundary at $r=R_{E,m}$, respectively. 
The resulting perturbation will turn out to be again negligible, so that screening takes place.

{\bf Mantle}. The approximate solution for $R_{E,n}\leq r \leq R_{E,m}$ is given by
\begin{eqnarray}
\eta(r) \approx \eta_{E,m} - \frac{1}{r}\left[ \frac{R_{E,m}(\eta_{E,m}-\eta_{E,c})}{1+\lambda_{E,c}\slash\lambda_{E,m}}\exp\left(\frac{r-R_{E,m}}{\lambda_{E,m}}\right) \right. \nonumber\\
-\left. \frac{\lambda_{E,m}\slash\lambda_{E,n}}{1+\lambda_{E,m}\slash\lambda_{E,n}} R_{E,n}(\eta_{E,n}-\eta_{E,m})\exp\left(\frac{R_{E,n}-r}{\lambda_{E,m}}\right)\right]. \nonumber\\
\end{eqnarray}
The properties of the solution in the mantle are analogous to the properties in the crust.

{\bf Core}. The approximate solution for $0<r \leq R_{E,n}$ is given by
\begin{eqnarray}
\eta(r) &\approx& \eta_{E,n} - 2\frac{R_{E,n}}{r}\frac{(\eta_{E,n}-\eta_{E,m})}{1+\lambda_{E,m}\slash\lambda_{E,n}}\exp\left(-\frac{R_{E,n}}{\lambda_{E,n}}\right)\nonumber\\
&\times& \sinh\left(\frac{r}{\lambda_{E,n}}\right).
\end{eqnarray}
The difference $|\eta(r)-\eta_{E,n}|$ between $\eta$ and the minimizer is exponentially suppressed for 
$R_{E,n}-r \gg \lambda_{E,n}$, hence in the whole core except in a thin shell of thickness of order $\lambda_{E,n}$ adjacent
to the boundary of the core.
The approximate solution in the core satisfies the boundary condition (\ref{bound-cond-0}).

Using expression (\ref{lambda-expression}) of $\lambda^2$ we see that, for given values of NMC gravity parameters $\alpha$ and $q$,
the approximate solution in the Earth's interior is completely determined in all the layers except the ocean layer where it
depends on the screening radius $r_E$, hence it is determined everywhere once the screening
radius is determined.

\subsection{Solution in the Moon's interior}

The model of the lunar interior is analogous to the Earth's model.
The Moon is modeled as a spherically symmetric distribution of matter with density $\rho_M=\rho_M(r)$, where $r$ is distance from the center,
and the satellite is divided into three homogeneous regions
separated by spherical surfaces of density discontinuities: crust, mantle and core. Numerical values of density and radii of 
discontinuity surfaces are reported in Section \ref{sec:Moon-density-model} of Appendix A.
The length $\lambda(\rho_M)$ turns out to be again a completely negligible quantity.

In the sequel $R_M$ is the Moon's radius, $R_{M,m}$ is the outher radius of the mantle, and $R_{M,n}$
is the radius of the core. The mass densities of crust, mantle and core are denoted by $\rho_{M,c},\rho_{M,m}$ and $\rho_{M,n}$, respectively.
The corresponding values of $\lambda(\rho_M)$ are denoted by $\lambda_{M,c},\lambda_{M,m}$ and $\lambda_{M,n}$. Analogously, the values of
$\eta$ minimizing the effective potential in the various layers are denoted by $\eta_{M,c},\eta_{M,m}$ and $\eta_{M,n}$.

The approximate solution for $\eta$ is analogous to the one found for Earth.

{\bf Crust}. The approximate solution for $R_{M,m}\leq r \leq r_M$ is given by
\begin{eqnarray}
& &\eta(r) \approx \eta_{M,c} - \frac{1}{r}\left[ \frac{1}{3}r_M \lambda^2_{M,c}\rho_{M,c}c^2\exp\left(\frac{r-r_M}{\lambda_{M,c}}\right) \right. \nonumber\\
&-&\left. \frac{\lambda_{M,c}\slash\lambda_{M,m}}{1+\lambda_{M,c}\slash\lambda_{M,m}} R_{M,m}(\eta_{M,m}-\eta_{M,c})\exp\left(\frac{R_{M,m}-r}{\lambda_{M,c}}\right)\right]. \nonumber\\
\end{eqnarray}
The properties of the solution are analogous to the ones in the ocean layer of Earth.
The value of $\eta$ at the lunar Moho is
\begin{equation}
\eta(R_{M,m}) \approx \eta_{M,c} + \frac{\lambda_{M,c}\slash\lambda_{M,m}}{1+\lambda_{M,c}\slash\lambda_{M,m}}(\eta_{M,m}-\eta_{M,c}).
\end{equation}

{\bf Mantle}. The approximate solution for $R_{M,n}\leq r \leq R_{M,m}$ is given by
\begin{eqnarray}
\eta(r) \approx \eta_{M,m} - \frac{1}{r}\left[ \frac{R_{M,m}(\eta_{M,m}-\eta_{M,c})}{1+\lambda_{M,c}\slash\lambda_{M,m}}\exp\left(\frac{r-R_{M,m}}{\lambda_{M,m}}\right) \right. \nonumber\\
-\left. \frac{\lambda_{M,m}\slash\lambda_{M,n}}{1+\lambda_{M,m}\slash\lambda_{M,n}} R_{M,n}(\eta_{M,n}-\eta_{M,m})\exp\left(\frac{R_{M,n}-r}{\lambda_{M,m}}\right)\right]. \nonumber\\
\end{eqnarray}

{\bf Core}. The approximate solution for $0<r \leq R_{M,n}$ is given by
\begin{eqnarray}\label{eta-Moon-core}
\eta(r) &\approx& \eta_{M,n} - 2\frac{R_{M,n}}{r}\frac{(\eta_{M,n}-\eta_{M,m})}{1+\lambda_{M,m}\slash\lambda_{M,n}}\exp\left(-\frac{R_{M,n}}{\lambda_{M,n}}\right)\nonumber\\
&\times& \sinh\left(\frac{r}{\lambda_{M,n}}\right).
\end{eqnarray}
The approximate solution in the core satisfies the boundary condition (\ref{bound-cond-0}).
Again, the approximate solution in the lunar interior is completely determined once the Moon's screening
radius $r_M$ is determined.

\subsection{Solution in the outskirts of the Solar System}\label{Sec:outskirts}

We assume that in the solar neighborhood of the Galaxy the field $\eta$ is close to the minimizer of the effective potential $V_{\rm eff}$,
so that the spacetime curvature $R$ is approximately given by the GR solution \cite{MBMGDeA}.
This assumption implies that the Milky Way is screened within a distance
of about $8\,{\rm kpc}$ from its center, where the Solar System is approximately located. Such a screening condition
may impose additional constraints on the NMC gravity model whose assessment requires the solution for
the gravitational field of the Milky Way, possibly taking also into account the effect of the other
galaxies in the local group. That will be the subject of a future paper.

The galactic mass density $\rho_g$ in the solar neighborhood of the Milky Way is $\rho_g \approx 6.9\times 10^{-24} \,{\rm g}/{\rm cm}^3$
\cite{HoFly}, so that we have $R\approx R_g$, with $R_g$ the GR solution $R_g=8\pi G \rho_g /c^2$.
The minimizer $\eta_g$ of the effective potential $V_{\rm eff}(\eta,\rho)$, corresponding to $\rho(\x,t)=\rho_g$, is given by
\begin{equation}\label{eta-g-minimizer}
\eta_g = \frac{c^4}{8\pi G} - 2\alpha q\left(\frac{8\pi G}{c^2}\right)^{\alpha-1}\rho_g^\alpha c^2.
\end{equation}
We denote by $r_g$ a distance from the Sun's center such that mass density is dominated by the galactic density component at points $\x$ such that $\vert\x-\x_S\vert>r_g$,
where $\x_S$ is the position vector of Sun's center.
We choose $r_g$ at the heliopause, the boundary between the solar wind and the interstellar medium \cite{MBMGDeA},
corresponding to a heliocentric radial distance of about $120\mbox{ AU }= 2.58\times 10^ 4R_\odot$, where $R_\odot$ is the Sun's radius.

Since $\eta$ approximately minimizes $V_{\rm eff}$ in the solar neighborhood of the Galaxy, 
then Eq. (\ref{eta-equation-potential}) becomes
\begin{equation}\label{Yukawa-outskirtsSolS}
\nabla^2\eta \approx \frac{1}{\lambda_g^2}(\eta-\eta_g), \qquad\mbox{for }\vert\x-\x_S\vert>r_g,
\end{equation}
\\
where $\lambda_g = \lambda(\rho_g)$.
The length $\lambda(\rho)$ increases as density decreases,
which is a typical property of the chameleon mechanism \cite{KW},
so that the length $\lambda_g$ is an upper bound for $\lambda$ in the Solar System.
The computations in the present paper will be made under the condition $\lambda_g \gg r_g$ which will permit us
to find analytic estimates of the results. We will find that such a condition is satisfied when the constraint from LLR measurements
is saturated. Then we have
\begin{equation}\label{Laplace-galaxy}
\nabla^2\eta = 0 \qquad\mbox{for } r_g<\vert\x-\x_S\vert\ll\lambda_g,
\end{equation}
and, for $\vert\x-\x_S\vert$ large enough, we assume
\begin{equation}\label{far-bound-cond}
\eta(\x,t) \approx \eta_g.
\end{equation}
Under our assumptions the solution $\eta$ is a harmonic function in the outskirts of the Solar System.

\subsection{Solution in interplanetary space}

In interplanetary space, where mass density $\rho$ is small and gradually approaches the galactic density $\rho_g$ as $\vert\x-\x_S\vert$ approaches $r_g$,
we proceed as in Ref. \cite{MBMGDeA} and we expand the derivative of the potential $V$ around the minimizer $\eta_g$:
\begin{equation}\label{potential-deriv-approx}
\frac{\partial V}{\partial\eta}(\eta,\rho) \approx \frac{\partial V}{\partial\eta}(\eta_g,\rho) +
\frac{\partial^2 V}{\partial\eta^2}(\eta_g,\rho)(\eta-\eta_g).
\end{equation}
Solving the expression (\ref{eta-definition}) of $\eta$,
\begin{equation}\label{eta-expression-solved}
\eta = \frac{c^4}{8\pi G} - 2 \alpha q R^{\alpha-1} \rho c^2,
\end{equation}
with respect to curvature $R=\omega(\eta,\rho)$ we find
\begin{equation}\label{R-omega-formula}
\omega(\eta,\rho) = \left( \frac{16\pi G}{c^2} \alpha q\rho \right)^{1\slash (1-\alpha)}
\left( 1 - \frac{8\pi G}{c^4} \eta \right)^{1\slash(\alpha-1)},
\end{equation}
from which, using the first equality in Eq. (\ref{potential-V-def}), we obtain the property
\begin{equation}
\frac{\partial V}{\partial\eta}(\eta,\rho) = \left( \frac{\rho}{\rho_g} \right)^{1\slash(1-\alpha)} \frac{\partial V}{\partial\eta}(\eta,\rho_g),
\end{equation}
where we note the explicit dependence of $\partial V/\partial\eta$ on $\rho$ due to the nonminimal coupling.
Taking now into account that at density $\rho_g$ (in the solar vicinity of the Galaxy) the field $\eta$ approximately
minimizes the effective potential $V_{\rm eff}$, so that
\begin{equation}
\frac{\partial V}{\partial\eta}(\eta_g,\rho_g) \approx \frac{1}{3}\, \rho_g c^2,
\end{equation}
we can compute the approximation (\ref{potential-deriv-approx}) of the derivative of the potential:
\begin{widetext}
\begin{eqnarray}\label{potential-deriv-computed}
\frac{\partial V}{\partial\eta}(\eta,\rho) &\approx& \left( \frac{\rho}{\rho_g} \right)^{1\slash(1-\alpha)}
\left[ \frac{\partial V}{\partial\eta}(\eta_g,\rho_g) +
\frac{\partial^2 V}{\partial\eta^2}(\eta_g,\rho_g)(\eta-\eta_g) \right]
\approx \left( \frac{\rho}{\rho_g} \right)^{1\slash(1-\alpha)} \left[ \frac{1}{3}\, \rho_g c^2 + \frac{1}{\lambda_g^2}(\eta-\eta_g) \right] \nonumber\\
&\approx& \frac{1}{3}\, \rho_g c^2 \left( \frac{\rho}{\rho_g} \right)^{1\slash(1-\alpha)},
\end{eqnarray}
\end{widetext}
where, taking into account that $\lambda_g$ is assumed large with respect to $r_g$, we have used the inequality
\begin{equation}\label{ineq-interplan-space}
\frac{1}{\lambda_g^2}(\eta-\eta_g) \ll \frac{1}{3}\rho_g c^2,
\end{equation}
that will be verified {\it a posteriori}. Then we have
\begin{equation}
\frac{\partial V_{\rm eff}}{\partial\eta} = \frac{\partial V}{\partial\eta} - \frac{1}{3}\, \rho c^2 \approx - \frac{1}{3}\, \widetilde\rho c^2,
\end{equation}
with
\begin{equation}\label{sq-bra-NMC}
\widetilde\rho = \rho\left[ 1 - \left(\frac{\rho_g}{\rho}\right)^{-\alpha\slash(1-\alpha)} \right].
\end{equation}
Then, in interplanetary space we look for the field $\eta$ which solves the Poisson equation
\begin{equation}
\nabla^2\eta = - \frac{1}{3}\, \widetilde\rho c^2.
\end{equation}
As $\vert\x-\x_S\vert$ tends to $r_g$, density $\rho$ tends to $\rho_g$ and the Poisson equation turns to Laplace equation (\ref{Laplace-galaxy})
which holds for $\vert\x-\x_S\vert\geq r_g$.

\section{Gravitational field of astronomical bodies}

In the thin shells inside the astronomical bodies, the function $\eta$ has to interpolate between the solution inside the screening radii computed in the previous section
and the solution in interplanetary space. By adapting to NMC gravity the chameleon mechanism developed in Ref. \cite{KW}, we require the interpolating function $\eta$
to satisfy the condition (see also Ref. \cite{MBMGDeA},\cite{HS})
\begin{equation}\label{deriv-potential-condit}
0 < \frac{\partial V}{\partial\eta}(\eta,\rho) \ll \frac{1}{3}\,\rho c^2,
\end{equation} 
inside the thin shells and in the vicinity of the astronomical bodies where mass density is significantly larger than the galactic density $\rho_g$ due to the presence of Earth's
atmosphere and solar wind. Hence the equation for $\eta$ becomes the Poisson equation
\begin{equation}
\nabla^2\eta = - \frac{1}{3}\,\rho c^2.
\end{equation}
The explicit dependence of $\partial V/\partial\eta$ on density $\rho$ in condition (\ref{deriv-potential-condit}) is a distinctive feature of the
application of the chameleon mechanism to NMC gravity with respect to $f(R)$ gravity.
Inequality (\ref{deriv-potential-condit}) will be verified a posteriori in Appendix E for $\alpha<-1\slash 2$. Since we have
\begin{equation}
\alpha < 0 \implies 0 < -\frac{\alpha}{1-\alpha} < 1,
\end{equation}
for $|\alpha|$ not too small, the second term inside the square bracket in Eq. (\ref{sq-bra-NMC}) is negligible in comparison to 1
in the thin shells inside the astronomical bodies and in zones where mass density is much larger than the galactic density $\rho_g$,
so that $\widetilde\rho \approx \rho$ and the Poisson equation,
\begin{equation}\label{Poisson-everywhere}
\nabla^2\eta = - \frac{1}{3}\, \widetilde\rho c^2,
\end{equation}
is valid with a good approximation everywhere outside of the screening radii. In the next sections we compute an approximate solution
of this equation.

\subsection{Boundary conditions}

We rewrite the Poisson equation (\ref{Poisson-everywhere}) in the form
\begin{equation}\label{Poisson-Eq-rewritten}
\nabla^2\left(\eta-\eta_g\right) = - \frac{1}{3}\, \widetilde\rho c^2,
\end{equation}
and we solve this equation in the unbounded domain $\Omega$ outside of the screening radii, with the boundary condition (\ref{far-bound-cond}) at large
distance from the Sun's center. We denote by $\partial\Omega$ the boundary of $\Omega$ and by $\x_S,\x_E,\x_M$ the centers of the Sun, Earth and Moon, respectively.
Then the boundary $\partial\Omega$ consists of the union of the three spherical surfaces with centers at $\x_S,\x_E,\x_M$ and radii given by
$r_S,r_E,r_M$, respectively.

We have to match the solution in $\Omega$ with the interior solution computed in Sec. \ref{sec:interior-bodies-solut}.
Using the value (\ref{eta-minimizer-Earth}) of $\eta$ minimizing the effective potential inside the Earth, and the solution (\ref{eta-sol-Earth-ocean}) for $\eta$ in the Earth's ocean layer,
the value of $\eta$ at the Earth's screening surface is given by
\begin{equation}
\eta \approx \frac{c^4}{8\pi G} - \frac{2-\alpha}{3(1-\alpha)}\lambda^2_{E,w}\rho_{E,w}c^2.
\end{equation}
Using the expression (\ref{eta-g-minimizer}) of the minimizer $\eta_g$ in the Galaxy, and formula (\ref{lambda-expression}) for $\lambda(\rho)$, we have
\begin{equation}
\eta_g =  \frac{c^4}{8\pi G} -\frac{\lambda^2_g}{3(1-\alpha)} \rho_g c^2.
\end{equation}
In the sequel we assume $|\alpha|$ not too close to zero in such a way that, using again the formula for $\lambda(\rho)$, we have
\begin{equation}
\frac{\lambda^2_{E,w}\rho_{E,w}}{\lambda^2_g \rho_g} = \left(\frac{\rho_{E,w}}{\rho_g}\right)^\alpha \ll 1,
\end{equation}
from which it follows that at the Earth's screening surface the boundary condition $\eta-\eta_g \approx c^4\slash(8\pi G)-\eta_g$ holds.
Note that the expression (\ref{eta-expression-solved}) of $\eta$, for $R>0$ and $\alpha<0$, implies $\eta\neq c^4\slash(8\pi G)$, so that the
above boundary condition is proposed as an approximate condition in the sense of the inequality $\vert (8\pi G\slash c^4)\eta - 1\vert \ll 1$.
Analogous considerations can be applied for the boundary conditions at the screening surfaces of the Sun and Moon.

Then we have the following Dirichlet boundary conditions for the Poisson equation (\ref{Poisson-Eq-rewritten}) in the set $\Omega$:
\begin{equation}\label{Dirichlet-bound-cond}
\begin{cases}
\eta(\x,t)-\eta_g \approx \frac{c^4}{8\pi G}-\eta_g \qquad\mbox{for $\x\in\partial\Omega$ and any $t$}, \\
\eta(\x,t)-\eta_g \approx 0 \quad\mbox{for $r_g<\vert\x-\x_S\vert$ large and any $t$}.
\end{cases}
\end{equation}
Moreover, the existence of $\nabla^2\eta$ requires the continuity of the first partial derivatives of $\eta$ across the screening surfaces.
The normal derivative of $\eta$ at the Earth's screening surface, computed from the interior by using solution (\ref{eta-sol-Earth-ocean}), is given by
\begin{equation}
\frac{d\eta}{dr}(r_E) \approx - \frac{1}{3}\lambda_{E,w}\rho_{E,w} c^2,
\end{equation}
and an analogous result holds for the Moon. Moreover, using the approximate solution (\ref{eta-inner-interior-complete}) in the Sun's interior, the normal derivative of $\eta$
at the Sun's screening surface is given by
\begin{equation}
\frac{d\eta}{dr}(r_S) \approx - \frac{\alpha\lambda^2(\rho_S(r_S))}{3(1-\alpha)}\frac{d\rho_S}{dr}(r_S) c^2.
\end{equation}
Then, taking into account the smallness of $\lambda(\rho)$ in the interior of the bodies,
we neglect the derivatives of $\eta$ from the interior of $\partial\Omega$ and we impose the Neumann boundary condition:
\begin{equation}\label{Neumann-bound-cond}
\nabla\eta(\x) \cdot \hat \n(\x) \approx 0 \qquad\mbox{for $\x\in\partial\Omega$ and any $t$},
\end{equation}
where $\hat\n(x)$ denotes a normal unit vector to the surface $\partial\Omega$ at point $\x$. For given screening radii $r_S,r_E,r_M$ the Dirichlet condition 
uniquely determines the solution of the Poisson equation, while the Neumann condition will be used to find the screening radii.

\subsection{Solution by means of Green's function}

In order to compute effects in the Sun-Earth-Moon system it is enough to compute the solution $\eta(\x,t)$ for $\vert\x-\x_S\vert$ of the order of Earth's distance from the Sun.
For a given screening radii and time instant $t$, the solution of the Dirichlet problem for the Poisson equation is represented by means of the Green's function $G(\x,\x_0)$ \cite{Haberm}:
\begin{eqnarray}\label{Green-representation}
\eta(\x,t) &=& \eta_g - \frac{c^2}{3}\int_\Omega \widetilde\rho(\x_0,t)G(\x,\x_0)d^3 x_0 \\
&+& \left(\frac{c^4}{8\pi G}-\eta_g\right)\int_{\partial\Omega}\nabla_{\x_0}G(\x,\x_0)\cdot\hat\n(\x_0) dS_0,\nonumber
\end{eqnarray}
provided that, for $\vert\x-\x_S\vert$ of the order of Earth's distance from the Sun, the following inequality is satisfied
\begin{equation}\label{Green-repr-inequality}
\vert\eta(\x,t)\vert \gg \vert I(\eta,G,R^\ast)\vert,
\end{equation}
where $I(\eta,G,R^\ast)$ is the following integral evaluated on a sphere $S(R^\ast)$ of large enough radius $R^\ast>r_g$ and center in $\x_S$,
\begin{equation}
I(\eta,G,R^\ast) = \int_{S(R^\ast)} \left[ \left(\eta-\eta_g\right)\nabla G - G\nabla\eta \right] \cdot \hat\n \,dS.
\end{equation}
In Green's representation formula (\ref{Green-representation}) the normal unit vector $\hat\n$ to $\partial\Omega$ points towards the interior of bodies.
In the next section we compute an analytical approximation of the Green's function.

\subsection{Method of images for a system of spheres}

If the set $\partial\Omega$ would consist of a single sphere, then the Green's function could be obtained by using the method of images.
Since in our case the set $\partial\Omega$ consists of three spheres, which are the three screening surfaces of Sun, Earth and Moon, respectively,
then we apply to our problem the extension of the method of images to a system of spheres that has been proposed in Ref. \cite{MetzLa}.
This is an iterative method that involves an infinite series of images so that it yields a representation of Green's function by means of an infinite series:
\begin{equation}
G(\x,\x_0) = G^{(0)}(\x,\x_0) + G^{(1)}(\x,\x_0) + G^{(2)}(\x,\x_0) + \cdots
\end{equation}
The first terms of the series are obtained as follows (see \cite{MetzLa} for further details). The zeroth-order term is
\begin{equation}\label{Greenf-zero-th-order-term}
G^{(0)}(\x,\x_0) = -\frac{1}{4\pi}\,\frac{1}{\vert\x - \x_0\vert}\,,
\end{equation}
with $\x_0\in\Omega$, so that $\x_0$ lies outside of the three screening spheres.

The first order term $G^{(1)}(\x,\x_0)$ involves three image points: one image inside each screening sphere.
In each sphere the image point is obtained by applying the usual method of electrostatics to the unitary source at $\x_0$:
\begin{eqnarray}
& & G^{(1)}(\x,\x_0) = \frac{1}{4\pi} \left( \frac{r_S}{\vert\x_S-\x_0\vert\cdot\vert\x-\widetilde\x_S\vert} \right. \\
&+& \left. \frac{r_E}{\vert\x_E-\x_0\vert\cdot\vert\x-\widetilde\x_E\vert}
+ \frac{r_M}{\vert\x_M-\x_0\vert\cdot\vert\x-\widetilde\x_M\vert} \right), \nonumber
\end{eqnarray}
where $\widetilde\x_S$ is the image of $\x_0$ inside the screening sphere of the Sun, which is given by
\begin{equation}
\widetilde\x_S = \x_S + r_S^2\, \frac{\x_0-\x_S}{\vert\x_0-\x_S\vert^2}\,,
\end{equation}
and the image points $\widetilde\x_E$ and $\widetilde\x_M$, inside the screening spheres of Earth and Moon, are obtained
by replacing in the expression of $\widetilde\x_S$ the subscript $S$ with $E$ and $M$, respectively.

The second order term $G^{(2)}(\x,\x_0)$ involves six image points: two images inside each screening sphere. The image points in each sphere are obtained by
iterating the procedure used for the first order term: consider the three sources at points $\widetilde\x_S,\widetilde\x_E,\widetilde\x_M$ with charges
\begin{equation}
-\frac{r_S}{\vert\x_S-\x_0\vert}, \quad -\frac{r_E}{\vert\x_E-\x_0\vert}, \quad -\frac{r_M}{\vert\x_M-\x_0\vert},
\end{equation}
respectively. Then, the six image points are obtained by applying the method of electrostatics to the above sources and they are given by:
\begin{itemize}
\item[{\rm (i)}] the images inside the Sun's screening sphere of the sources at $\widetilde\x_E$ and $\widetilde\x_M$;
\item[{\rm (ii)}] the images inside the Earth's screening sphere of the sources at $\widetilde\x_S$ and $\widetilde\x_M$;
\item[{\rm (iii)}] the images inside the Moon's screening sphere of  the sources at $\widetilde\x_S$ and $\widetilde\x_E$.
\end{itemize}
The resulting expression of $G^{(2)}(\x,\x_0)$  is given in Appendix B.
This procedure is repeated iteratively giving rise to an infinite series of images and terms in the Green's function. The convergence of the series is discussed in Ref. \cite{MetzLa}.
If the distances between the spheres are much greater than the radii, a condition which is satisfied for the Sun-Earth-Moon system, then the contribution
of the higher order terms decreases quickly, and we find that the Green's function up to the second order will suffice while
higher order terms will not be necessary. Eventually, we observe that notwithstanding the ratio of Sun's radius and the Earth-Moon distance is not small,
such a ratio never appears in the computations.

\subsection{Solution for the scalar field $\eta$}\label{sec:Green-funct-solution}

In this section we give the solution $\eta$ in $\Omega$ by using the Green's function up to the first order, while the second order terms are given in Appendix B.
Then the function $G^{(0)}(\x,\x_0)+G^{(1)}(\x,\x_0)$ is substituted in Green's representation formula (\ref{Green-representation}), 
and the volume integral over $\Omega$ and the surface integral over $\partial\Omega$ are computed according to the following scheme:
\begin{itemize}
\item[{\rm (i)}] we have $\widetilde\rho\approx\rho$ inside the thin shells, and density becomes immediately much smaller outside the thin shells in solar atmosphere,
in terrestrial atmosphere and outside of the Moon; then in the volume integral over $\Omega$ the contribution outside of the thin shells can be safely neglected, so that
we have to add the integrals over the three shells; these integrals are then evaluated in closed-form by resorting to spherical coordinates for each of the three astronomical
bodies;
\item[{\rm (ii)}] the surface integrals over the three screening spheres that constitute $\partial\Omega$ are evaluated by using Gauss theorem.
\end{itemize}
The result of the computation is the following. For each astronomical body we introduce the effective mass which is a function of the screening radius.
The effective mass of the Sun is
\begin{equation}
M_{\odot,{\rm eff}}(r_S) = 4\pi \int_{r_S}^{R_\odot} \rho_S(r)r^2dr\,,
\end{equation}
and the effective masses $M_{\oplus,{\rm eff}}(r_E)$ and $M_{M,{\rm eff}}(r_M)$ of Earth and Moon are defined analogously. The solution $\eta$ is given by
\begin{equation}\label{eta-green-total}
\eta(\x,t) = \eta_S(\x,t) + \eta_E(\x,t) + \eta_M(\x,t) + \eta_g,
\end{equation}
where $\eta_S,\eta_E$ and $\eta_M$ are the contributions from the thin shells of Sun, Earth and Moon, respectively.
The contribution from the Earth's shell is
\begin{eqnarray}\label{eta-green-Earth}
\eta_E(\x,t) &=& \left( \frac{c^4}{8\pi G} - \eta_g \right) \frac{r_E}{\vert\x-\x_E\vert} + \I_E(\x,t) + \J_E(\x,t) \nonumber\\
&-& \frac{c^2}{3}\,\frac{r_E}{\vert\x-\x_E\vert}\int_{r_E}^{R_\oplus}\rho_E(r)r\,dr\,,
\end{eqnarray}
where the term $\I_E$ is given by
\begin{eqnarray}\label{I-Earth-first}
\I_E(\x,t) &=& \frac{c^2}{3\vert\x-\x_E\vert}\int_{r_E}^{\vert\x-\x_E\vert}\rho_E(r)r^2dr \\
&+& \frac{c^2}{3}\int_{\vert\x-\x_E\vert}^{R_\oplus}\rho_E(r)r\,dr \qquad\mbox{for } \vert\x-\x_E\vert < R_\oplus, \nonumber
\end{eqnarray}
and
\begin{equation}\label{I-Earth-second}
\I_E(\x,t) = \frac{c^2}{12\pi}\,\frac{M_{\oplus,{\rm eff}}(r_E)}{\vert\x-\x_E\vert} \qquad\mbox{for } \vert\x-\x_E\vert \geq R_\oplus,
\end{equation}
and the term $\J_E=\J_{E,S}+\J_{E,M}$ is given by
\begin{eqnarray}\label{J-Earth}
\J_{E,S}(\x,t) &=& -\frac{c^2}{12\pi}\,\frac{r_S M_{\oplus,{\rm eff}}(r_E)}{\left\vert \vert\x-\x_S\vert(\x_S-\x_E)+r_S^2\hat\n_S\right\vert}, \\
\J_{E,M}(\x,t) &=& -\frac{c^2}{12\pi}\,\frac{r_M M_{\oplus,{\rm eff}}(r_E)}{\left\vert \vert\x-\x_M\vert(\x_M-\x_E)+r_M^2\hat\n_M\right\vert}, \nonumber
\end{eqnarray}
where $\hat\n_S$ and $\hat\n_M$ are the outward unit normal vectors to the screening surfaces of the Sun and Moon, respectively.

The function $\eta_E(\x,t)$ depends on time through the centers $\x_S(t),\x_E(t),\x_M(t)$ of the bodies, which vary with time along the respective orbits.

The meaning of the terms in the expression (\ref{eta-green-Earth}) of $\eta_E$ is the following: the first term is the surface integral over Earth's screening surface
of the term in $G^{(1)}(\x,\x_0)$ corresponding to the image of $\x_0$ inside Earth's screening sphere (the other three surface integrals vanish); 
$\I_E$ is the volume integral over the Earth's shell of the zeroth-order term of Green's function; $\J_{E,S}$ and $\J_{E,M}$
are the volume integrals of the terms in $G^{(1)}(\x,\x_0)$ corresponding to the images of $\x_0$ inside the screening spheres of the Sun and the Moon, respectively; 
the last term is the volume integral corresponding to the image of $\x_0$ inside Earth's screening sphere.

The solution $\eta$ is formally symmetric with respect to the three bodies:
\begin{itemize}
\item[{\rm (i)}] the contribution $\eta_S$ from the Sun's shell is obtained by replacing in Eqs. (\ref{eta-green-Earth}-\ref{I-Earth-second}) 
the subscript $E$ with $S$, the radius $R_\oplus$ with $R_\odot$, and the effective mass $M_{\oplus,{\rm eff}}$ with $M_{\odot,{\rm eff}}$; then
\begin{equation}
\J_E=\J_{E,S}+\J_{E,M} \,\mbox{is replaced with}\, \J_S=\J_{S,E}+\J_{S,M}, 
\end{equation}
in the first of Eqs. (\ref{J-Earth}) the subscripts $E$ and $S$ are exchanged and $M_{\oplus,{\rm eff}}$ is replaced with $M_{\odot,{\rm eff}}$,
and in the second of Eqs. (\ref{J-Earth}) the subscript $E$ is replaced with $S$ and $M_{\oplus,{\rm eff}}$ with $M_{\odot,{\rm eff}}$;
\item[{\rm (ii)}]  the contribution $\eta_M$ from the Moon's shell is obtained with analogous changes.
\end{itemize}
We observe that the terms $\J_S,\J_E$ and $\J_M$ originate from the term $G^{(1)}(\x,\x_0)$ in the Green's function which involves the first order images.

Eventually, since the various terms in $\eta(\x,t)$ decrease as $1\slash\vert\x-\x_S\vert$, $1\slash\vert\x-\x_E\vert$ and $1\slash\vert\x-\x_M\vert$ in interplanetary space,
and the integral $I(\eta,G,R^\ast)$ decreases as $1/R^\ast$, then inequality (\ref{Green-repr-inequality}) is satisfied for
$\vert\x-\x_S\vert$ of the order of Earth's distance from the Sun and $R^\ast>r_g$ large enough.

In Appendix C we report the error in the verification of the Dirichlet condition on $\partial\Omega$ when the solution $\eta$ is computed
by using the Green's function up to the second order. We also show that increasing the number of image points, hence the order of Green's function, the approximation
of the Dirichlet condition improves.

\subsection{Determination of the screening radii}\label{sec:screen-radii}

Now we impose the Neumann boundary condition (\ref{Neumann-bound-cond}) on $\partial\Omega$ and we use such a condition to find integral equations
that determine the three screening radii $r_S,r_E,r_M$.

Let us consider Earth's screening sphere where we have to compute the scalar product $\nabla\eta\cdot\hat\n_E$. We compute the contribution to the scalar product
given by the leading terms, the other ones being negligible because suppressed by factors involving the small radius-to-distance ratios (and their powers)
of the astronomical bodies. 

Let us first consider the contribution from the solar term $\eta_S$.
Since $\I_S$ is the volume integral over the Sun's shell of the zeroth-order term (\ref{Greenf-zero-th-order-term}) of Green's function, and $\J_{S,E}$
is the volume integral of the term in $G^{(1)}(\x,\x_0)$ corresponding to the image of $\x_0$ inside the Earth's screening sphere, then, by the properties of image points
we have $\I_S(\x,t)+\J_{S,E}(\x,t)=0$ on the screening sphere, so that the vector $\nabla\left(\I_S+\J_{S,E}\right)$ is orthogonal to the sphere.
By means of a Taylor approximation we have
\begin{eqnarray}\label{Is-Jse-Taylor-2nd}
& &\nabla\left(\I_S+\J_{S,E}\right)\cdot\hat\n_E = \frac{c^2}{12\pi}\left\{1-3\frac{r_E}{\vert\x_E-\x_S\vert}\hat\n_{ES}\cdot\hat\n_E \right. \nonumber\\
&-&\left.\frac{5}{2}\left[1-3\left(\hat\n_{ES}\cdot\hat\n_E\right)^2\right]\left(\frac{r_E}{\vert\x_E-\x_S\vert}\right)^2\right\}
\frac{M_{\odot,{\rm eff}}(r_S)}{r_E\vert\x_E-\x_S\vert} \nonumber\\
&+& \OO\left(\left(\frac{r_E}{\vert\x_E-\x_S\vert}\right)^3\right),
\end{eqnarray}
where $\hat\n_{ES}=(\x_E-\x_S)\slash\vert\x_E-\x_S\vert$, from which, since
\begin{equation}\label{geom-suppress}
\frac{r_E}{\vert\x_E-\x_S\vert} < 10^{-4}, \qquad \frac{r_E^2}{\vert\x_E-\x_S\vert^2} < 10^{-8},
\end{equation}
it follows that the scalar product has the constant dominant term
\begin{equation}
\nabla\left(\I_S+\J_{S,E}\right)\cdot\hat\n_E \approx 
\frac{c^2}{12\pi}\,\frac{M_{\odot,{\rm eff}}(r_S)}{r_E\vert\x_E-\x_S\vert},
\end{equation}
plus a variable part on the sphere suppressed by the small geometric factor $r_E\slash\vert\x_E-\x_S\vert$ and its powers. 

Let us now consider the integral term in $\eta_S$,
\begin{equation}
-\frac{c^2}{3}\,\frac{r_S}{\vert\x-\x_S\vert}\int_{r_S}^{R_\odot}\rho_S(r)r\,dr\,,
\end{equation}
obtained by replacing $E$ with $S$ in the last term of the solution (\ref{eta-green-Earth}) for $\eta_E$.
By using the method of images it is shown in Appendix C that such an integral term cancels on the Earth's screening sphere with a term in $\eta_S$ resulting from
the second order Green's function $G^{(2)}(\x,\x_0)$. Hence the gradient of the sum of these two terms is orthogonal to the sphere and the contribution to the scalar product
$\nabla\eta_S\cdot\hat\n_E$, by means of a computation analogous to Eq. (\ref{Is-Jse-Taylor-2nd}), has the constant dominant term
\begin{equation}
-\frac{c^2}{3}\,\frac{r_S}{r_E\vert\x_E-\x_S\vert}\int_{r_S}^{R_\odot}\rho_S(r)r\,dr\,,
\end{equation}
plus a small variable part on the sphere. The surface term in $\eta_S$,
\begin{equation}
\left( \frac{c^4}{8\pi G} - \eta_g \right) \frac{r_S}{\vert\x-\x_S\vert},
\end{equation}
obtained by replacing $E$ with $S$ in the first term of Eq. (\ref{eta-green-Earth}), gives rise to an analogous cancellation discussed in Appendix C.
Then, arguing as before, we find a contribution to the scalar product with the constant dominant term
\begin{equation}
\frac{r_S}{r_E\vert\x_E-\x_S\vert}\left( \frac{c^4}{8\pi G} - \eta_g \right).
\end{equation}
Further contributions from $\eta_S$ turn out to be negligible. The contribution from the lunar term $\eta_M$ is obtained by replacing $S$ with $M$ in the previous expressions.
Eventually, the contribution from the terrestrial term $\eta_E$ is given by
\begin{equation}
\nabla\eta_E\cdot\hat\n_E \approx \frac{c^2}{3r_E}\int_{r_E}^{R_\oplus}\rho_E(r)r\,dr - \frac{1}{r_E}\left( \frac{c^4}{8\pi G} - \eta_g \right).
\end{equation}
By substituting all these contributions in the scalar product $\nabla\eta\cdot\hat\n_E$, and imposing
the Neumann boundary condition (\ref{Neumann-bound-cond}) on Earth's screening sphere, we obtain the following integral equation:
\begin{widetext}
\begin{eqnarray}\label{radii-integral-equations}
& &\frac{c^2}{3}\int_{r_E}^{R_\oplus}\rho_E(r)r\,dr - \left(1-\frac{r_S}{\vert\x_E-\x_S\vert} - \frac{r_M}{\vert\x_E-\x_M\vert}\right)\left( \frac{c^4}{8\pi G} - \eta_g \right) \\
&+&\frac{c^2}{3}\left[\frac{1}{\vert\x_E-\x_S\vert}\left(\frac{M_{\odot,{\rm eff}}(r_S)}{4\pi}-r_S\int_{r_S}^{R_\odot}\rho_S(r)r\,dr\right)
+\frac{1}{\vert\x_E-\x_M\vert}\left(\frac{M_{M,{\rm eff}}(r_M)}{4\pi}-r_M\int_{r_M}^{R_M}\rho_M(r)r\,dr\right)\right] = 0. \nonumber
\end{eqnarray}
\end{widetext}
Then, repeating the computation on the screening spheres of the Sun and Moon, we obtain a system of integral equations. 
Such equations are obtained from Eq. (\ref{radii-integral-equations}) by exchanging the subscripts $S,E,M$.
Using formula (\ref{eta-g-minimizer}) for the minimizer $\eta_g$,
the solution to the resulting system of integral equations determines the three screening radii $r_E,r_S,r_M$ for given values of the NMC gravity parameters $\alpha$ and $q$.
The system of equations generalizes the integral equation found in Ref. \cite{MBMGDeA} to a system of three gravitationally interacting extended bodies.
 
Neglecting factors involving radius-to-distance ratios, the following relations follow from Eq. (\ref{radii-integral-equations}) and the other integral equations, as a first approximation:
 \begin{equation}\label{integrals-r-density-relat}
\int_{r_S}^{R_\odot}\rho_S(r)r\,dr \approx \int_{r_E}^{R_\oplus}\rho_E(r)r\,dr \approx \int_{r_M}^{R_M}\rho_M(r)r\,dr\,.
 \end{equation}
 If we also assume that all bodies have a thin shell, so that we have $R_\oplus-r_E\ll R_\oplus$ for the Earth, and analogous inequalities for the Sun and Moon,
 then the above approximate relations are equivalent to the following relations between effective masses:
 \begin{equation}\label{effective-masses-relat}
 \frac{M_{\odot,{\rm eff}}(r_S)}{R_\odot} \approx  \frac{M_{\oplus,{\rm eff}}(r_E)}{R_\oplus} \approx  \frac{M_{M,{\rm eff}}(r_M)}{R_M}\,.
 \end{equation}
Eventually, we observe that the results obtained in this section show that the normal derivative of $\eta$ on the screening spheres has a main part which is constant on each sphere,
and a much smaller variable part which can be neglected thanks to inequalities of type (\ref{geom-suppress}). 
Since the interior solution computed in Sec. \ref{sec:interior-bodies-solut} is spherically symmetric inside each body,
then it has a constant normal derivative on each screening sphere, so that matching the solution outside of the screening radii
with the interior solution yields a consistent approximation, as it was anticipated at the beginning of Sec. \ref{sec:interior-bodies-solut}.

\subsection{Verification of inequalities}\label{sec:verif-ineq}

The solution for $\eta$ has been computed by assuming inequalities (\ref{cond-f1-f2}-\ref{cond-f1R-f2R}), (\ref{ineq-interplan-space}) and (\ref{deriv-potential-condit}), 
necessary in order to find an analytic approximation of the solution, that have to be verified a posteriori. In this section we show that the computed solution
satisfies inequalities (\ref{cond-f1-f2}-\ref{cond-f1R-f2R}) and (\ref{ineq-interplan-space}), while inequality (\ref{deriv-potential-condit}) is verified in Appendix E.

Let us first consider inequality (\ref{cond-f1R-f2R}) and the solution inside the screening radii. In the case of Earth,
using the value (\ref{eta-minimizer-Earth}) of $\eta$ minimizing the effective potential, the solution for $\eta$ found in Sec. \ref{sec:sol-inside-Earth} satisfies
\begin{equation}
\left\vert \frac{8\pi G}{c^4}\eta -1 \right\vert \sim \frac{G}{c^2}\lambda^2(\rho_E)\rho_E \sim \left(\frac{\lambda(\rho_E)}{R_\oplus}\right)^2 \frac{GM_\oplus}{c^2R_\oplus}\ll 1.
\end{equation}
Analogous results can be found for the Moon and the Sun. Then, using now the solution outside of the screening radii found in Sec. \ref{sec:Green-funct-solution},
for $\vert\x-\x_S\vert < r_g$ the expression $\vert (8\pi G/c^4)\eta-1\vert$ is bounded by a sum of terms each of which is bounded by a quantity of type
\begin{equation}
\mbox{either}\quad \frac{GM_{\rm eff}}{c^2R} \ll 1, \quad\mbox{or}\quad \frac{GM_{\rm eff}}{c^2d} \ll 1,
\end{equation}
where $M_{\rm eff}$ is an effective mass, $R$ is the radius of a body, and $d$ is a distance between the astronomical bodies. Moreover, using formula (\ref{eta-g-minimizer})
for the minimizer $\eta_g$ and the integral equations (\ref{radii-integral-equations}), we have $\vert (8\pi G/c^4)\eta_g-1\vert\ll 1$. Since $\eta$ is a harmonic function
for $\vert\x-\x_S\vert>r_g$ by Eq. (\ref{Laplace-galaxy}), and $\eta\approx\eta_g$ at large distance from the Sun by the boundary condition (\ref{far-bound-cond}),
then by the maximum principle for harmonic functions $\eta$ satisfies the desired inequality also for $\vert\x-\x_S\vert > r_g$. Hence the computed solution $\eta$
satisfies inequality (\ref{cond-f1R-f2R}) everywhere.

We now consider the second of inequalities (\ref{cond-f1-f2}). We set $R_{\rm GR}=8\pi G\rho\slash c^2$ and we observe that $R\approx R_{\rm GR}$ both inside the screening radii
and in the solar neighborhood of the Galaxy, $R\ll R_{\rm GR}$ in the thin shells (see Appendix E), and $R$ interpolates between such values in interplanetary space, so that 
$R\slash R_{\rm GR} \lesssim 1$ everywhere. Then, using formula (\ref{eta-expression-solved}), we have
\begin{equation}\label{inequality-f2}
\left\vert f^2(R)\right\vert = \vert q \vert R^\alpha = \frac{1}{2\vert\alpha\vert}\,\frac{R}{R_{\rm GR}}\left\vert \frac{8\pi G}{c^4}\eta -1 \right\vert \ll 1,
\end{equation}
for $\vert\alpha\vert$ not too close to zero, which is the case of interest for applications to astrophysics \cite{drkmattgal} and cosmology \cite{curraccel}.
Then the second of inequalities (\ref{cond-f1-f2}) is satisfied, the first being trivial.

Eventually we verify inequality (\ref{ineq-interplan-space}) under the assumption that the thin shell condition is satisfied for all the astronomical bodies,
a condition that will follow from the constraint from LLR measurements. First we observe that the leading terms of the solution $\eta$ in interplanetary space are given by
\begin{equation}
\eta-\eta_g \approx \frac{c^2}{12\pi}\left( \frac{M_{\odot,{\rm eff}}}{\vert\mathbf{x}-\mathbf{x_S}\vert} + \frac{M_{\oplus,{\rm eff}}}{\vert\mathbf{x}-\mathbf{x_E}\vert} 
+ \frac{M_{M,{\rm eff}}}{\vert\mathbf{x}-\mathbf{x_M}\vert} \right),
\end{equation}
where we have dropped the dependence of the effective masses on the screening radii for simplicity. Then, using the integral equations that determine the screening radii
under the approximation (\ref{integrals-r-density-relat}), the thin shell condition and the definition (\ref{lambda-expression}) of $\lambda_g$, we find
\begin{equation}
M_{\oplus,{\rm eff}} \approx 4\pi R_\oplus \int_{r_E}^{R_\oplus}\rho_E(r)rdr \approx \frac{4\pi}{1-\alpha} R_\oplus \lambda_g^2 \rho_g.
\end{equation}
Using the approximate relations (\ref{effective-masses-relat}) between effective masses, we have
\begin{eqnarray}
\eta-\eta_g \approx \frac{c^2\lambda_g^2\rho_g}{3(1-\alpha)}\left( \frac{R_\odot}{\vert\mathbf{x}-\mathbf{x_S}\vert}
+ \frac{R_\oplus}{\vert\mathbf{x}-\mathbf{x_E}\vert} + \frac{R_M}{\vert\mathbf{x}-\mathbf{x_M}\vert} \right), \nonumber\\
\end{eqnarray}
from which, for $\alpha<0$, inequality (\ref{ineq-interplan-space}) is satisfied if
\begin{equation}
\frac{R_\odot}{\vert\mathbf{x}-\mathbf{x_S}\vert} + \frac{R_\oplus}{\vert\mathbf{x}-\mathbf{x_E}\vert} + \frac{R_M}{\vert\mathbf{x}-\mathbf{x_M}\vert} \ll 1,
\end{equation}
which is satisfied in interplanetary space at large enough distance from the astronomical bodies.

\subsection{Solution for the potentials $\Phi$ and $\Psi$}

We have assumed the field $\eta$ is close to the minimizer of the effective potential $V_{\rm eff}$ in the
solar neighbourhood of the Galaxy for $\vert\x-\x_S\vert > r_g$, so that GR is approximately satisfied. 

In what follows we denote by $U$ the Newtonian potential of the mass distribution with density $\rho$,
\begin{equation}\label{Newt-potential}
U(\x,t) = G \int \frac{\rho(\y,t)}{\vert\x-\y\vert}d^3y,
\end{equation}
which satisfies the Poisson equation $\nabla^2U = - 4\pi G \rho$.

Using Eqs. (\ref{Psi-equation-approx}-\ref{Phi-equation-approx}) it follows that the potential $\Psi$ of the metric is related to the deviation from GR, then
we impose the following boundary conditions in the Galaxy at large distance $\vert\x-\x_S\vert$ from the Sun's center, where GR is satisfied by our assumptions:
\begin{equation}\label{Phi-Psi-boundcond-infty}
\Phi(\x,t) \approx \frac{1}{c^2}U(\x,t), \qquad \Psi(\x,t) \approx 0.
\end{equation}
Combining equations (\ref{Psi-equation-approx}) and (\ref{trace-approx}) for $\Psi$ and $\eta$ we have
\begin{equation}
\nabla^2 \left( \Psi + \frac{8\pi G}{c^4}\, \eta \right) = 0.
\end{equation}
Using now the second of boundary conditions (\ref{Phi-Psi-boundcond-infty}) for $\Psi$ and the boundary condition (\ref{far-bound-cond}) for $\eta$,
for all points $\x$ on a sphere with center in $\x_S$ and large enough radius $R^\ast>r_g$, and for any time $t$, we have
\begin{equation}
\Psi(\x,t) + \frac{8\pi G}{c^4}\, \eta(\x,t) \approx \frac{8\pi G}{c^4}\, \eta_g.
\end{equation}
Hence, by the maximum principle for harmonic functions, it follows that the harmonic function (with respect to the spatial variables $\x$)
\begin{equation}
\Psi(\x,t) + \frac{8\pi G}{c^4}\, \eta(\x,t)
\end{equation}
is constant inside the sphere of radius $R^\ast$, so that the solution for the potential $\Psi$ is given by
\begin{equation}\label{potential-Psi-sol}
\Psi(\x,t) = - \frac{8\pi G}{c^4}\left[ \eta(\x,t) - \eta_g \right].
\end{equation}
The solution for $\Psi$ then follows immediately from the solution for $\eta$ found in the previous sections.
Combining now equations (\ref{Psi-equation-approx}) and (\ref{Phi-equation-approx}) for $\Psi$ and $\Phi$ we have
\begin{equation}
\nabla^2\left( \Phi - \frac{U}{c^2} -\frac{1}{2}\, \Psi \right) = 0.
\end{equation}
Then, applying both boundary conditions (\ref{Phi-Psi-boundcond-infty}) and using again the maximum principle for harmonic functions,
the solution $\Phi$ of this equation is given by
\begin{equation}\label{Phi-solution-everywhere}
\Phi(\x,t) = \frac{1}{c^2}\, U(\x,t) + \frac{1}{2}\, \Psi(\x,t).
\end{equation}
The solutions found for $\Phi$ and $\Psi$ define the space-time metric (\ref{metric}).

\section{Dynamics of continuous bodies}

We consider the motion of Earth and Moon in the gravitational field of the Sun. The equations describing the dynamics of the system
are obtained by taking the covariant divergence of the energy-momentum tensor and applying Bianchi identities to the gravitational field equations (see Ref. \cite{BBHL}),
as given by Eq. (\ref{covar-div-1}) that we repeat for convenience:
\begin{equation}\label{covar-div-2}
\nabla_\mu T^{\mu\nu} = \frac{f^2_R }{ 1 + f^2} ( g^{\mu\nu} \LL_m - T^{\mu\nu} ) \nabla_\mu R.
\end{equation}
In the following computation we neglect mass density of solar and terrestrial atmospheres, and of solar wind in interplanetary space, so that density $\rho(\x,t)$
has a compact support consisting of the three spheres of radii $R_\odot,R_\oplus$ and $R_M$. We assume densities assigned inside the bodies 
according to the profiles given in Appendix A, and we assume all the astronomical bodies in hydrostatic equilibrium.
According to Sec. \ref{sec:metric-stress-tensor} the Sun is considered as a perfect fluid, while the Earth and Moon are approximately described 
as continuous bodies in a hydrostatic state of stress, so that inside the bodies the equations describing the dynamics of continuous media formally coincide with the equations 
of hydrodynamics of a perfect fluid.

We begin by computing the 0th component of Eq. (\ref{covar-div-2}): using the components of the energy-momentum tensor
given by Eqs. (\ref{T-00}) and (\ref{T-0i}), the left-hand side of this equation yields
\begin{equation}\label{covar-div-0th}
\nabla_\mu T^{\mu 0} = c\frac{\partial\rho}{\partial t}+c\frac{\partial}{\partial x^i}(\rho v^i) + \OO\left(\frac{1}{c}\right).
\end{equation}
Now we compute the right-hand side of Eq. (\ref{covar-div-2}). First we observe that inside the screening surfaces we have $R = \OO(1\slash c^2)$.
Then, the integral equations (\ref{radii-integral-equations}) imply
\begin{equation}\label{q-O-property}
\vert q \vert = \rho_g^{-\alpha} \cdot \OO\left(\frac{1}{c^{2-2\alpha}}\right),
\end{equation}
being $\alpha<0$, from which, using formula (\ref{R-omega-formula}) for curvature $R=\omega(\eta,\rho)$
and the property $\vert(8\pi G/c^4)\eta-1\vert=\OO(1/c^2)$ found in Sec. \ref{sec:verif-ineq},
we have $R = \OO(1\slash c^2)$ also in the thin shells of the astronomical bodies.
Using now property (\ref{q-O-property}) and the definition $f^2(R)=q R^\alpha$, it follows
\begin{equation}
f^2(R) = \rho_g^{|\alpha|} \cdot \OO\left(\frac{1}{c^2}\right), \quad
f^2_R = \rho_g^{|\alpha|} \cdot \OO(1),
\end{equation}
from which the evaluation of the right-hand side of Eq. (\ref{covar-div-2}) yields
\begin{equation}
\frac{f^2_R }{ 1 + f^2} ( g^{\mu 0} \LL_m - T^{\mu 0} ) \frac{\partial R}{\partial x^\mu} = \OO\left(\frac{1}{c}\right).
\end{equation}
Using Eq. (\ref{covar-div-0th}) and neglecting terms of order $\OO(1\slash c^2)$, the continuity equation then follows
in the nonrelativistic limit as usual:
\begin{equation}\label{contin-eq}
\frac{\partial\rho}{\partial t}+\frac{\partial}{\partial x^i}(\rho v^i) = 0.
\end{equation}
The NMC term on the right-hand side of Eq. (\ref{covar-div-2}) gives a distinctive contribution
to the spatial part of this equation that now we compute. Using the components (\ref{T-00})-(\ref{T-ij})
of the energy-momentum tensor, for $i=1,2,3$ the left-hand side yields
\begin{eqnarray}
\nabla_\mu T^{\mu i} &=& \frac{\partial}{\partial t}(\rho v^i)+\frac{\partial}{\partial x^j}(\rho v^i v^j)
-\rho\frac{\partial U}{\partial x_i}+\frac{c^2}{2}\rho\frac{\partial\Psi}{\partial x_i}  \nonumber\\
&+&\frac{\partial p}{\partial x_i} + \OO\left(\frac{1}{c^2}\right).
\end{eqnarray}
Using now the continuity equation Eq. (\ref{contin-eq}), at order $O(1)$, we get
\begin{equation}\label{covar-div-ij}
\nabla_\mu T^{\mu i} = \rho\frac{dv^i}{dt} +\frac{\partial p}{\partial x_i} -\rho\frac{\partial U}{\partial x_i}
+\frac{c^2}{2}\rho\frac{\partial\Psi}{\partial x_i} + \OO\left(\frac{1}{c^2}\right),
\end{equation}
where $d/dt=\partial/\partial t+v^i\partial/\partial x^i$ is the {\it material derivative} of continuum mechanics.

For $i=1,2,3$, taking into account that $\vert f^2 \vert \ll 1$, the right-hand side of Eq. (\ref{covar-div-2}) yields
\begin{equation}\label{extra-f-ij}
\frac{f^2_R }{ 1 + f^2} ( g^{\mu i} \LL_m - T^{\mu i} ) \frac{\partial R}{\partial x^\mu} = -c^2 f^2_R\,\rho\frac{\partial R}{\partial x_i} + \OO\left(\frac{1}{c^2}\right).
\end{equation}
Combining equations (\ref{covar-div-ij}) and (\ref{extra-f-ij}), and neglecting terms of order $\OO(1\slash c^2)$,
we obtain the equations of NMC dynamics of continuous bodies in hydrostatic state of stress and in the nonrelativistic limit,
\begin{equation}\label{NMC-contin-bodies}
\rho\frac{d\vv}{dt} = \rho\nabla U - \nabla p - \frac{1}{2} \rho c^2 \nabla\Psi - c^2 f^2_R\,\rho\nabla R,
\end{equation}
where the vector notation has been used. These equations are the Eulerian equations of Newtonian hydrodynamics with the presence of two additional terms:
\begin{itemize}
\item[{\rm (i)}] a fifth force density proportional to the gradient of the metric potential $\Psi$;
\item[{\rm (ii)}] an extra force density proportional to the product of $f^2_R$ by the gradient of curvature $R$.
\end{itemize}
The extra force density in (ii) has been extensively discussed in Ref. \cite{BBHL},
and for relativistic perfect fluids in Ref. \cite{BLP}.
While the fifth force is typical of $f(R)$ gravity theory, the extra force is specific of NMC gravity for the choie Eq. (\ref{f1-f2-specific}).

\subsection{Fifth force inside the screening spheres}\label{sec:fifth-force-inside-sr}

The fifth force density is given by
\begin{equation}\label{F-force-density}
-\frac{1}{2} \rho c^2 \nabla\Psi = \frac{4\pi G}{c^2}\rho\nabla\eta,
\end{equation}
where the expression (\ref{potential-Psi-sol}) of $\Psi$ in terms of the function $\eta$ has been used.
In the interior of the screening spheres the magnitude of fifth force is expected to be largest at the surfaces of density discontinuity where the largest
deviations from GR take place. Let us compute such a force at the Moho, the crust-mantle discontinuity in the Earth's interior.

The radial derivative of $\eta$ is continuous, and using the solution for $\eta$ inside the Earth's screening radius, found in Sec. \ref{sec:sol-inside-Earth}, we have
\begin{equation}
\frac{d\eta}{dr}(R_{E,m}) \approx \frac{\eta_{E,c}-\eta_{E,m}}{\lambda_{E,c}+\lambda_{E,m}}.
\end{equation}
The quantity $\rho_E d\eta/dr$ is discontinuous at the Moho and the maximum value of its magnitude is $\rho_{E,m}\vert d\eta/dr(R_{E,m})\vert$.
Then, using the value (\ref{eta-minimizer-Earth}) of $\eta$ minimizing the effective potential $V_{\rm eff}$ and formula (\ref{lambda-expression}) for $\lambda$, 
the magnitude of the fifth force density is estimated by
\begin{eqnarray}
\left\vert F_{\rm f}\right\vert &\approx& \frac{4}{3}\,\frac{\pi G}{1+|\alpha|}\rho_{E,m}\frac{\vert\lambda^2_{E,c}\rho_{E,c}-\lambda^2_{E,m}\rho_{E,m}\vert}{\lambda_{E,c}+\lambda_{E,m}}
\nonumber\\
&\leq& \frac{8\pi}{3} G\lambda_{E,c}\rho_{E,m}\rho_{E,c}.
\end{eqnarray}
Now the magnitude of the Newtonian force density at the Moho is 
\begin{equation}
\left\vert F_N\right\vert = \rho_{E,m}\left\vert\frac{dU}{dr}(R_{E,m})\right\vert = \frac{4\pi}{3} G R_{E,m}\rho_{E,m}\overline\rho_{E,c},
\end{equation}
where $\overline\rho_{E,c}$ is the average Earth density below the crust.
Since we have $\lambda(\rho_E)\ll R_{E,m}$ with $\lambda(\rho_E)$ completely negligible ($\lambda(\rho_E)\lesssim 10^{-7}$ m for $\alpha=-1$ according to 
Section \ref{sec:Earth-density-model} of Appendix A), it then follows $\left\vert F_{\rm f}\right\vert \ll \left\vert F_N\right\vert$.
The same behavior is found at the other density discontinuities, moreover, far from discontinuities the fifth force is further decreased by exponential suppression
with decay constant $1/\lambda(\rho_E)$ in the various layers.
Eventually, the perturbation of the Newtonian gravitational force, hence of hydrostatic equilibrium, is completely negligible inside the Earth's screening radius,
confirming the effectiveness of the screening mechanism.

Inside Earth's thin shell, hence for $r_E<r<R_\oplus$, the fifth force exerted by the Sun and Moon contributes to the motion of Earth's center of mass (see Section \ref{sec:Earth-Moon-accel}),
while the leading contribution from Earth itself is radial, so that it does not contribute to the motion of center mass but only to hydrostatic equilibrium.
Such a contribution to hydrostatic equilibrium is computed in Section \ref{F-force-t-shell} of Appendix E where it is shown that the resulting perturbation is again negligible in
comparison with Newtonian force.

Analogous results hold for the Moon, while for the Sun see Ref. \cite{MBMGDeA}.

\subsection{Jump conditions for the pressure}\label{sec:pressure-jumps}

The expression (\ref{R-omega-formula}) for curvature $R=\omega(\eta,\rho)$, which we rewrite in the form
\begin{equation}\label{curvR-as-fraction}
R = \left(\frac{c^2\slash(8\pi G)-\eta\slash c^2}{2\alpha q\rho}\right)^{\frac{1}{\alpha-1}},
\end{equation}
shows that $R$ is discontinuous at surfaces across which mass density is discontinuous, such as the external surfaces of Earth and Moon or the Mohorovi\v{c}i\'c discontinuity.
Since the extra force is proportional to $\nabla R$, then such a force is concentrated at the surfaces of density discontinuity and the concentration
gives rise to a jump of pressure $p$ across these surfaces that we now compute. We will find that such a jump of pressure is undetectable for interesting values
of parameter $\alpha$.

In the following the astronomical body considered is either the Earth or the Moon. Let $R_{\rm d}$ be the radius of a discontinuity surface, and let $\rho^+$ and $\rho^-$
be the values of density on the two sides of the discontinuity. We adopt the method of Ref. \cite{Obrien-Sy} and we introduce
a boundary layer across which density changes continuously, then we proceed to the limit in which the thickness of the layer tends to zero. 
Let $\eps$ be a positive small parameter and let $r$ be the distance from the center of the body, then we introduce a family of mass density functions $\rho_\eps$ defined by
\begin{equation}
\begin{cases}
\rho_\eps(r) = \rho(r) \quad\mbox{for } r < R_{\rm d}-\eps \mbox{ and } r > R_{\rm d}+\eps, \\
\rho_\eps(R_{\rm d}+\eps) = \rho^+, \quad \rho_\eps(R_{\rm d}-\eps) = \rho^-,
\end{cases}
\end{equation}
where $\rho_\eps(r)$ is continuously differentiable in the interval $(R_{\rm d}-\eps,R_{\rm d}+\eps)$.
Moreover, we denote by means of the subscript $\eps$ the functions which are solution of the field equations and equations of motion for assigned density $\rho_\eps$, and we assume
that such functions converge as $\eps\to 0$ to the solution of the problem with density $\rho$.

Let now $\hat\n=(\x-\x_o)\slash\vert\x-\x_o\vert$, where $\x_o$ is the position vector of the center of the body, then replacing $\rho$ with $\rho_\eps$ in Eqs. (\ref{NMC-contin-bodies}),
taking the scalar product of the resulting equations by the unit vector $\hat\n$, and integrating radially over the boundary layer of thickness $2\eps$,
at the time instant $t$ we have
\begin{eqnarray}
\Delta p_\eps &=& -c^2\int_{R_{\rm d}-\eps}^{R_{\rm d}+\eps}\rho_\eps f^2_{R_\eps}\frac{\partial R_\eps}{\partial r}dr \\
&+& \int_{R_{\rm d}-\eps}^{R_{\rm d}+\eps}\rho_\eps\left[-\frac{d\vv_\eps}{dt}\cdot\hat\n+\frac{\partial U_\eps}{\partial r}-\frac{c^2}{2}\frac{\partial\Psi_\eps}{\partial r}\right]dr, \nonumber
\end{eqnarray}
where $\Delta p_\eps=p_\eps(R_{\rm d}+\eps,\theta,\varphi,t) - p_\eps(R_{\rm d}-\eps,\theta,\varphi,t)$ and
$(r,\theta,\varphi)$ are spherical coordinates on the discontinuity surface.
Taking the limit as $\eps\rightarrow 0$, and observing that the derivatives in the integrand in the second row of the above equation are bounded
above and below for all $\eps$, we find
\begin{equation}
\Delta p = -c^2\lim_{\eps\rightarrow 0}\int_{R_{\rm d}-\eps}^{R_{\rm d}+\eps}\rho_\eps f^2_{R_\eps}\frac{\partial R_\eps}{\partial r}dr,
\end{equation}
where $\Delta p=p(R_{\rm d}^+,\theta,\varphi,t)-p(R_{\rm d}^-,\theta,\varphi,t)$ is the pressure jump.
In order to compute the pressure jump we now compute the above limit of the integral. Using $f^2(R_\eps)=q R_\eps^\alpha$ and Eq. (\ref{curvR-as-fraction}),
the integrand is given by
\begin{equation}
-\left(2\vert\alpha\vert\right)^{\frac{\alpha}{1+\vert\alpha\vert}}\vert q\vert^{\frac{1}{1+\vert\alpha\vert}}
\rho_\eps\frac{\partial}{\partial r}\left[ \rho_\eps^{\frac{\alpha}{1+\vert\alpha\vert}}\left(\frac{c^2}{8\pi G}-\frac{\eta_\eps}{c^2}\right)^{\frac{\alpha}{\alpha-1}}\right],
\end{equation}
which can be decomposed into the sum of a bounded term whose contribution vanishes in the limit as $\eps\rightarrow 0$, and an unbounded term which in the
limit gives rise to the pressure jump. The result of the computation is
\begin{equation}\label{pressure-jump}
\Delta p = \mathcal{C}(\alpha,q)
\left(\frac{c^2}{8\pi G}-\frac{\eta}{c^2}\right)^{\frac{\alpha}{\alpha-1}}
\left({\rho^-}^{\frac{1}{1+\vert\alpha\vert}} - {\rho^+}^{\frac{1}{1+\vert\alpha\vert}} \right),
\end{equation}
where the coefficient $\mathcal{C}(\alpha,q)$ is given by
\begin{equation}
\mathcal{C}(\alpha,q) = c^2 2^{\frac{\alpha}{1-\alpha}}\left(\alpha q\right)^{\frac{1}{1+\vert\alpha\vert}},
\end{equation}
and the function 
$\eta=\eta(R_{\rm d},\theta,\varphi,t)$ is evaluated on the discontinuity surface at the point with angular coordinates $(\theta,\varphi)$. The jump $\Delta p$
does not depend on the angular coordinates on the discontinuity surfaces which are located inside the screening spheres, while the jump varies with position on the 
external surfaces of Earth and Moon.

We see that the jump of pressure depends both on the solution $\eta$ and, explicitly, on the NMC gravity parameters $\alpha$ and $q$
(take into account that in Sec. \ref{sec:scalar-eta-equation} we have found that $\alpha<0$ implies $q<0$).

\subsection{Extra force inside the screening spheres}\label{sec:extra-force-inside-sr}

We consider the perturbation of the hydrostatic equilibrium due to extra force in the interior of astronomical bodies. Let us apply the jump condition for pressure
at the mantle-core discontinuity inside the Moon. Separating the motion of the center of mass from the internal condition of hydrostatic equilibrium in 
Eqs. (\ref{NMC-contin-bodies}), integrating radially and taking into account the pressure jump, we find
\begin{eqnarray}
& & p(R_{\rm d}^+) - p(0) -\Delta p = \int_0^{R_{\rm d}} \rho_M\frac{\partial U_M}{\partial r}dr \\
& & - \frac{c^2}{2}\int_0^{R_{\rm d}} \rho_M\frac{\partial \Psi_M}{\partial r}dr - c^2 \int_0^{R_{\rm d}} \rho_M f^2_R\left[\frac{\partial R}{\partial r}\right]dr, \nonumber
\end{eqnarray}
where $R_d=R_{M,n}$ is the radius of the mantle-core interface,
$p(0)$ is pressure at the Moon's center, $U_M$ is the Newtonian potential sourced by Moon's mass density, $\Psi_M$ is the potential inside the Moon's screening radius,
and $[\partial R/\partial r]$ denotes the derivative $\partial R/\partial r$ outside of the discontinuity surface.

In order to compute the pressure jump $\Delta p$ we use Eq. (\ref{pressure-jump}) and the solution (\ref{eta-Moon-core}) for $\eta$ inside the Moon's core:
\begin{eqnarray}
& &\frac{c^2}{8\pi G} - \frac{1}{c^2}\eta(R_{M,n}) = \frac{1}{3(1+|\alpha|)}\,\frac{\lambda_{M,m}\lambda_{M,n}}{\lambda_{M,m}+\lambda_{M,n}} \nonumber\\
&\times& \left(\lambda_{M,m}\rho_{M,m}+\lambda_{M,n}\rho_{M,n}\right) < \frac{2}{3(1+|\alpha|)}\,\lambda_{M,m}^2\rho_{M,m}, \nonumber\\
\end{eqnarray}
from which we obtain the following estimate for the pressure jump:
\begin{equation}
\vert\Delta p\vert < \frac{16}{3}\,\pi G \lambda_{M,m}^2\rho_{M,n}^2.
\end{equation}
For the Newtonian term we have
\begin{equation}\label{pressure-Newt-term-inside}
\int_0^{R_{\rm d}} \rho_M\frac{\partial U_M}{\partial r}dr = - \frac{2}{3}\,\pi G R_{M,n}^2\rho_{M,n}^2,
\end{equation}
from which, being $\lambda_{M,m}\ll R_{M,n}$ with $\lambda_{M,m}$ completely negligible ($\lambda(\rho_M)\lesssim 10^{-7}$ m for $\alpha=-1$ according to 
Section \ref{sec:Moon-density-model} of Appendix A), it then follows
\begin{equation}
\vert\Delta p\vert \ll \left\vert \int_0^{R_{\rm d}} \rho_M\frac{\partial U_M}{\partial r}dr \right\vert.
\end{equation}
The volume contribution of the extra force inside Moon's core is given by
\begin{equation}
- c^2 \int_0^{R_{\rm d}} \rho_M f^2_R\left[\frac{\partial R}{\partial r}\right]dr = -c^2\rho_{M,n}\left[ f^2(R_d^-) - f^2(0) \right],
\end{equation}
Using expression (\ref{curvR-as-fraction}) of curvature $R$, the solution (\ref{eta-Moon-core}) for $\eta$ inside Moon's core, taking into account
that $\lambda_{M,n}\ll R_d$ implies $\eta(0)\approx \eta_{M,n}$, 
we have
\begin{equation}
\vert f^2(R_d^-)\vert > \vert f^2(0)\vert,
\end{equation}
then we compute the following estimate:
\begin{eqnarray}
& & c^2 \left\vert \int_0^{R_{\rm d}} \rho_M f^2_R\left[\frac{\partial R}{\partial r}\right]dr \right\vert < 2 c^2\rho_{M,n}\vert f^2(R_d^-)\vert \nonumber\\
& & < \frac{16}{3} 2^{\frac{1}{\alpha-1}} \pi G \left(\frac{\rho_{M,n}}{\rho_{M,m}}\right)^{\frac{1}{1+|\alpha|}}\frac{\lambda_{M,m}^2}{\vert\alpha\vert}\rho^2_{M,m}.
\end{eqnarray}
Since $\lambda_{M,m}^2/|\alpha| \ll R_{M,n}^2$ for values of $\alpha$ of interest for application to astrophysics and cosmology (see Refs. \cite{drkmattgal},\cite{curraccel}
and Section \ref{sec:Moon-density-model} of Appendix A), then for such values the volume contribution of the extra force turns out to be completely negligible
in comparison with the Newtonian term (\ref{pressure-Newt-term-inside}).

The same behavior is found in other regions of the Moon's interior and in the Earth's interior, while for the Sun see Ref. \cite{MBMGDeA}.
The effectiveness of the screening mechanism is then confirmed by taking into account also the extra force.

Inside the thin shell of both the Moon and Earth the extra force contributes to the motion of centers of mass (see Section \ref{sec:Earth-Moon-accel-extra}) while the extra force exerted
by a body on itself has a leading contribution which is radial, so that it contributes only to hydrostatic equilibrium.
The contribution to hydrostatic equilibrium is computed in Earth's thin shell in Section \ref{Extra-force-t-shell} of Appendix E where it is shown that the resulting perturbation is 
again negligible in comparison with Newtonian force. An analogous result holds for the Moon, while for the Sun see Ref. \cite{MBMGDeA}.

\subsection{Motion of centers of mass}

In the following we denote by $\VV_E(t)$ and $\VV_M(t)$ the regions of space occupied by Earth and Moon, respectively, at the time instant $t$.
By using the continuity equation and Reynolds transport theorem of continuum mechanics, we have
\begin{equation}\label{Rey-mass-center}
M_\oplus\frac{d^2\x_E}{dt^2} = \int_{\VV_E(t)} \frac{d\vv}{dt}\rho_E(\x,t)d^3x,
\end{equation}
where $M_\oplus$ is the mass of Earth, and an analogous equation holds for the Moon. By substituting the expression of $\rho d\vv/dt$ given by Eq. (\ref{NMC-contin-bodies})
we obtain the various contributions to the acceleration of Earth and Moon. Because of the presence of discontinuities, the integrals involving the pressure gradient and the extra
force are computed as follows. We consider the Earth motion. The results of Section \ref{sec:pressure-jumps} show that the extra force makes pressure discontinuous
across the external surface of Earth, and the pressure jump,
\begin{equation}
\Delta p = p_{\rm atm} - p(R_\oplus^-,\theta,\varphi,t),
\end{equation}
where $p_{\rm atm}$ is atmospheric pressure at sea level, is given by formula (\ref{pressure-jump}) with $\rho^-=\rho_{E,w}$ and $\rho^+=\rho_{\rm atm}$, the atmospheric
density at sea level.

The contribution of the extra force to the integral on the right-hand side of Eq. (\ref{Rey-mass-center}) is then given by
\begin{equation}\label{extra-contrib-p-volume}
-c^2 \int_{\VV_E(t)}\rho_E f^2_R\left[\nabla R\right]d^3x - \int_{\partial\VV_E(t)}p(R_\oplus^-,\theta,\varphi,t)\hat\n_E d\sigma,
\end{equation}
where $[\nabla R]$ denotes the vector function $\nabla R$ outside of the discontinuity surfaces, and
we have taken into account that the contributions of pressure jumps across the discontinuity surfaces in the Earth's interior vanish because of spherical symmetry
of curvature $R$, assuming such surfaces contained in the screening sphere.
The surface integral of $p(R_\oplus^-,\theta,\varphi,t)$ is computed by using the expression (\ref{pressure-jump}) of $\Delta p$ taking into account that
the integral of $p_{\rm atm}$ vanishes because of spherical symmetry, assuming $p_{\rm atm}$ uniform on the Earth surface.

Then, substituting the expression of $\rho d\vv/dt$ given by Eq. (\ref{NMC-contin-bodies}) into the integral in Eq. (\ref{Rey-mass-center}),
and using Eq. (\ref{extra-contrib-p-volume}), we obtain
\begin{eqnarray}\label{total-Earth-accelerat}
M_\oplus\frac{d^2\x_E}{dt^2} &=& \int_{\VV_E(t)}\rho_E\nabla U\, d^3x - \frac{c^2}{2} \int_{\VV_E(t)}\rho_E\nabla\Psi\, d^3x \nonumber\\
&-& \int_{\partial\VV_E(t)}p^-\hat\n_E d\sigma - c^2 \int_{\VV_E(t)}\rho_E f^2_R\left[\nabla R\right]d^3x, \nonumber\\
\end{eqnarray}
where $p^- = p(R_\oplus^-,\theta,\varphi,t)$. The first integral is the contribution of Newtonian gravity to the acceleration of Earth, the second integral is the
contribution of fifth force, the surface integral is the contribution of the extra pressure on Earth's surface, and the last integral is the volume contribution of the extra force.
An analogous formula holds for the Moon.

In the next section we evaluate the integral which gives the contribution of the fifth force.

\section{Acceleration of Earth and Moon due to the fifth force}\label{sec:Earth-Moon-accel}

The fifth force contribution to the acceleration of Earth is given by
\begin{equation}
M_\oplus\left(\frac{d^2\x_E}{dt^2}\right)_{\rm f} =  -\frac{c^2}{2}\int_{\VV_E(t)} \rho_E(\x,t)\nabla\Psi(\x,t)d^3x,
\end{equation}
and an analogous integral holds for the Moon. In Sec. \ref{sec:fifth-force-inside-sr} we have found that the contribution to the fifth force from the interior of the screening sphere
is negligible due to the smallness of $\lambda(\rho_E)$, 
so that the contribution to the integral over $\VV_E(t)$ only comes from the thin shell defined by $r_E < \vert\x-\x_E\vert < R_\oplus$.

The integral over the thin shell is evaluated by using the expression (\ref{potential-Psi-sol}) of the potential $\Psi$ in terms of $\eta$,
which gives $\nabla\Psi=-(8\pi G/c^4)\nabla\eta$, and using the solution for the function $\eta$ computed by means of the method of images.

Let us first consider the contribution from the solar term $\eta_S$.
Using spherical coordinates  we find the values of the following integrals:
\begin{equation}
\frac{4\pi G}{c^2}\int_{\VV_E(t)} \rho_E\nabla\I_S d^3x = \frac{G}{3}M_{\odot,{\rm eff}}M_{\oplus,{\rm eff}}\frac{\x_S-\x_E}{\vert\x_S-\x_E\vert^3} \,,
\end{equation}
where we have dropped the dependence of effective masses on the screening radii, and
\begin{equation}\label{int-J_SE-vanish}
\int_{\VV_E(t)} \rho_E \nabla\J_{S,E} \,d^3x = 0.
\end{equation}
Let us now consider the contribution from the following terms of $\eta_S$:
\begin{equation}\label{eta_S-integral-plus-surf-term}
\frac{r_S}{\vert\x-\x_S\vert}\left(\frac{c^4}{8\pi G} - \eta_g - \frac{c^2}{3}\int_{r_S}^{R_\odot}\rho_S(r)r\,dr\right).
\end{equation}
Using the integral equation on the Sun's screening sphere, which is obtained from Eq. (\ref{radii-integral-equations}) by exchanging $E$ with $S$, we have
\begin{equation}\label{Sun-integr-eq-ORd}
\frac{c^2}{3}\int_{r_S}^{R_\odot}\rho_S(r)r\,dr = \frac{c^4}{8\pi G} - \eta_g + \OO\left(\frac{R}{d}\right),
\end{equation}
where $\OO(R/d)$ denotes terms multiplied by a factor of type $R\slash d$, where $R$ is a radius and $d$ is a distance between the astronomical bodies.
Hence, by taking the gradient of the expression (\ref{eta_S-integral-plus-surf-term}) and integrating over the Earth's thin shell, we find contributions to the
fifth force multiplied by factors $R\slash d$.

The contribution from the lunar term $\eta_M$ is obtained by replacing $S$ with $M$ in the previous expressions.

For the contribution from the terrestrial term $\eta_E$ we find
\begin{equation}\label{Earth-first-integral}
\int_{\VV_E(t)} \rho_E(\x,t)\nabla\I_E(\x,t)d^3x = 0,
\end{equation}
\begin{eqnarray}
& &\left(\frac{c^4}{8\pi G} - \eta_g - \frac{c^2}{3}\int_{r_E}^{R_\oplus}\rho_E(r)r\,dr\right) \times \nonumber\\
&\times& \int_{\VV_E(t)} \rho_E(\x,t)\nabla\frac{r_E}{\vert\x-\x_E\vert}d^3x = 0,
\end{eqnarray}
because of spherical symmetry.

All other contributions from $\eta$ resulting from Green's function $G(\x,\x_0)$ up to second order are discussed in Appendix D where it is argued that they either vanish
or are multiplied by factors $R\slash d$.
Moreover, it turns out that all contributions to the fifth force of the order of $R\slash d$ cancel each other. Since in the Sun-Earth-Moon system the ratios $R\slash d$ are small, then
corrections of the order of $(R\slash d)^2$ have not been computed because exceedingly small to give rise to observable effects.

Combining all these results, we obtain the acceleration of the Earth due to fifth force and computed by using the Green's function up to second order:
\begin{eqnarray}\label{Earth-fifth-force-result}
M_\oplus\left(\frac{d^2\x_E}{dt^2}\right)_{\rm f} &=& \frac{G}{3}M_{\oplus,{\rm eff}}\left[ M_{\odot,{\rm eff}}\frac{\x_S-\x_E}{\vert\x_S-\x_E\vert^3} \right. \nonumber\\
&+& \left. M_{M,{\rm eff}}\frac{\x_M-\x_E}{\vert\x_M-\x_E\vert^3} \right].
\end{eqnarray}
The acceleration of the Moon due to fifth force is analogous:
\begin{eqnarray}\label{Moon-fifth-force-result}
M_M\left(\frac{d^2\x_M}{dt^2}\right)_{\rm f} &=& \frac{G}{3}M_{M,{\rm eff}}\left[ M_{\odot,{\rm eff}}\frac{\x_S-\x_M}{\vert\x_S-\x_M\vert^3} \right. 
\nonumber\\
&+& \left. M_{\oplus,{\rm eff}}\frac{\x_E-\x_M}{\vert\x_E-\x_M\vert^3} \right].
\end{eqnarray}
In the case of astronomical bodies with uniform mass density
the above expressions coincide with the accelerations of Earth and Moon found in \cite{KW} for chameleon gravity and in
\cite{Capoz-Tsu} for $f(R)$ gravity theory. In the more realistic case of bodies with varying density such expressions give different results.
Moreover, the equations that determine the screening radii in NMC gravity in general are different from the corresponding equations in $f(R)$ gravity \cite{Burrage}.

Since the accelerations of the Earth and Moon depend on the effective masses of the bodies, which are the masses of the respective thin shells
and these depend on the internal structure of the bodies through density and size, then a violation of the Weak Equivalence Principle
takes place. Such a violation can be looked for in the Earth-Moon system by means of LLR measurements.

\section{Acceleration of Earth and Moon due to the extra force}\label{sec:Earth-Moon-accel-extra}

The acceleration due to extra force takes contributions from the extra pressure on the surface of the moving astronomical body
and from the volume part inside the body. In the sequel we compute the acceleration of Earth, while the acceleration of the Moon
is obtained by an analogous computation. Using Eqs. (\ref{total-Earth-accelerat}) and (\ref{pressure-jump}), the contribution of the surface integral of pressure
to the acceleration is given by
\begin{eqnarray}\label{Earth-accel-extra-force-surface}
& &M_\oplus\left(\frac{d^2\x_E}{dt^2}\right)_{\rm e,s} = - \int_{\partial\VV_E(t)}p(R_\oplus^-,\theta,\varphi,t)\hat\n_E d\sigma \nonumber\\
&=& \mathcal{C}(\alpha,q)\rho_{E,w}^{\frac{1}{1+\vert\alpha\vert}}
\int_{\partial\VV_E(t)}\left(\frac{c^2}{8\pi G}-\frac{\eta}{c^2}\right)^{\frac{\alpha}{\alpha-1}}\hat\n_E d\sigma, \nonumber\\
\end{eqnarray}
where we have neglected the atmospheric density with respect to the density of seawater. Using again Eq. (\ref{total-Earth-accelerat}) the volume part of the extra force,
taking into account that the contribution from the interior of Earth's screening sphere vanishes because of spherical symmetry, is given by
\begin{eqnarray}
& &M_\oplus\left(\frac{d^2\x_E}{dt^2}\right)_{\rm e,v} = - c^2 \int_{\VV_E(t)}\rho_E f^2_R\left[\nabla R\right]d^3x \nonumber\\
&=& - c^2 \int_{\Omega_E}\rho_E f^2_R\nabla R\,d^3x = -c^2\rho_{E,w}\int_{\Omega_E} \nabla f^2\, d^3x,\nonumber\\
\end{eqnarray}
where $\Omega_E=\{r_E<\vert\x-\x_E\vert<R_\oplus\}$ is the thin shell of Earth, which is assumed to lie inside seawater.
Using now Gauss theorem, the formula $f^2(R)=qR^\alpha$, the expression (\ref{curvR-as-fraction}) of curvature, and taking into account that $R$ is constant on the
screening surface, we have
\begin{eqnarray}\label{Earth-accel-extra-force-volume}
& &M_\oplus\left(\frac{d^2\x_E}{dt^2}\right)_{\rm e,v} = -c^2\rho_{E,w} \int_{\partial\VV_E(t)} f^2(R)\hat\n_E d\sigma \nonumber\\
&=& -\frac{1}{\alpha}\mathcal{C}(\alpha,q)\rho_{E,w}^{\frac{1}{1+\vert\alpha\vert}}
\int_{\partial\VV_E(t)}\left(\frac{c^2}{8\pi G}-\frac{\eta}{c^2}\right)^{\frac{\alpha}{\alpha-1}}\hat\n_E d\sigma, \nonumber\\
\end{eqnarray}
so that we have the following relation between the volume and surface contributions to Earth's acceleration:
\begin{equation}
\left(\frac{d^2\x_E}{dt^2}\right)_{\rm e,v} = -\frac{1}{\alpha}\left(\frac{d^2\x_E}{dt^2}\right)_{\rm e,s}.
\end{equation}
For the function $\eta$ inside the integral (\ref{Earth-accel-extra-force-volume}) we use the solution computed by means of the Green's function up to the second order:
note that such a solution was computed by neglecting curvature $R$ inside the thin shell according to inequality (\ref{deriv-potential-condit}), 
so that with this procedure we compute a first correction to $R=0$. This correction satisfies itself inequality (\ref{deriv-potential-condit}) for $\alpha<-1\slash 2$ as it is shown in Appendix E.

In order to approximate the surface integral (\ref{Earth-accel-extra-force-volume}) we introduce the small radius-to-distance ratios $\eps_1,\eps_2,\dots$, where
\begin{equation}\label{small-geo-ratios}
\eps_1 = \frac{r_E}{\vert\x_E-\x_S\vert}, \qquad \eps_2 = \frac{r_E}{\vert\x_E-\x_M\vert},
\end{equation}
the other geometric ratios are defined analogously, and the thin-shell parameter of Earth,
\begin{equation}\label{Earth-TS-delta-parameter}
\delta_E= \frac{\Delta R_\oplus}{R_\oplus} = \frac{R_\oplus-r_E}{R_\oplus},
\end{equation}
so that the thin-shell condition for Earth reads $\delta_E\ll 1$. If such a condition is satisfied, then Earth's screening surface lies inside seawater
and, at first order in $\delta_E$, we have the relations
\begin{equation}\label{Earth-eff-mass-deltaE}
M_{\oplus,{\rm eff}}(r_E) \approx 4\pi\rho_{E,w}R_\oplus^3\delta_E,
\end{equation}
and
\begin{equation}\label{integral-eff-mass-deltaE}
\int_{r_E}^{R_\oplus}\rho_E(r)rdr - \frac{M_{\oplus,{\rm eff}}(r_E)}{4\pi R_\oplus} \approx \frac{M_{\oplus,{\rm eff}}(r_E)}{8\pi R_\oplus}\,\delta_E.
\end{equation}
Now we observe that the Dirichlet condition (\ref{Dirichlet-bound-cond}) requires the function $(c^2\slash(8\pi G)-\eta\slash c^2)$ to vanish on Earth's screening surface,
so that such a function, when evaluated on the external surface of Earth, is infinitesimal with respect to $\delta_E$. Such an infinitesimal property has to be satisfied
by the approximation of function $\eta$ in order to have a valid approximation of the surface integral (\ref{Earth-accel-extra-force-volume}). Then, if we consider the solution $\eta$
computed with the Green's function up to the first order, we have argued in Sec. \ref{sec:screen-radii} that the sums $\I_S+\J_{S,E}$ and $\I_M+\J_{M,E}$ vanish
on Earth's screening surface, so that they are infinitesimal with respect to $\delta_E$ on $\partial\VV_E$.
Moreover, using Eqs. (\ref{eta-green-total})-(\ref{eta-green-Earth}), the radial terms of $(c^2\slash(8\pi G)-(\eta_E+\eta_g)\slash c^2)$, which depend on $\vert\x-\x_E\vert$, also
vanish on the screening surface.
The other terms in the first order solution do not vanish on the screening surface, so that the solution computed with the Green's function up to the second order has to be used.
By using the method of images for a system of spheres, we show in Appendix C that the remaining terms in the first order solution cancel on the screening surface
with terms of the second order solution, so that the sum of all such terms is infinitesimal with respect to $\delta_E$ on $\partial\VV_E$.

Using these results we make a second order Taylor approximation on the external surface of Earth, with respect to $\delta_E$, and $\eps_1,\eps_2,\dots$,
of the terms in the function $(c^2\slash(8\pi G)-\eta\slash c^2)$ which vanish on Earth's screening surface. By the above arguments
such a Taylor approximation has the overall multiplicative factor $\delta_E$.
By using the method of images one can check that the
remaining terms in the second order solution which do not vanish on the screening surface, and which are not considered in the Taylor approximation,
cancel with terms of the third order solution giving rise to smaller corrections, and this procedure can be iterated at higher orders.
The leading terms in the Taylor approximation are computed in Appendix D and they are the following:
\begin{eqnarray}\label{eta-difference-Taylor-expand}
& & \frac{c^2}{8\pi G}-\frac{1}{c^2}\eta(R_\oplus,\theta,\varphi,t) \approx \frac{\delta_E}{12\pi} \left[ \frac{M_{\oplus,{\rm eff}}}{2R_\oplus} + 3\times \right. \nonumber\\
&\times& \left. \left(\eps_1\frac{M_{\odot,{\rm eff}}}{\vert\x_E-\x_S\vert} \hat\n_{ES} + \eps_2\frac{M_{M,{\rm eff}}}{\vert\x_E-\x_M\vert} \hat\n_{EM} \right)\cdot\hat\n_E +\dots\right],
\nonumber\\
\end{eqnarray}
where $\theta,\varphi$ are angular coordinates on Earth's surface,
$\hat\n_E=\hat\n_E(R_\oplus,\theta,\varphi)$, the term multiplied by $\eps_1$ comes from the Taylor approximation of $\I_S+\J_{S,E}$, 
the term multiplied by $\eps_2$ comes from the approximation of $\I_M+\J_{M,E}$, and the dots represent all terms that involve the other geometric ratios.
We will find that these further terms give a negligible contribution to the extra force.

The further first order Taylor approximation with respect to $\eps_1,\eps_2,\dots$ then follows:
\begin{eqnarray}\label{eta-Earth-Taylor-delta-eps}
& &\left[ \frac{c^2}{8\pi G}-\frac{1}{c^2}\eta(R_\oplus,\theta,\varphi,t) \right]^{\frac{\alpha}{\alpha-1}} \approx \left(\frac{\delta_E}{12\pi}\right)^{\frac{\vert\alpha\vert}{1+\vert\alpha\vert}}
\nonumber\\
&\times&\left[ \left(\frac{M_{\oplus,{\rm eff}}}{2R_\oplus}\right)^{\frac{\vert\alpha\vert}{1+\vert\alpha\vert}} + \frac{3\vert\alpha\vert}{1+\vert\alpha\vert}
\left(\frac{M_{\oplus,{\rm eff}}}{2R_\oplus}\right)^{-\frac{1}{1+\vert\alpha\vert}} \right. \\
&\times&\left. \left( \eps_1\frac{M_{\odot,{\rm eff}}}{\vert\x_E-\x_S\vert} \hat\n_{ES} + \eps_2\frac{M_{M,{\rm eff}}}{\vert\x_E-\x_M\vert} \hat\n_{EM} \right)\cdot\hat\n_E +\dots\right].
\nonumber
\end{eqnarray}
Now, using the integral equation (\ref{radii-integral-equations}), substituting the expression (\ref{eta-g-minimizer}) of $\eta_g$ into the integral equation, 
neglecting factors involving radius-to-distance ratios in the integral equation and using the relation (\ref{integral-eff-mass-deltaE}), 
the term $\alpha q$ inside the coefficient $\mathcal{C}(\alpha,q)$ is approximated by
\begin{equation}
\alpha q \approx \left(\frac{8\pi G}{c^2}\right)^{1+\vert\alpha\vert}\rho_g^{-\alpha}\frac{M_{\oplus,{\rm eff}}}{24\pi R_\oplus}.
\end{equation}
Substituting the Taylor approximation (\ref{eta-Earth-Taylor-delta-eps}) and the above expression of $\alpha q$ inside the 
contributions (\ref{Earth-accel-extra-force-surface}) and (\ref{Earth-accel-extra-force-volume}) to the Earth acceleration,
using the relation (\ref{Earth-eff-mass-deltaE}) for the Earth's effective mass, executing the surface integrals, and adding the surface and volume parts,
we obtain the total expression of the extra force on Earth. We find that a computation identical to the one executed for the fifth force shows that
the terms represented by dots in the Taylor approximation (\ref{eta-Earth-Taylor-delta-eps}) give contributions to the extra force of the order of $(R\slash d)^2$, so
that they have been neglected. The final expression of the extra force is given by
\begin{eqnarray}
& &M_\oplus\left(\frac{d^2\x_E}{dt^2}\right)_{\rm e} \approx \left(\frac{2R_\oplus}{\Delta R_\oplus}\right)^{\frac{1}{1+\vert\alpha\vert}}\frac{G}{3}\,\frac{r_E}{R_\oplus}
\left(\frac{\rho_g}{\rho_{E,w}}\right)^{\frac{\vert\alpha\vert}{1+\vert\alpha\vert}} \nonumber\\
&\times& M_{\oplus,{\rm eff}} \left[ M_{\odot,{\rm eff}}\frac{\x_E-\x_S}{\vert\x_E-\x_S\vert^3} + M_{M,{\rm eff}}\frac{\x_E-\x_M}{\vert\x_E-\x_M\vert^3} \right]. \nonumber\\
\end{eqnarray}
Comparing with Eq. (\ref{Earth-fifth-force-result}), we see that the extra force and the fifth force are parallel vectors which point in opposite directions.
The ratio of magnitude of the two vectors is given by
\begin{equation}\label{Q-Earth-ratio-extra-fifth}
Q_\oplus \approx \left(\frac{2R_\oplus}{\Delta R_\oplus}\right)^{\frac{1}{1+\vert\alpha\vert}}\frac{r_E}{R_\oplus}
\left(\frac{\rho_g}{\rho_{E,w}}\right)^{\frac{\vert\alpha\vert}{1+\vert\alpha\vert}}.
\end{equation}
The contribution of the extra force to the Moon's acceleration is analogous:
\begin{eqnarray}
& &M_M\left(\frac{d^2\x_M}{dt^2}\right)_{\rm e} \approx \left(\frac{2R_M}{\Delta R_M}\right)^{\frac{1}{1+\vert\alpha\vert}}\frac{G}{3}\,\frac{r_M}{R_M}
\left(\frac{\rho_g}{\rho_{M,c}}\right)^{\frac{\vert\alpha\vert}{1+\vert\alpha\vert}} \nonumber\\
&\times& M_{M,{\rm eff}} \left[ M_{\odot,{\rm eff}}\frac{\x_M-\x_S}{\vert\x_M-\x_S\vert^3} + M_{\oplus,{\rm eff}}\frac{\x_M-\x_E}{\vert\x_M-\x_E\vert^3} \right], \nonumber\\
\end{eqnarray}
and the corresponding ratio of magnitude of this vector and the fifth force on the Moon is given by
\begin{equation}\label{Q-Moon-ratio-extra-fifth}
Q_M \approx \left(\frac{2R_M}{\Delta R_M}\right)^{\frac{1}{1+\vert\alpha\vert}}\frac{r_M}{R_M}
\left(\frac{\rho_g}{\rho_{M,c}}\right)^{\frac{\vert\alpha\vert}{1+\vert\alpha\vert}}.
\end{equation}
If the thin-shell condition for the Moon is satisfied, then the Moon's screening surface lies inside the Moon's crust so that the crustal density $\rho_{M,c}$ appears in the above formulae.

\section{Equivalence principle violation}

The results of the previous sections imply that the Earth and Moon fall toward the Sun with different accelerations, hence a violation of the universality of free fall (UFF) takes place.
The UFF can be satisfied only if the astronomical bodies are completely screened (the effective masses vanish), a condition that follow from the equations (\ref{radii-integral-equations})
that determine the screening radii only if $\eta_g=c^4/(8\pi G)$, hence $q=0$ from Eq. (\ref{eta-g-minimizer}), so that NMC gravity reduces to GR.

Using the expressions of the acceleration of Earth and Moon due to fifth force and extra force, computed in Secs. \ref{sec:Earth-Moon-accel} and \ref{sec:Earth-Moon-accel-extra},
the leading terms of the relative Earth-Moon acceleration are given by
\begin{eqnarray}
& &\a_M - \a_\oplus = -GM^\ast\frac{\x_M-\x_E}{\vert\x_M-\x_E\vert^3} + \Delta_{\rm ESM}GM_\odot\frac{\x_S-\x_E}{\vert\x_S-\x_E\vert^3} \nonumber\\
& & +\left( 1+ \Delta_{\rm tidal}\right)GM_\odot\left(\frac{\x_E-\x_S}{\vert\x_E-\x_S\vert^3} - \frac{\x_M-\x_S}{\vert\x_M-\x_S\vert^3}\right), \nonumber\\
\end{eqnarray}
where
\begin{eqnarray}
\Delta_{\rm ESM} &=& \frac{1}{3}\frac{M_{\odot,{\rm eff}}}{M_\odot}\left[ (1-Q_M)\frac{M_{M,{\rm eff}}}{M_M} -
(1-Q_\oplus)\frac{M_{\oplus,{\rm eff}}}{M_\oplus} \right], \nonumber\\
\label{Delta-ESM}\\
M^\ast &=& M_\oplus + M_M \nonumber\\
&+& \frac{1}{3}\frac{M_{\oplus,{\rm eff}}}{M_\oplus}\frac{M_{M,{\rm eff}}}{M_M}\left[M_\oplus(1-Q_M)+M_M(1-Q_\oplus)\right], \nonumber\\
\\
\Delta_{\rm tidal} &=& \frac{1}{3}(1-Q_M)\frac{M_{M,{\rm eff}}}{M_M}\frac{M_{\odot,{\rm eff}}}{M_\odot}.
\end{eqnarray}
The meaning of the terms in the expression of $\a_M-\a_\oplus$ is the following \cite{Wi}:
\begin{itemize}
\item[{\rm (i)}] the first term is the relative acceleration due to the gravitational attraction between the Earth and Moon;
\item[{\rm (ii)}] the second term, which can be written in the form,
\begin{equation}\label{UFF-accell}
\Delta_{\rm ESM}\, \mathbf{g}_S,
\end{equation}
$\mathbf{g}_S$ being the Newtonian acceleration of Earth due to the Sun, is the UFF violation-related difference between the Earth and the Moon accelerations toward the Sun, hence,
in the framework of NMC gravity, this term gives rise to a violation of the WEP;
\item[{\rm (iii)}] the third term is the solar tidal perturbation of the Moon's orbit, $\Delta_{\rm tidal}$ being the NMC gravity correction to the Newtonian perturbation.
\end{itemize}
The size of the UFF violation is represented by the parameter $\Delta_{\rm ESM}$, where ESM stands for Sun, Earth and Moon, since, by definition of the effective mass,
such a parameter depends on the composition (density) and size of all the three astronomical bodies \cite{Visw-Fienga}. Particularly, the WEP violation depends on
size and composition of the Sun, in addition to the more usual dependence on size and composition of the Earth and Moon.

If the astronomical bodies are screened with screening radii close enough to radii of the bodies, then
the effective masses and, consequently, $\Delta_{\rm ESM}$ can be made small enough in such a way that an experimental bound on WEP can be satisfied.

If we denote $d_{\rm EM}$ the Earth-Moon distance, then, the UFF acceleration (\ref{UFF-accell}) gives rise to a polarization of the Moon's orbit in the direction of the Sun with 
a periodic perturbation $\Delta d_{\rm EM}$ of the form \cite{Nordtvedt-1}
\begin{equation}
\Delta d_{\rm EM} = S\Delta_{\rm ESM}\cos D,
\end{equation}
where $D$ is the synodic phase measured from the new Moon, and $S$ is a scaling factor whose theoretical computation gives $S=2.9\times 10^{12}$ cm \cite{Nordtvedt-2}.

The Newtonian solar tidal perturbation of the Moon's orbit also produces a periodic perturbation of $d_{\rm EM}$ that has a component at the synodic frequency;
nevertheless, such a perturbation can be accounted for to a very small uncertainty \cite{Williams09}, so that the corresponding NMC gravity correction can be neglected for
$|\Delta_{\rm tidal}|\ll 1$.

In the next section constraints on the NMC gravity model will be obtained by means of the test of WEP performed in Ref. \cite{Visw-Fienga} using lunar LLR data.

\section{LLR constraints on NMC gravity parameters}\label{sec:LLR-constraints}

In Ref. \cite{Visw-Fienga} the authors give a general constraint in terms of difference between the Earth and the Moon accelerations toward the Sun, without assuming metric theories
or other types of modified gravity theories. In order to test UFF violations, a supplementary acceleration of the form (\ref{UFF-accell}) is introduced in the geocentric equation of
motion of the Moon. The parameter $\Delta_{\rm ESM}$ is estimated in the LLR adjustment together with a set of parameters of the lunar ephemerides listed in \cite{Visw-Fienga}.

Then, using the expression (\ref{Delta-ESM}), specific of our framework, the estimate of $\Delta_{\rm ESM}$ obtained in \cite{Visw-Fienga} can be directly translated into a
constraint on NMC gravity parameters.
The result on the WEP violation parameter in \cite{Visw-Fienga}, based on 48 years of LLR data, is given by
\begin{equation}\label{Delta-ESM-estimate}
\Delta_{\rm ESM} = \left(3.8\pm 7.1\right)\times 10 ^{-14}.
\end{equation}
In our framework $\Delta_{\rm ESM}$ is a function of $\alpha$ and of the screening radii of the Sun, Earth and Moon, 
\begin{equation}
\Delta_{\rm ESM} = \Delta_{\rm ESM}(\alpha,r_S,r_E,r_M),
\end{equation}
which are determined by Eqs. (\ref{radii-integral-equations}) as functions of the parameters $\alpha$ and $q$ of the NMC gravity model:
\begin{equation}
r_S=r_S(\alpha,q),\quad r_E=r_E(\alpha,q),\quad r_M=r_M(\alpha,q).
\end{equation}
Substituting such functions in the constraint (\ref{Delta-ESM-estimate}) we obtain the set of the admissible values of parameters $\alpha$ and $q$.

In order to avoid either too small or too large numbers we replace parameter $q$ with the following rescaled, dimensionless parameter:
\begin{equation}\label{q-rescaled}
\widetilde q = q R_g^\alpha,
\end{equation}
with $R_g=8\pi G\rho_g\slash c^2$. With this substitution, the function $f^2(R)$ can be written in the form
\begin{equation}
f^2(R) = \widetilde q \left(\frac{R}{R_g}\right)^\alpha.
\end{equation}
We represent the constraints from LLR by means of a two-dimensional exclusion plot in the plane with coordinates $\alpha,q$.
The resulting admissible region in the plane is restricted by means of the condition $\lambda_g\gg r_g$ introduced in Section \ref{Sec:outskirts} which,
expressed in terms of parameters $\alpha,\widetilde q$ and using Eq. (\ref{lambda-expression}) with $\rho=\rho_g$, becomes
\begin{equation}
\label{q2lowerbound}
\left[ \frac{3}{4\pi}\alpha(1-\alpha)\frac{c^2}{G\rho_g}\widetilde q \right]^{1\slash 2} > 10^2 r_g,
\end{equation}
where we have required $\lambda_g>10^2 r_g$.

In the numerical computation the screening radii have been determined by Eqs. (\ref{radii-integral-equations}) neglecting the factors involving radius-to-distance ratios,
since they give rise to corrections that are not visible at the scale of the following exclusion plots.

\begin{widetext}
	
\begin{figure}[H]
\centering
\begin{minipage}{.49\columnwidth}
	\centering
	\includegraphics[width=1.01\textwidth]{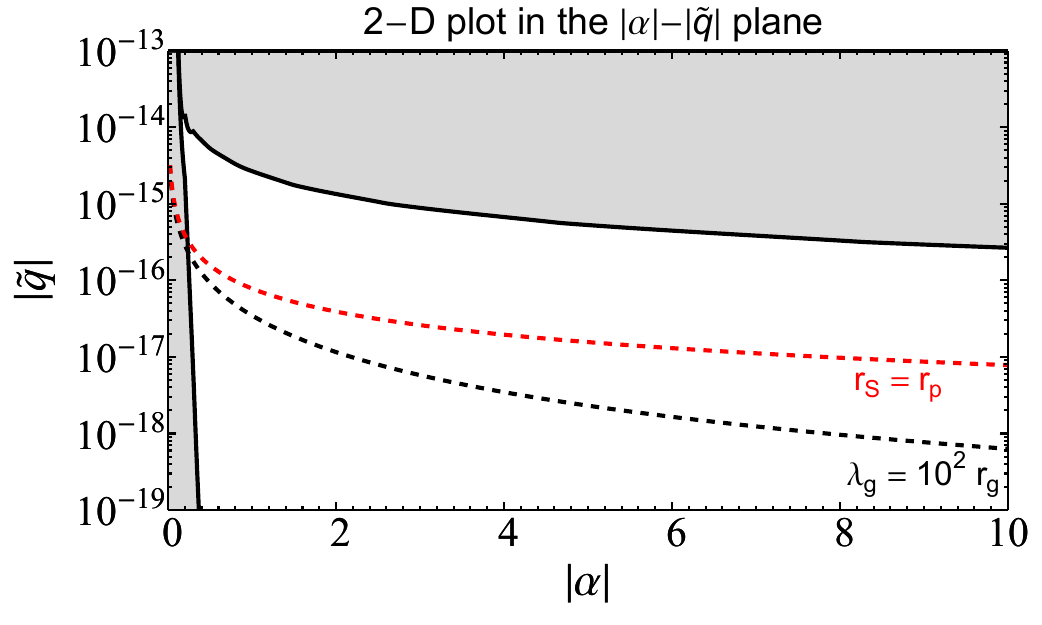}
	\caption{LLR constraints on the parameter quarter plane $|\alpha|,|\widetilde q|$ for $0>\alpha> -10$. The solid black line yields the upper bound on $|\widetilde q|$
	from LLR data, the dashed black line yields the lower bound on $|\widetilde q|$ from inequality $\lambda_g > 10^2 r_g$. The dashed red line represents the condition
	Sun's screening radius $r_S$ equal to the radius $r_p$ at the base of the solar photosphere.}
	\label{fig:alpha-q-extended}
\end{minipage}\hfill
\begin{minipage}{.49\columnwidth}
        \vspace{-0.8 cm}
	\centering
	\includegraphics[width=1.01\textwidth]{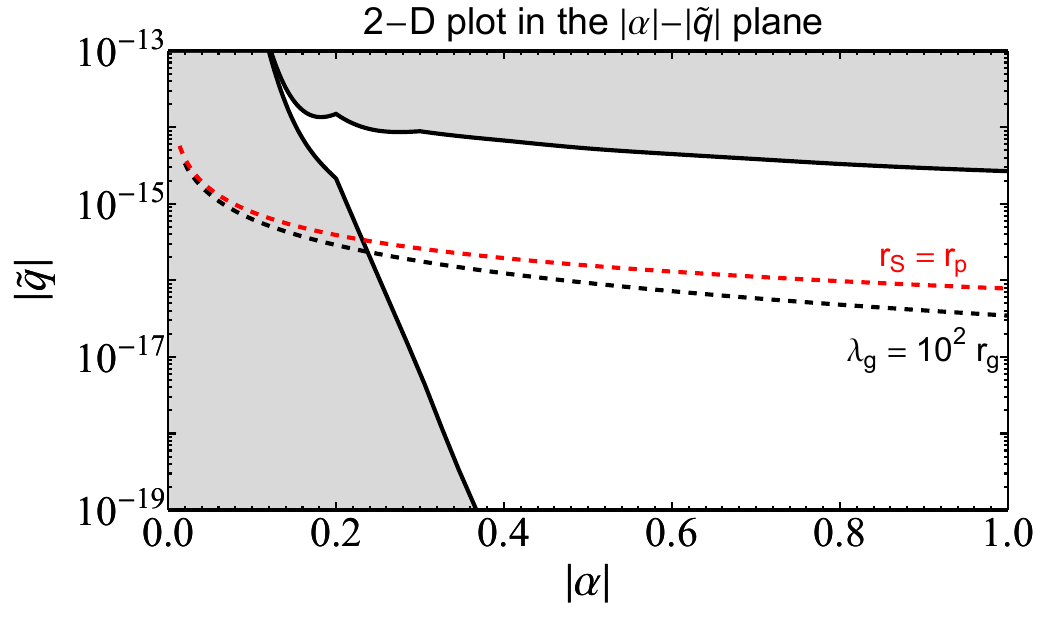}
	\caption{LLR constraints on the parameter quarter plane $|\alpha|,|\widetilde q|$ for $0>\alpha> -1$. The solid black line yields the upper bound on $|\widetilde q|$
	from LLR data. The meaning of the dashed black line and of the dashed red line is the same as in Fig. \ref{fig:alpha-q-extended}.}
	\label{fig:alpha-q}
\end{minipage}
\end{figure}
\end{widetext}

The LLR constraints are graphically reported in Figures \ref{fig:alpha-q-extended} and \ref{fig:alpha-q}: admissible regions for parameters are plotted in white,
while the excluded regions are plotted in grey. 
Fig. \ref{fig:alpha-q-extended} shows the admissible region for values of parameter $\alpha$ in the range $(-10,0)$;
since both $\alpha$ and $\widetilde q$ are negative the admissible region is plotted in the quarter plane with coordinates $(|\alpha|,|\widetilde q|)$.
Fig. \ref{fig:alpha-q} shows the admissible region for values of $\alpha$ in the range $(-1,0)$. The portion of the admissible region which lies above the
dashed red line $r_S=r_p$ corresponds to values of $r_S$ such that the Sun's screening sphere lies in the solar convection zone.

In Ref. \cite {MBMGDeA} it has been found that the Cassini measurement of PPN parameter $\gamma$ constrains the parameters $\alpha,\widetilde q$
to be of the order $|\widetilde q| < 10^{-12}$ for $-1> \alpha > -10$. Hence, the constraints from WEP violation and LLR data provide tighter bounds on
model parameters compared to bounds from Cassini measurement.

\begin{widetext}
	
\begin{figure}[H]
\centering
\begin{minipage}{.49\columnwidth}
	\centering
	\includegraphics[width=1.01\textwidth]{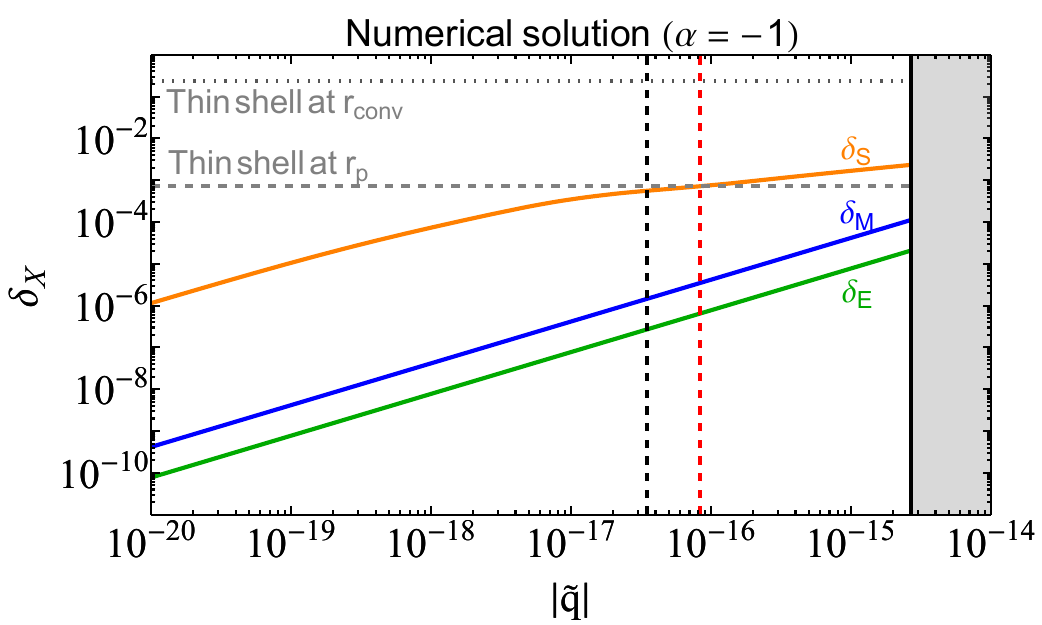}
	\caption{Plot of thin-shell parameters $\delta_S,\delta_M$ and $\delta_E$ versus $\vert\widetilde q\vert$ for $\alpha=-1$. The vertical solid black line on the right
	corresponds to the upper bound on $|\widetilde q|$ for $\alpha=-1$. The vertical dashed black line corresponds to the condition $\lambda_g=10^2 r_g$ and the red one
	corresponds to $r_S=r_p$. The horizontal dashed and dotted lines correspond to $r_S=r_p$ and $r_S=r_{\rm conv}$, respectively.}
	\label{fig:delta-param-alpha=-1}
\end{minipage}\hfill
\begin{minipage}{.49\columnwidth}
        \vspace{-0.4 cm}
	\centering
	\includegraphics[width=1.01\textwidth]{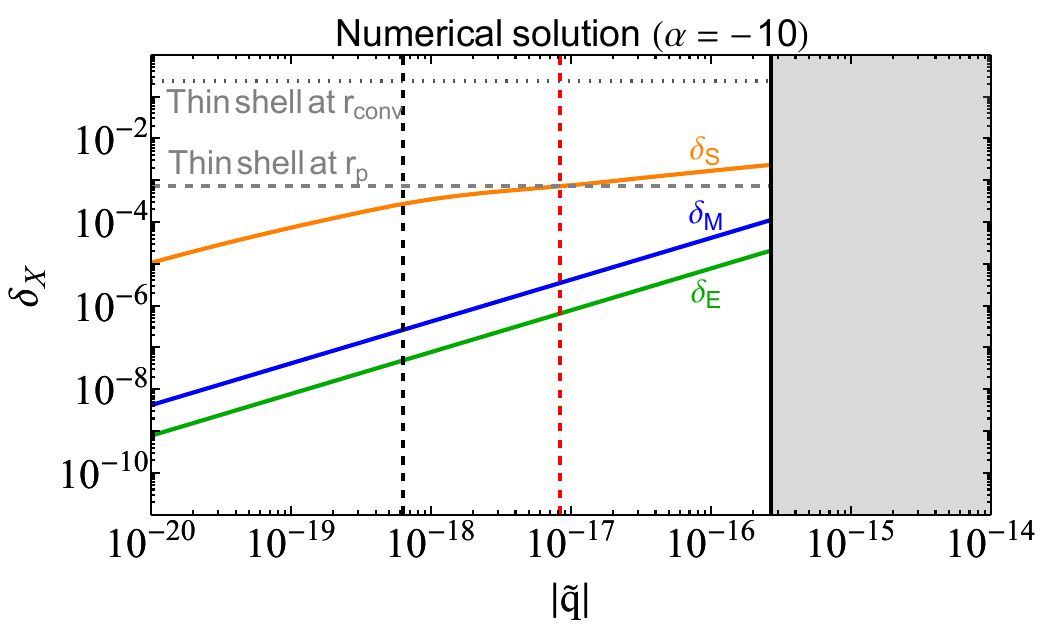}
	\caption{Plot of thin-shell parameters $\delta_S,\delta_M$ and $\delta_E$ versus $\vert\widetilde q\vert$ for $\alpha=-10$. The vertical solid black line on the right
	corresponds to the upper bound on $|\widetilde q|$ for $\alpha=-10$. The meaning of the vertical dashed black and red lines, and of the horizontal dashed and
	dotted lines, is the same as in Fig \ref{fig:delta-param-alpha=-1}.}
	\label{fig:delta-param-alpha=-10}
\end{minipage}
\end{figure}
\end{widetext}

Figures \ref{fig:delta-param-alpha=-1} and \ref{fig:delta-param-alpha=-10} show the thin-shell parameter of Earth $\delta_E$ and the corresponding parameters
$\delta_S$ and $\delta_M$ of Sun and Moon, which are defined analogously to formula (\ref{Earth-TS-delta-parameter}), plotted versus $\vert\widetilde q\vert$ for a fixed
value of $\alpha$. Fig. \ref{fig:delta-param-alpha=-1} shows the plot for $\alpha=-1$ and Fig. \ref{fig:delta-param-alpha=-10} for $\alpha=-10$.
The excluded regions are colored in grey. The figures show that increasing $\vert\widetilde q\vert$ the Sun's screening radius enters into the convection zone.

The numerical results show that for values $\vert\alpha\vert >1/2$, such that the thin-shell approximation is self-consistent, we have for the Earth
the upper bound $\delta_E<2\times 10^{-5}$ which, using Earth's density profile in Section \ref{sec:Earth-density-model} of Appendix A, 
corresponds to a thin shell $\Delta R_\oplus\approx 127$ m in seawater. Since the Earth has been modeled by means of a sphere (as it is usually done for this kind of problems),
it is interesting to compare the thin shell with topographic variation on Earth's surface. First we observe that the ocean and seas cover $70.8\%$ of the surface of the Earth, then, because of
currents, tides and other dynamic effects, the surface of oceans departs by roughly $\pm 2$ m from the {\em geoid}, which is the equipotential surface of the Earth's gravity field
(including centrifugal force) going through the ocean surfaces in average \cite{Vanicek}. The geoid, in turn, locally differs from the reference ellipsoid used in geodesy which has a small
flattening $f\approx 1/300$, so that it is very close to a sphere. The geoid undulations, which are the local deviations between the geoid and the ellipsoid, range worldwide from
-107 m (North Central Indian Ocean) to 85 m (Western Pacific, east of New Guinea) relative to the ellipsoid, but over the large majority of the ocean surface the undulations range
from -20 m to 20 m \cite{Lemoine}. Hence, on the majority of Earth's areas, the deviation between the Earth's topographic surface and a sphere is less than $1/6$ of the value of 
$\Delta R_\oplus$ which saturates the LLR constraint,  so that the use of the spherical approximation of the thin shell is justified in order to find analytical order of magnitude estimates
of the constraints on the NMC gravity parameters.

For the Moon we have the upper bound $\delta M<10^{-4}$ which, using Moon's density profile in Section \ref{sec:Moon-density-model} of Appendix A, 
corresponds to a thin shell $\Delta R_M\approx 174$ m in the lunar crust.

\begin{figure}[H]
\centering
\includegraphics[width=0.49\textwidth]{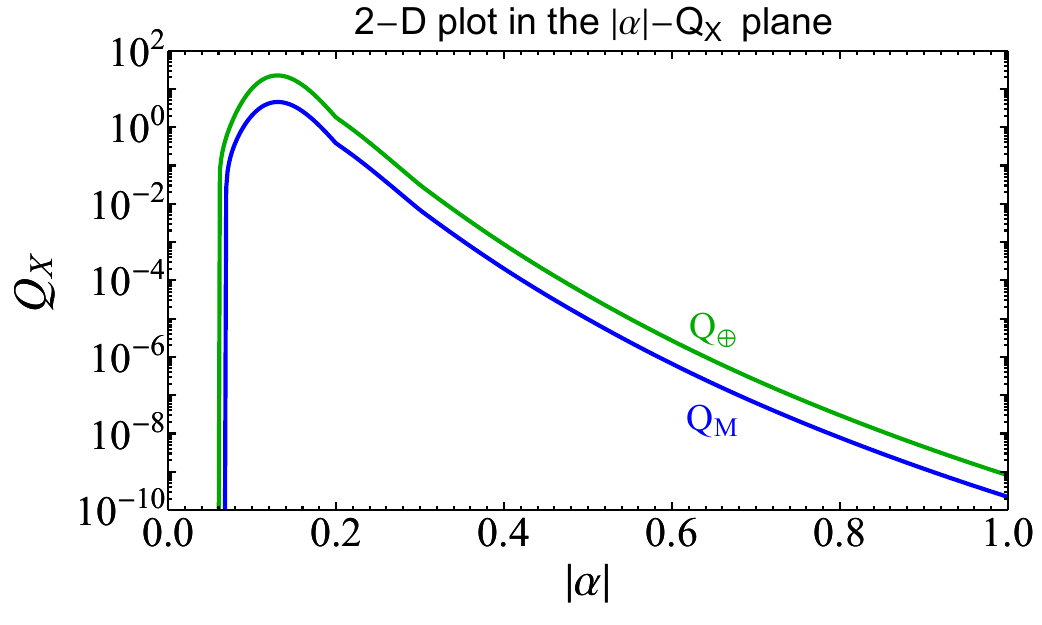}
\caption{Plot of the ratios $Q_\oplus$ and $Q_M$ of magnitude of extra force and fifth force versus $\vert\alpha\vert$ for $-1<\alpha<0$.}
\label{fig:Q_E-Q_M}
\end{figure}

Figure 5 shows the ratios $Q_\oplus$ and $Q_M$ of magnitude of extra force and fifth force for the Earth and Moon, respectively, versus $\vert\alpha\vert$ for $-1<\alpha<0$.
The ratios $Q_\oplus$ and $Q_M$ are of order unity for $\alpha\approx -1\slash 5$, however, in Appendix E it is shown that the thin shell inequality (\ref{deriv-potential-condit}) is satisfied for 
$\alpha<-1\slash 2$ so that the solution for $\eta$ is self-consistent for such values of $\alpha$. For $\alpha<-1\slash 2$ both $Q_\oplus$ and $Q_M$ are less than $10^{-4}$ so that
the extra force is negligible in comparison with the fifth force. The smallness of the ratios $Q_\oplus$ and $Q_M$ is due to the presence in Eqs. (\ref{Q-Earth-ratio-extra-fifth}) and
(\ref{Q-Moon-ratio-extra-fifth}) of the term
\begin{equation}
\left(\frac{\rho_g}{\rho}\right)^{\frac{\vert\alpha\vert}{1+\vert\alpha\vert}},
\end{equation}
where $\rho$ is density in the thin shell of the astronomical body: either the density of seawater on Earth or density of the lunar crust.
Eventually, at the galactic scale where density is of the order of $\rho_g$ we expect the extra force to have relevant effects.

We conclude this section with the evaluation of the pressure jump on Earth at the interface between seawater and atmosphere. Using expression (\ref{pressure-jump}) of the
pressure jump with $\rho^-=\rho_{E,w}$ and $\rho^+=\rho_{\rm atm}$, the Taylor approximation (\ref{eta-Earth-Taylor-delta-eps}) and the thin shell approximation, we find the
following leading term for the pressure jump:
\begin{equation}
\Delta p \approx \frac{G}{3}\frac{M_{\oplus,{\rm eff}}}{R_\oplus}\rho_{E,w}\left(\frac{\delta_E}{2}\frac{\rho_g}{\rho_{E,w}}\right)^{\frac{\vert\alpha\vert}{1+\vert\alpha\vert}},
\end{equation}
where we have neglected atmospheric density with respect to density of seawater. The pressure jump increases by decreasing $\vert\alpha\vert$ and the value of $\Delta p$
for $\alpha=-1/2$ at the saturation of the LLR constraint, which corresponds to $\delta_E=2\times 10^{-5}$, is given by $\Delta p\approx 6\times 10^{-10}p_{\rm atm}$, where
$p_{\rm atm}$ is atmospheric pressure at sea level, hence a negligible quantity. By increasing $\vert\alpha\vert$ the jump $\Delta p$ decreases quickly.

\section{Conclusions}

In this work we have examined the NMC theories of gravity specified by Eq. (\ref{f1-f2-specific}). These theories for certain values of $q$ and $\alpha$ are potentially inconsistent with the Equivalence Principle given that the fifth force and the extra non-Newtonian force they give rise could imply that Earth and Moon fall differently towards the Sun. We have shown that the screening chameleon-type mechanism, discussed for the first time in the context of these theories in Ref. \cite{MBMGDeA}, can be used to shield any violation of the Equivalence Principle provided the astronomical bodies are screened in such a way that their screening radii are close enough to the radii of the bodies as discussed in section IX. Furthermore, we have shown in section X for which range of values of the parameters $\alpha$ and $q$,  compatibility with 48 years of LLR data can be ensured. Hence, the screening mechanism here discussed does allow for compatibility with data even of versions of the NMC gravity theories that are potentially problematic. We find that constraints from WEP compatible with the  LLR data yield tighter bounds on the model parameters compared than the corresponding bounds from the Cassini measurement of the parametrized post-Newtonian parameter, $\gamma$ \cite{MBMGDeA}. 
Indeed, WEP and LLR data, compatible with lunar ephemerides, do provide stringent bounds on the parameters $q$ and $\alpha$, more specifically, as depicted in
Figs. \ref{fig:alpha-q-extended} and \ref{fig:alpha-q}, for $\vert \alpha \vert$ and the rescaled value $\vert \widetilde  q \vert$ given by Eq. (\ref{q-rescaled}). Compatibility with data is found for
$\vert\widetilde q\vert < 10^{-14}$ and for $0.5 < \vert \alpha \vert < 10$. As for the robustness of the screening mechanism, the thin-shell conditions were tested for different model parameters for Earth, Moon and Sun, as shown in Figs. \ref{fig:delta-param-alpha=-1} and \ref{fig:delta-param-alpha=-10}. As can one inspect, consistency is found for the range of model parameters that are compatible with data.

Recently, WEP has been tested with a precision of $10^{-15}$ by the MICROSCOPE mission \cite{Microscope}, nevertheless, the resulting bounds on chameleon gravity are not
competitive with state-of-the-art constraints \cite{Pernot-Borras} since MICROSCOPE was not designed for testing modified gravity theories of this type. 

As a final remark, we can state, on general grounds, that our results provide yet another confirmation that the NMC gravity model under study remains a workable alternative to address certain issues for which GR is not fully adequate. Furthermore, given that the particular subclass of NMC model under study requires the development of various techniques for the implementation of the screening mechanism, its relevance goes beyond the specificities of the model as the technical issues addressed in our work can, in principle, be useful in any model of gravity. 
 
\section*{Acknowledgments}

The work of R.M. is partially supported, and the work of M.M. and S.DA is fully supported, by 
INFN (Istituto Nazionale di Fisica Nucleare, Italy), as part of the MoonLIGHT-2 experiment in the framework 
of the research activities of the Commissione Scientifica Nazionale n. 2 (CSN2).
S.DA and M.M. would like to thank ASI (Agenzia Spaziale Italiana) for the support granted to them with ASI-INFN Agreement n. 2019-15-HH.0
and ESA (European Space Agency) for the support granted to them with ESA-INFN Contract n. 4000133721/21/NL/CR.

O.B. acknowledges the partial support from Funda\c{c}\~ao para 
a Ci\^encia e a Tecnologia (Portugal) through the research project
CERN/FIS-PAR/0027/2021. 

\appendix
\section*{Appendix A: interior density profiles}

We report on models of mass density profiles for the Sun, Earth and Moon which have been
used in order to find analytical estimates of the constraints on the NMC gravity parameters at a suitable order of magnitude.

\subsection{Sun density profile}\label{sec:Sun-density-model}

The complete model of mass density of the Sun is reported in Ref. \cite{MBMGDeA}. In this appendix we report the density profiles of the convection zone
and photosphere since the screening radius $r_S$ that saturates the LLR bound lies in such zones. For the details see Ref. \cite{MBMGDeA}.

The radius of the Sun is $R_\odot=6.9634\times 10^5\,{\rm km}$, 
the solar atmosphere begins below the spherical surface of radius $R_\odot$ and center in the origin, 
at a depth of about 500 km, and extends outward from the Sun. Then $r_p=R_\odot-500\,{\rm km}$ is the radius at the base of the photosphere.
Matter in the Sun is modeled as a perfect gas in hydrostatic equilibrium.

{\bf Convection zone}. We use a polytrope model with an effective polytropic index $n_c=2.33$ \cite{Mullan}:
\begin{equation}\label{conv-density-prof}
\rho_S(r) = K_c\left[ T_S(r) \right]^{n_c},
\end{equation}
with $K_c=3.44\times 10^{-16}$, and the radius $r$ varying in the range
\begin{equation}
r_{\rm conv} \leq r < r_p, \qquad r_{\rm conv} = 5.3185\times 10^5 \,\mbox{km}.
\end{equation}
The temperature profile is approximated by \cite{Mullan}
\begin{equation}\label{conv-temp-prof}
T_S(r) = \frac{GM_\odot}{C_p r} - T_0,
\end{equation}
with $M_\odot=1.989\times 10^{33}\,{\rm g}$,
$C_p=2.95\times 10^8 \,{\rm erg\,g^{-1}\,K^{-1}}$ is an averaged value of the specific heat at constant pressure,
and $T_0=6.461\times 10^6 \,{\rm K}$.

{\bf Photosphere}.
For the density in the photosphere we use the following model, adapted from \cite{DeJager}, for $r_p\leq r < R_\odot$:
\begin{equation}
\rho_S(r) = \frac{\mu m_p p_m}{k_B[T_m+A(R_\odot-r)^2])}\exp\left(\frac{R_\odot-r}{H_p}\right),
\end{equation}
where $\mu=1.26$, $m_p=1.66\times 10^{-24}\,{\rm g}$ is the proton mass, $k_B=1.3806\times 10^{-16}\,{\rm erg \, K^{-1}}$ is the Boltzmann constant,
$H_p=117\,{\rm km}$, $A=8.8\times 10^{-3}\,{\rm km^{-2}K}$, $T_m=4.4\times 10^3\,{\rm K}$ is the temperature minimum at the top of the photosphere,
$p_m$ is pressure corresponding at the temperature minimum, such that $p_m\slash k_B=1.2\times 10^{19}\,{\rm K\,cm^{-3}}$.

The Sun density profile permits us to compute the values of the length $\lambda(\rho_S)$. By using the definition (\ref{lambda-expression}) of $\lambda$ we have
\begin{equation}\label{lambda-ratio-rho_g}
\lambda(\rho_S) = \lambda_g\left(\frac{\rho_S}{\rho_g}\right)^{\frac{\alpha-1}{2}},
\end{equation}
and by the discussion in Sec. \ref{Sec:outskirts} we require $\lambda_g \gg r_g$.
For instance, if $\alpha=-1$ and $\lambda_g=10^3 r_g \sim 10^5$ AU,
at the bottom of the solar convection zone where $\rho_S\approx 1.65\times 10^{-1} \,{\rm g}/{\rm cm}^3$, we have
$\lambda(\rho_S)\approx 10^{-6} \,{\rm m}$. At the top of the convection zone where $\rho_S\approx 2.73\times 10^{-7} \,{\rm g}/{\rm cm}^3$, we have
$\lambda(\rho_S)\approx 7\times 10^{-1} \,{\rm m}$.

\subsection{Earth density profile}\label{sec:Earth-density-model}

We consider an average Earth model in the sense of \cite{PREM}. The Earth is divided into four regions with constant mass density
separated by spherical surfaces of density discontinuities: ocean layer, crust, mantle and core. The following numerical values of radii and densities
are taken from \cite{PREM,Stacey}. The radius of the Earth is $R_\oplus = 6371$ km.

{\bf Ocean layer}. Since the ocean and seas cover $70.8\%$ of the surface of the Earth, the uppermost layer of the average Earth model consists of seawater with
a depth of 3 km. The layer corresponds to distances $r$ from the Earth's center $R_{E,c}<r\leq R_\oplus$:
\begin{equation}
R_{E,c} = 6368\,\,{\rm km}, \qquad \rho_{E,w}=1.02\,\,{\rm g\,cm^{-3}}.
\end{equation}

{\bf Crust}. The layer corresponds to $R_{E,m}<r\leq R_{E,c}$:
\begin{equation}
R_{E,m} = 6346.6\,\,{\rm km}, \qquad \rho_{E,c}=2.7\,\,{\rm g\,cm^{-3}}.
\end{equation}
At radius $R_{E,m}$ there is the Mohorovi\v{c}i\'c discontinuity.

{\bf Mantle}. The layer corresponds to $R_{E,n}<r\leq R_{E,m}$:
\begin{equation}
R_{E,n} = 3480\,\,{\rm km}, \qquad \rho_{E,m}=4.5\,\,{\rm g\,cm^{-3}}.
\end{equation}
At radius $R_{E,n}$ there is the Gutenberg discontinuity.

{\bf Core}. The layer corresponds to $0<r\leq R_{E,n}$:
\begin{equation}
\rho_{E,n}=11\,\,{\rm g\,cm^{-3}}.
\end{equation}
Using Eq. (\ref{lambda-ratio-rho_g}) we may compute $\lambda(\rho_E)$ in the various Earth's layers. We have $\lambda(\rho_E) \leq \lambda(\rho_{E,w})$, then,
using for instance $\alpha=-1$ and $\lambda_g= 10^5$ AU, we find $\lambda(\rho_{E,w})\approx 10^{-7} \,{\rm m}$, hence a completely negligible quantity.

\subsection{Moon density profile}\label{sec:Moon-density-model}

The Moon is divided into three regions with constant mass density
separated by spherical surfaces of density discontinuities: crust, mantle and core.
The following numerical values of radii and densities are taken from \cite{Garcia,Wieczor,Matsumoto}. The radius of the Moon is $R_M = 1737$ km.

{\bf Crust}. The layer corresponds to distances $r$ from the Moon's center $R_{M,m}<r\leq R_M$ \cite{Wieczor}:
\begin{equation}
R_{M,m} = 1697\,\,{\rm km}, \qquad \rho_{M,c}=2.55\,\,{\rm g\,cm^{-3}}.
\end{equation}
At radius $R_{M,m}$ there is the lunar Moho.

{\bf Mantle}. The layer corresponds to $R_{M,n}<r\leq R_{M,m}$:
\begin{equation}
\rho_{M,m}=3.4\,\,{\rm g\,cm^{-3}}.
\end{equation}
{\bf Core}. The layer corresponds to $0<r\leq R_{M,n}$ \cite{Garcia,Matsumoto}:
\begin{equation}
R_{M,n} = 330\,\mbox{--}\,400\,\,{\rm km}, \qquad \rho_{M,n}=3.9\,\mbox{--}\,5.5\,\,{\rm g\,cm^{-3}}.
\end{equation}
Considerable uncertainty is connected with the radius and physical state of a metallic core.

In the Moon's layers, using again $\alpha=-1$ and $\lambda_g= 10^5$ AU, we have $\lambda(\rho_M)\lesssim 10^{-7} \,{\rm m}$.
\smallskip

\appendix
\section*{Appendix B: second order Green's function}

First we give the six second order images inside the screening spheres.
The position vectors of the two image points inside the Sun's screening sphere are given by
\begin{eqnarray}
\widehat\x_{SE} = \x_S + r_S^2\frac{\widetilde\x_E-\x_S}{\vert\widetilde\x_E-\x_S\vert^2}, \nonumber\\
\widehat\x_{SM} = \x_S + r_S^2\frac{\widetilde\x_M-\x_S}{\vert\widetilde\x_M-\x_S\vert^2},
\end{eqnarray}
where $\widehat\x_{SE},\widehat\x_{SM}$ are the images of the sources at $\widetilde\x_E,\widetilde\x_M$, respectively.
The other four second order images are obtained by changing the subscripts $S,E,M$.

The second order term of Green's function contains six terms and its expression is then given by
\begin{widetext}
\begin{eqnarray}
G^{(2)}(\x,\x_0) &=& -\frac{1}{4\pi} \left[ \frac{r_E}{\vert\x_E-\x_0\vert}\,\frac{r_S}{\vert\widetilde\x_E-\x_S\vert}\,\frac{1}{\vert\x-\widehat\x_{SE}\vert}
+ \frac{r_M}{\vert\x_M-\x_0\vert}\,\frac{r_S}{\vert\widetilde\x_M-\x_S\vert}\,\frac{1}{\vert\x-\widehat\x_{SM}\vert} \right. \nonumber\\
&+& \frac{r_S}{\vert\x_S-\x_0\vert}\,\frac{r_E}{\vert\widetilde\x_S-\x_E\vert}\,\frac{1}{\vert\x-\widehat\x_{ES}\vert}
+ \frac{r_M}{\vert\x_M-\x_0\vert}\,\frac{r_E}{\vert\widetilde\x_M-\x_E\vert}\,\frac{1}{\vert\x-\widehat\x_{EM}\vert} \\
&+& \left. \frac{r_S}{\vert\x_S-\x_0\vert}\,\frac{r_M}{\vert\widetilde\x_S-\x_M\vert}\,\frac{1}{\vert\x-\widehat\x_{MS}\vert}
+ \frac{r_E}{\vert\x_E-\x_0\vert}\,\frac{r_M}{\vert\widetilde\x_E-\x_M\vert}\,\frac{1}{\vert\x-\widehat\x_{ME}\vert} \right]. \nonumber
\end{eqnarray}
\end{widetext}
We now give the contribution $\eta_E^{(2)}(\x,t)$ from the Earth's thin shell to the solution $\eta(\x,t)$ obtained by adding the second order term of Green's function.
The following expression has to be added to formula (\ref{eta-green-Earth}). We drop the dependence on $(\x,t)$ for simplicity:
\begin{widetext}
\begin{eqnarray}\label{etaE-solut-order-2-a}
\eta_E^{(2)} &=& - \left( \frac{c^4}{8\pi G} - \eta_g \right) r_E \left[\frac{r_S}{\left\vert \vert\x-\x_S\vert(\x_S-\x_E)+r_S^2\hat\n_S\right\vert}
+ \frac{r_M}{\left\vert \vert\x-\x_M\vert(\x_M-\x_E)+r_M^2\hat\n_M\right\vert} \right] \nonumber\\
&+& \F_{E,ES} + \F_{E,SE} + \F_{E,EM} + \F_{E,ME} + \F_{E,SM} + \F_{E,MS},
\end{eqnarray}
\begin{equation}\label{etaE-solut-order-2-b}
\F_{E,ES} + \F_{E,SE} = \frac{c^2}{3}\,r_E r_S \left\{\frac{\int_{r_E}^{R_\oplus}\rho_E(r)r\,dr}{\left\vert \vert\x-\x_S\vert(\x_S-\x_E)+r_S^2\hat\n_S\right\vert} + 
\frac{M_{\oplus,{\rm eff}}(r_E)\slash 4\pi}{\left\vert r_E^2(\x_E-\x_S) + \vert\x-\x_E\vert\left(\vert\x_E-\x_S\vert^2-r_S^2\right)\hat\n_E \right\vert} \right\},
\end{equation}
\begin{eqnarray}
\F_{E,SM} + \F_{E,MS} &=& \frac{c^2}{12\pi}\,r_M r_S \left(A_{SM} + A_{MS}\right) M_{\oplus,{\rm eff}}(r_E), \nonumber\\
A_{SM} &=& \left\vert \vert\x-\x_M\vert \left[ \vert\x_S-\x_E\vert(\x_M-\x_S)+r_S^2\hat\n_{SE} \right] + r_M^2\vert\x_S-\x_E\vert\hat\n_M \right\vert^{-1},
\end{eqnarray}
\end{widetext}
where $\F_{E,EM} + \F_{E,ME}$ is obtained by replacing subscript $S$ with $M$ in Eq. (\ref{etaE-solut-order-2-b}),
$A_{MS}$ is obtained by exchanging subscripts $S$ and $M$ in the expression of $A_{SM}$, and $\hat\n_{SE}=(\x_S-\x_E)\slash\vert\x_S-\x_E\vert$. 

The first two terms in the expression (\ref{etaE-solut-order-2-a}) of $\eta_E^{(2)}$ are the contributions from the surface integrals over
Earth's screening surface; the terms $\F_{E,ES} + \F_{E,SE}$ in Eq. (\ref{etaE-solut-order-2-b}) are the volume integrals over the Earth's thin shell of the terms in $G^{(2)}(\x,\x_0)$
corresponding to the second order images inside Sun and Earth of the sources at $\widetilde\x_E$ and $\widetilde\x_S$, respectively; the other terms in $\eta_E^{(2)}$ have
analogous meanings. The contributions $\eta_S^{(2)},\eta_M^{(2)}$ from the Sun and Moon are found by exchanging the subscripts $S,E,M$.
The sum of all the resulting terms yield the second order correction $\eta^{(2)}$ to the solution $\eta$.

\appendix
\section*{Appendix C: verification of Dirichlet condition}

We compute the solution $\eta$ on $\partial\Omega$ by using the Green's function up to the second order.
We report the computation for the Earth's screening sphere, the results for the Sun and the Moon being analogous. 

Let us first consider the solution up to the first order and the contribution from the solar term $\eta^{(1)}_S$.
In Section \ref{sec:screen-radii} we have argued that $\I_S(\x,t)+\J_{S,E}(\x,t)=0$ on the Earth's screening sphere.
The remaining terms give the following contribution on the screening sphere, which is approximated at the leading order with respect to ratios $R\slash d$, where $R$ is a radius
and $d$ is a distance between the astronomical bodies:
\begin{eqnarray}\label{order1-eta_S-onEarth}
& &\eta^{(1)}_S(\x,t) \approx \left(\frac{c^4}{8\pi G}-\eta_g\right)\frac{r_S}{\vert\x-\x_S\vert} \\
&-& \frac{c^2}{3}\frac{r_S}{\vert\x-\x_S\vert}\int_{r_S}^{R_\odot}\rho_S(r)rdr - \frac{c^2}{12\pi}\frac{r_M}{\vert\x_M-\x_S\vert}\frac{M_{\odot,{\rm eff}}(r_S)}{\vert\x-\x_M\vert}. \nonumber
\end{eqnarray}
The first order contribution from the Moon is obtained by exchanging $S$ with $M$ in the previous expression.
The first order contribution from the Earth is given by
\begin{eqnarray}\label{order1-eta_E-onEarth}
\eta^{(1)}_E(\x,t) &\approx& \frac{c^4}{8\pi G}-\eta_g
- \frac{c^2}{12\pi} \frac{r_S}{\vert\x_S-\x_E\vert}\frac{M_{\oplus,{\rm eff}}(r_E)}{\vert\x-\x_S\vert} \nonumber\\
&-& \frac{c^2}{12\pi}\frac{r_M}{\vert\x_M-\x_E\vert}\frac{M_{\oplus,{\rm eff}}(r_E)}{\vert\x-\x_M\vert}.
\end{eqnarray}
By adding the contributions from the three astronomical bodies and $\eta_g$ according to formula (\ref{eta-green-total}), we find that the difference $\eta^{(1)}(\x,t)-c^4\slash(8\pi G)$,
between the first order solution and the Dirichlet datum (\ref{Dirichlet-bound-cond}) on Earth's screening sphere, contains terms multiplied by factors of the order of $R\slash d$.

We now consider the terms given by the Green's function up to the second order and we start again with the contribution from the solar term $\eta^{(2)}_S$.
The integral term in $\eta^{(1)}_S(\x,t)$ (which is exact),
\begin{equation}\label{eta_S-integral-term}
-\frac{c^2}{3}\,\frac{r_S}{\vert\x-\x_S\vert}\int_{r_S}^{R_\odot}\rho_S(r)r\,dr\,,
\end{equation}
obtained by replacing $E$ with $S$ in the last term of the solution (\ref{eta-green-Earth}) for $\eta_E$, is the volume integral over the Sun's thin shell of the term in $G^{(1)}(\x,\x_0)$
corresponding to the image point $\widetilde\x_S$, which is the image of $\x_0$ inside Sun's screening sphere. Now exchange $E$ with $S$ in the solution
(\ref{etaE-solut-order-2-a}-\ref{etaE-solut-order-2-b})
for $\eta_E^{(2)}$ of Appendix B, then consider the resulting term $\F_{S,SE}$ which turns out to be proportional to the integral in expression (\ref{eta_S-integral-term}).
The term $\F_{S,SE}$ is the volume integral over the Sun's shell of the term in $G^{(2)}(\x,\x_0)$ corresponding to the image of $\widetilde \x_S$ inside the Earth's screening sphere,
so that, by the properties of image points the sum of the integral term (\ref{eta_S-integral-term}) plus $\F_{S,SE}$ vanishes on the screening sphere. 

The surface term in $\eta_S^{(1)}$,
\begin{equation}\label{eta_S-surface-term}
\left( \frac{c^4}{8\pi G} - \eta_g \right) \frac{r_S}{\vert\x-\x_S\vert},
\end{equation}
obtained by replacing $E$ with $S$ in the first term of Eq. (\ref{eta-green-Earth}), is the surface integral over Sun's screening surface
of the term in $G^{(1)}(\x,\x_0)$ corresponding to the image point $\widetilde\x_S$. Now exchange $E$ with $S$ in Eq. (\ref{etaE-solut-order-2-a})
of Appendix B, then consider the first term in the square bracket which turns out to be proportional to the expression $c^4/(8\pi G)-\eta_g$.
This term is the surface integral of the term in $G^{(2)}(\x,\x_0)$ corresponding to the image of $\widetilde \x_S$ inside the Earth's screening sphere,
so that, the sum of this term plus the surface term (\ref{eta_S-surface-term}) again vanishes on the screening sphere.

The term $\J_{S,M}$ in $\eta_S^{(1)}$ (which is approximated on Earth's screening surface by the last term in Eq. (\ref{order1-eta_S-onEarth})),
is the volume integral over the Sun's thin shell of the term in $G^{(1)}(\x,\x_0)$ corresponding to the image point $\widetilde\x_M$. Then exchange again $E$ with $S$ in the solution
for $\eta_E^{(2)}$ and consider the resulting term $\F_{S,ME}$ which is the volume integral over the Sun's shell of the term in $G^{(2)}(\x,\x_0)$ corresponding 
to the image of $\widetilde \x_M$ inside the Earth's screening sphere, then one can check that the sum of $\J_{S,M}$ plus $\F_{S,ME}$ vanishes on the screening sphere.

Hence, all the first order terms in $\eta_S^{(1)}$ cancel with some second order terms in $\eta_S^{(2)}$ and the remaining terms are approximated on Earth's screening
surface as follows:
\begin{widetext}
\begin{eqnarray}
\eta^{(1)}_S(\x,t)+\eta^{(2)}_S(\x,t) &\approx& \frac{c^2}{12\pi}\left(\frac{r_S r_E}{\vert\x_S-\x_E\vert^2}\frac{1}{\vert\x-\x_S\vert}
+ \frac{r_E r_M}{\vert\x_E-\x_S\vert\cdot\vert\x_M-\x_E\vert}\frac{1}{\vert\x-\x_M\vert}
+ \frac{r_S r_M}{\vert\x_S-\x_M\vert^2}\frac{1}{\vert\x-\x_S\vert}\right)M_{\odot,{\rm eff}}(r_S) \nonumber\\
&+& \frac{r_M r_S}{\vert\x_M-\x_S\vert\cdot\vert\x-\x_M\vert}\left( \frac{c^2}{3}\int_{r_S}^{R_\odot}\rho_S(r)rdr-\frac{c^4}{8\pi G}+\eta_g \right).
\end{eqnarray}
\end{widetext}

The contribution from the lunar term $\eta_M$ is obtained by replacing $S$ with $M$ in the previous expressions.

Now we consider the second order contribution from Earth.
The terms $\J_{E,S}$ and $\J_{E,M}$ in $\eta_E^{(1)}$ (which are approximated on Earth's screening surface by the last two terms in Eq. (\ref{order1-eta_E-onEarth})),
are the volume integrals over the Earth's thin shell of the terms in $G^{(1)}(\x,\x_0)$ corresponding to the image points $\widetilde\x_S$ and $\widetilde\x_M$, respectively.
Then consider the term $\F_{E,SE}$ in $\eta_E^{(2)}$ which is the volume integral over the Earth's shell of the term in $G^{(2)}(\x,\x_0)$ corresponding 
to the image of $\widetilde \x_S$ inside the Earth's screening sphere, then one can check that the sum of $\J_{E,S}$ plus $\F_{E,SE}$ vanishes on the screening sphere.
Analogously, the sum of $\J_{E,M}$ plus $\F_{E,ME}$ vanishes on the screening sphere.

Hence, all the first order terms in $\eta_E^{(1)}$, with the exception of $c^4/(8\pi G)-\eta_g$,
cancel with some second order terms in $\eta_E^{(2)}$ and the remaining terms are approximated on Earth's screening surface as follows:
\begin{widetext}
\begin{eqnarray}
\eta^{(1)}_E(\x,t)+\eta^{(2)}_E(\x,t) &\approx& \frac{c^4}{8\pi G} - \eta_g + \frac{c^2}{3}\left(\frac{r_S}{\vert\x_S-\x_E\vert}\frac{r_E}{\vert\x-\x_S\vert}
+\frac{r_M}{\vert\x_M-\x_E\vert}\frac{r_E}{\vert\x-\x_M\vert}\right)\int_{r_E}^{R_\oplus}\rho_E(r)rdr \nonumber\\
&+& \frac{c^2}{12\pi}\left( \frac{r_M r_S}{\vert\x_S-\x_E\vert\cdot\vert\x_M-\x_S\vert}\frac{1}{\vert\x-\x_M\vert}
+\frac{r_S r_M}{\vert\x_M-\x_E\vert\cdot\vert\x_S-\x_M\vert}\frac{1}{\vert\x-\x_S\vert}\right)M_{\oplus,{\rm eff}}(r_E) \nonumber\\
&-& \left(\frac{r_E r_S}{\vert\x_S-\x_E\vert\cdot\vert\x-\x_S\vert} + \frac{r_E r_M}{\vert\x_M-\x_E\vert\cdot\vert\x-\x_M\vert}\right)\left(\frac{c^4}{8\pi G}-\eta_g\right).
\end{eqnarray}
\end{widetext}
If we compare the above result with the first order computation we find that the difference
\begin{equation}
\eta^{(1)}(\x,t)+\eta^{(2)}(\x,t)-c^4\slash(8\pi G),
\end{equation}
between the solution computed up to second order and the Dirichlet datum (\ref{Dirichlet-bound-cond}) on Earth's screening sphere, contains terms multiplied by factors
of order of $(R\slash d)^2$. Since the ratios of the type of $R\slash d$ are small, it then follows that the approximation of the Dirichlet condition improves
by increasing the number of image points.

\appendix
\section*{Appendix D: fifth force and extra force at second order}

We discuss the contributions to fifth force and extra force resulting from the solution $\eta$ computed by resorting to the second order Green's function $G^{(2)}(\x,\x_0)$.

\subsection{Fifth force}

We consider the contributions from the solar term $\eta_S$. By exchanging $E$ with $S$ in the second order solution
(\ref{etaE-solut-order-2-a}-\ref{etaE-solut-order-2-b}) for $\eta_E$ of Appendix B, we observe that the terms
\begin{equation}
\frac{c^4}{8\pi G} - \eta_g - \frac{c^2}{3}\int_{r_S}^{R_\odot}\rho_S(r)r\,dr
\end{equation}
appear in the second order solution multiplied by a geometric factor of the same form of $\J_{S,E}$. Hence, the contribution of such terms 
to the fifth force is proportional to
\begin{equation}
\left(\frac{c^4}{8\pi G} - \eta_g - \frac{c^2}{3}\int_{r_S}^{R_\odot}\rho_S(r)r\,dr\right)\int_{\VV_E(t)} \rho_E \nabla\J_{S,E} \,d^3x,
\end{equation}
which vanishes because of formula (\ref{int-J_SE-vanish}). Furthermore, using also the integral equation on the Sun's screening sphere,
we find that the sum of the contribution from $\nabla\J_{S,M}$ plus all other second order terms in $\nabla\eta_S$ gives contributions multiplied by factors $R\slash d$.
That completes the contributions from $\eta_S$.

The contribution from the lunar term $\eta_M$ is obtained by replacing $S$ with $M$ in the previous results.

Now we consider the second order solution Eq. (\ref{etaE-solut-order-2-a}) for $\eta_E$ of Appendix B. 
By using the integral equation (\ref{radii-integral-equations}) on the
Earth's screening sphere, we observe that the terms in the first row of Eq. (\ref{etaE-solut-order-2-a})
cancel with the terms $\F_{E,ES}+\F_{E,EM}$ in the second row, given in Eq. (\ref{etaE-solut-order-2-b}), except for $\OO(R\slash d)$.
Furthermore, we find that the sum of the contribution from $\nabla\J_E$ plus all other second order terms in $\nabla\eta_E$ gives contributions multiplied by factors $R\slash d$.

Eventually, by using the thin shell assumption for all bodies and the approximate relations (\ref{effective-masses-relat}) between effective masses, 
it turns out that all contributions to the fifth force of the order of $R\slash d$ cancel each other.

\subsection{Extra force}

We compute the leading terms in the Taylor approximation (\ref{eta-difference-Taylor-expand}). Let us first consider the contribution from the solar term $\eta_S$.
We have $\I_S(\x,t)+\J_{S,E}(\x,t)=0$ on the Earth's screening sphere.
By exchanging $E$ with $S$ in the second order solution (\ref{etaE-solut-order-2-a}-\ref{etaE-solut-order-2-b}) for $\eta_E$ of Appendix B, we obtain the approximation $\eta_S^{(2)}$
and we observe that the terms in the sum
\begin{equation}
-\frac{c^2}{3}\,\frac{r_S}{\vert\x-\x_S\vert}\int_{r_S}^{R_\odot}\rho_S(r)r\,dr + \F_{S,SE}(\x,t)
\end{equation}
have the same geometric factors of $\I_S(\x,t)+\J_{S,E}(\x,t)$. The surface term (\ref{eta_S-surface-term}) and the corresponding term in $\eta_S^{(2)}$
identified in Appendix C have the same property. Then all these sums have a common second order Taylor approximation on the external surface of Earth with respect to
$\delta_E$ and $\eps_1$, and the contribution of this Taylor approximation to $-\eta_S\slash c^2$ is given by
\begin{eqnarray}
&-&\frac{\delta_E}{\vert\x_E-\x_S\vert}\left( 1-3\eps_1\hat\n_{ES}\cdot\hat\n_E \right) \\
&\times&\left[ \frac{M_{\odot,{\rm eff}}}{12\pi} - \frac{r_S}{3}\int_{r_S}^{R_\odot}\rho_S(r)dr + r_S\left(\frac{c^2}{8\pi G}-\frac{\eta_g}{c^2}\right) \right]. \nonumber
\end{eqnarray}
The contribution from the lunar term $\eta_M$ is obtained by replacing $S$ with $M$ in the previous expression.

Using then Eqs. (\ref{eta-green-total})-(\ref{eta-green-Earth}) and (\ref{integral-eff-mass-deltaE}),
the radial terms of $(c^2\slash(8\pi G)-(\eta_E+\eta_g)\slash c^2)$, which depend on $\vert\x-\x_E\vert$, evaluated on Earth's external surface, yield
\begin{equation}
\delta_E \left( \frac{M_{\oplus,{\rm eff}}}{24\pi R_\oplus} - \frac{1}{3}\int_{r_E}^{R_\oplus}\rho_E(r)rdr + \frac{c^2}{8\pi G}-\frac{\eta_g}{c^2} \right).
\end{equation}
Adding all above terms from $\eta_S,\eta_M$ and $\eta_E$, and eliminating $(c^2\slash{8\pi G}-\eta_g\slash c^2)$ from the Earth's contribution
by means of the integral equation (\ref{radii-integral-equations}) on the Earth's screening sphere, we obtain
\begin{eqnarray}
& &\frac{c^2}{8\pi G} - \frac{1}{c^2}\eta(R_\oplus,\theta,\varphi,t) \approx \delta_E \left\{  \frac{M_{\oplus,{\rm eff}}}{24\pi R_\oplus}
+ \frac{\hat\n_{ES}\cdot\hat\n_E}{\vert\x_E-\x_S\vert} \right. \nonumber\\
&\times& 3\eps_1 \left[ \frac{M_{\odot,{\rm eff}}}{12\pi} - \frac{r_S}{3}\int_{r_S}^{R_\odot}\rho_S(r)dr + r_S\left(\frac{c^2}{8\pi G}-\frac{\eta_g}{c^2}\right) \right] \nonumber \\
&+& \left. \text{lunar terms} \right\},
\end{eqnarray}
where the lunar terms are obtained by replacing $S$ with $M$ in the terms depending on solar quantities.
Eventually, substituting Eq. (\ref{Sun-integr-eq-ORd}) we obtain the leading terms of the approximation (\ref{eta-difference-Taylor-expand}).

\appendix
\section*{Appendix E: thin shell inequality}

We give a proof {\it a posteriori} of inequality (\ref{deriv-potential-condit}) that has been used in order to compute the solution for function $\eta$.
Particularly, we prove that such a solution is consistent with the inequality for $\alpha<-1\slash 2$. We give the proof for the Earth's thin shell, the case of the Moon
being similar, while the case of the Sun has been proved in Ref. \cite{{MBMGDeA}}. Since $\eta$ is a solution of Eq. (\ref{trace-approx}),
we have to prove that for $r>r_E$ and $r$ close to $r_E$, the solution $\eta$ in the thin shell is such that curvature $R=\omega(\eta,\rho)$
quickly becomes much smaller than the GR curvature, so that inequality (\ref{deriv-potential-condit}) is verified a posteriori.

We use formula (\ref{curvR-as-fraction}) for curvature which we write in the form
\begin{equation}\label{TS-curvature-R}
R = (2\alpha q\rho)^{\frac{1}{1+\vert\alpha\vert}}\left(\frac{c^2}{8\pi G}-\frac{\eta}{c^2}\right)^{-\frac{1}{1+\vert\alpha\vert}}.
\end{equation}
The term $(c^2\slash{8\pi G}-\eta\slash c^2)$ is computed on the sphere of radius $r$, with $r_E<r<R_\oplus$, with the same method used to compute the
same expression for $r=R_\oplus$ in formula (\ref{eta-difference-Taylor-expand}) and reported in Appendix D. The leading term depends on $r$ and it is given by
\begin{equation}\label{eta-diff-Taylor-r}
\frac{c^2}{8\pi G} - \frac{1}{c^2}\eta(r) \approx \frac{\delta_r}{6r}\int_{r_E}^r \rho_E(r^\prime)(r^\prime)^2 dr^\prime,
\end{equation}
where
\begin{equation}\label{delta_r-delta_E}
\delta_r = \frac{r-r_E}{r} = \frac{1-r_E\slash r}{1-r_E\slash R_\oplus}\,\delta_E.
\end{equation}
Now, using the thin shell approximation so that the Earth's screening surface lies in the ocean layer, we have
\begin{equation}
\int_{r_E}^{R_\oplus}\rho_E(r)rdr = \frac{1}{2}\rho_{E,w}\left(R_\oplus^2-r_E^2\right) \approx \rho_{E,w}R_\oplus^2\delta_E.
\end{equation}
Using then the integral equation (\ref{radii-integral-equations}), substituting the expression (\ref{eta-g-minimizer}) of $\eta_g$ into the integral equation, 
and neglecting factors involving radius-to-distance ratios in the integral equation, we find the approximation
\begin{equation}\label{TS-delta-E-approx}
\delta_E \approx \frac{6\alpha q}{\rho_{E,w}R_\oplus^2}\left(\frac{8\pi G}{c^2}\right)^{\alpha-1}\rho_g^\alpha.
\end{equation}
Using again the thin shell approximation we have
\begin{equation}\label{M_E(r)-eff-approx}
\frac{1}{6r}\int_{r_E}^r \rho_E(r^\prime)(r^\prime)^2 dr^\prime \approx \frac{1}{6}\rho_{E,w}r^2 \delta r.
\end{equation}
Moreover, using definition (\ref{lambda-expression}) of $\lambda$, we have
\begin{equation}\label{TS-alpha-q-approx}
\alpha q = \frac{\lambda_g^2}{6(1+\vert\alpha\vert)}\left(\frac{8\pi G}{c^2}\rho_g\right)^{1+\vert\alpha\vert}.
\end{equation}
Substituting formulae (\ref{eta-diff-Taylor-r})-(\ref{delta_r-delta_E}) and (\ref{TS-delta-E-approx}-\ref{TS-alpha-q-approx}) in expression (\ref{TS-curvature-R}) of curvature, we find
for $r_E<r<R_\oplus$,
\begin{eqnarray}
R &\approx& \left[2(1+\vert\alpha\vert)\right]^{\frac{1}{1+\vert\alpha\vert}}\left(\frac{R_\oplus-r_E}{r-r_E}\,\frac{R_\oplus}{\lambda_g}\right)^{\frac{2}{1+\vert\alpha\vert}} \nonumber\\
&\times& \left(\frac{\rho_g}{\rho_{E,w}}\right)^{\frac{\vert\alpha\vert-1}{\vert\alpha\vert+1}}R_{\rm GR},
\end{eqnarray}
where $R_{\rm GR}= (8\pi G\slash c^2)\rho_{E,w}$ is the value of GR curvature in the ocean layer. The expression of curvature diverges as $r\rightarrow r_E$, which means
that the approximation is not valid in this limit. Nevertheless, there exist $\alpha_0$ and $r^\ast\in(r_E,R_\oplus)$, with $r^\ast=r^\ast(\alpha)$ close to $r_E$,
such that for any $\alpha<\alpha_0$ and any $r\in(r^\ast,R_\oplus)$ we have $R\ll R_{\rm GR}$, so that the desired inequality is verified in almost all the thin shell.
In the following we give a sample of values.

Let us set $\lambda_g=10^2 r_g$. We have $R<10^{-3}R_{\rm GR}$
\begin{eqnarray}
& &\mbox{for}\quad \alpha=-2\quad \mbox{and}\quad r^\ast-r_E=10^{-15}(R_\oplus-r_E), \nonumber\\
& &\mbox{for}\quad \alpha=-1\quad \mbox{and}\quad r^\ast-r_E=10^{-5}(R_\oplus-r_E). \nonumber
\end{eqnarray}
We have $R<R_{\rm GR}\slash 20$
\begin{equation}
\mbox{for}\quad  \alpha=-1\slash 2\quad \mbox{and}\quad r^\ast-r_E=3.6\times 10^{-2}(R_\oplus-r_E). \nonumber
\end{equation}
If we increase $\lambda_g$ by setting $\lambda_g=7.07\times 10^2 r_g$, which is the upper bound permitted by the LLR constraint for $\alpha=-1\slash 3$, 
then we have $R<10^{-1}R_{\rm GR}$
\begin{equation}
\mbox{for}\quad  \alpha=-1\slash 3\quad \mbox{and}\quad r^\ast-r_E=0.2(R_\oplus-r_E). \nonumber
\end{equation}
We see that $r^\ast$ increases as $\alpha$ increases, and we consider the inequality verified for $\alpha<-1\slash 2$.

\subsection{Fifth force in the thin shell}\label{F-force-t-shell}

The fifth force density is given by Eq. (\ref{F-force-density}) and the leading term in Earth's thin shell is computed by using Eqs. (\ref{eta-diff-Taylor-r}) and (\ref{M_E(r)-eff-approx}):
for $r^\ast(\alpha)<r<R_\oplus$ the fifth force density is given by
\begin{equation}
F_{\rm f}(r) \approx -\frac{4}{3}\pi G(r-r_E)\left(\rho_{E,w}\frac{R_\oplus\delta_E}{R_\oplus-r_E}\right)^2,
\end{equation}
and it achieves maximum magnitude at $r=R_\oplus$. The Newtonian force density at $r=R_\oplus$ has magnitude
$\vert F_{\rm N}\vert = (4/3)\pi G\rho_{E,w}\langle\rho_\oplus\rangle R_\oplus$, where $\langle\rho_\oplus\rangle$ is Earth's mean density, so that the ratio of magnitude is
\begin{equation}
\frac{\vert F_{\rm f}(R_\oplus)\vert}{\vert F_{\rm N}(R_\oplus)\vert} \approx \frac{\rho_{E,w}}{\langle\rho_\oplus\rangle}\delta_E.
\end{equation}
In Section \ref{sec:LLR-constraints} we have found the upper bound $\delta_E<2\cdot 10^{-5}$ for $\vert\alpha\vert >1/2$ which corresponds to the value $3.6\cdot 10^{-6}$
for the ratio of magnitude of fifth force and Newtonian force. Since Earth's thin shell is contained in seawater, then
the fifth force is negligible (see the ocean experiment in \cite{Zum}) in the thin shell for $\alpha<-1/2$ and $r>r^\ast(\alpha)$.

For values of $r$ such that $r_E<r<r^\ast(\alpha)$ the above approximation of $\eta$ cannot be used, nevertheless, the radial integral of fifth force density can be evaluated
by using the boundary conditions. Using the Dirichlet condition $\eta(r_E)\approx c^4\slash(8\pi G)$, and considering the leading term of $\eta(r^\ast)$, we find
\begin{eqnarray}
\int_{r_E}^{r^\ast}F_{\rm f}(r)dr &\approx& \frac{4\pi G}{c^2}\rho_{E,w}\left[\eta(r^\ast)-\eta(r_E)\right] \nonumber\\
&=& -\frac{2}{3}\pi G\rho^2_{E,w}\left(r^\ast-r_E\right)^2.
\end{eqnarray}
The evaluation of the corresponding integral of Newtonian force density yields
\begin{equation}\label{radial-integr-Newt}
\int_{r_E}^{r^\ast}F_{\rm N}(r)dr \approx -\frac{4}{3}\pi G\rho^2_{E,w}\left(3+\frac{\langle\rho_\oplus\rangle}{\rho_{E,w}}\right)r_E\left(r^\ast-r_E\right).
\end{equation}
Since $3+\langle\rho_\oplus\rangle\slash\rho_{E,w}> 8$, the ratio of the integral of fifth force and the integral of Newtonian force is bounded by the ratio
\begin{equation}
\frac{r^\ast-r_E}{r_E}=\frac{r^\ast-r_E}{R_\oplus-r_E}\frac{R_\oplus}{r_E}\delta_E,
\end{equation}
so that, by using the values of $r^\ast-r_E$ found in the previous subsection and $\delta_E<2\cdot 10^{-5}$, such a ratio is less than $10^{-6}$ for $\alpha<-1/2$ and
the contribution of the fifth force turns out to be again negligible. 

Analogous results can be found for the Moon.

\subsection{Extra force in the thin shell}\label{Extra-force-t-shell}

Using formula (\ref{curvR-as-fraction}) for curvature $R$ and taking into account that density is constant in the thin shell of both Earth and Moon, 
we find for the extra force density $F_{\rm e}=-\rho c^2 f^2_R\nabla R$ the expression
\begin{eqnarray}
F_{\rm e} &=& -\frac{1}{2}\frac{R}{1+\vert\alpha\vert}\nabla\eta = -\frac{R}{(8\pi G\slash c^2)\rho}\frac{F_{\rm f}}{1+\vert\alpha\vert}, \nonumber\\
&=& -\frac{R}{R_{\rm GR}}\frac{F_{\rm f}}{1+\vert\alpha\vert}.
\end{eqnarray}
Since we have $R\ll R_{\rm GR}$ for $r^\ast(\alpha)<r<R_\oplus$ it follows $\vert F_{\rm e}\vert \ll \vert F_{\rm f}\vert$ for such values of radius in Earth,
so that the extra force is also negligible in the thin shell.

For values of $r$ such that $r_E<r<r^\ast(\alpha)$ the radial integral of the extra force density can be evaluated by using the boundary conditions:
\begin{eqnarray}
\int_{r_E}^{r^\ast}F_{\rm e}(r)dr &=& -\rho_{E,w}c^2\int_{r_E}^{r^\ast}f^2_R\frac{dR}{dr}dr \\
&=& \rho_{E,w}c^2\vert\widetilde q\vert\left[\left(\frac{R(r^\ast)}{R_g}\right)^\alpha-\left(\frac{R(r_E)}{R_g}\right)^\alpha\right], \nonumber
\end{eqnarray}
with $\widetilde q=q R_g^\alpha$. Using formula (\ref{TS-delta-E-approx}) we have
\begin{equation}
c^2\vert\widetilde q\vert \approx \frac{4}{3\vert\alpha\vert}\pi G \rho_{E,w}R_\oplus^2\delta_E,
\end{equation}
from which, being $R(r^\ast)\ll R_{\rm GR}(r^\ast)=R(r_E)$ for the boundary condition on Earth's screening sphere, and $\alpha<0$, we have the inequality
\begin{equation}
\int_{r_E}^{r^\ast}F_{\rm e}(r)dr < \frac{4}{3\vert\alpha\vert}\pi G \rho^2_{E,w}R_\oplus^2\delta_E\left(\frac{R(r^\ast)}{R_g}\right)^\alpha.
\end{equation}
Using now the integral (\ref{radial-integr-Newt}), the absolute value of the ratio of the integral of the extra force and the integral of Newtonian force is bounded by the ratio
\begin{equation}
\frac{R_\oplus^2\delta_E}{\vert\alpha\vert r_E(3+\langle\rho_\oplus\rangle/\rho_{E,w})}\frac{1}{r^\ast-r_E}\left(\frac{R(r^\ast)}{R_g}\right)^\alpha.
\end{equation}
The ratio increases by decreasing $\vert\alpha\vert$ so that it is bounded by the value achieved for $\alpha=-1/2$ and, 
using the values of $r^\ast=r^\ast(\alpha)$ which satisfy the thin shell inequality, we find that the above ratio is less than $10^{-10}$, hence the contribution of the extra force
turns out to be again negligible. 

Analogous results can be found for the Moon.



\end{document}